\documentclass[acmsmall,authorversion]{acmart}
\usepackage{xspace}
\usepackage{url}
\usepackage{listings}
\usepackage[shortcuts]{glossaries-extra}

\definecolor{codegreen}{rgb}{0,0.6,0}
\definecolor{codegray}{rgb}{0.5,0.5,0.5}
\definecolor{codeblue}{rgb}{0.2,0.2,0.8}
\definecolor{backcolour}{rgb}{1,1,1}

\lstdefinestyle{mystyle}{
    backgroundcolor=\color{backcolour},   
    commentstyle=\color{codegray},
    keywordstyle=\color{codeblue},
    numberstyle=\tiny\color{codegray},
    stringstyle=\color{codegreen},
    basicstyle=\ttfamily\footnotesize,
    breakatwhitespace=false,         
    breaklines=true,                 
    captionpos=b,                    
    keepspaces=true,                 
    numbers=left,             
    numbersep=5pt,                  
    showspaces=false,                
    showstringspaces=false,
    showtabs=false,                  
    tabsize=2
}

\AtBeginDocument{%
  \providecommand\BibTeX{{%
    \normalfont B\kern-0.5em{\scshape i\kern-0.25em b}\kern-0.8em\TeX}}}

\setcopyright{cc}
\setcctype{by-nc}
\acmDOI{10.1145/3715762}
\acmYear{2025}
\acmJournal{PACMSE}
\acmVolume{2}
\acmNumber{FSE}
\acmArticle{FSE045}
\acmMonth{7}
\received{2024-09-13}
\received[accepted]{2025-01-14}

\title{How Do Programming Students Use Generative AI?}

\author{Christian Rahe}
\email{christian.rahe@uni-hamburg.de}
\orcid{0009-0007-2110-6663}
\affiliation{%
  \institution{University of Hamburg}
  \city{Hamburg}
 \state{Hamburg}
 \country{Germany}
}

\author{Walid Maalej}
\email{walid.maalej@uni-hamburg.de}
\orcid{0000-0002-6899-4393}
\affiliation{%
 \institution{University of Hamburg}
 \city{Hamburg}
 \state{Hamburg}
 \country{Germany}
}

\begin{document}



\newcommand{\rqeval}{RQ1\xspace}
\newcommand{\rqstrat}{RQ2\xspace}
\newcommand{\rqauto}{RQ3\xspace}


\newcommand{\exfind}{P1\xspace}
\newcommand{\exfix}{P2\xspace}


\newcommand{\mcts}{CallsTS\xspace}
\newcommand{\mcds}{CallsDS\xspace}
\newcommand{\mcin}{CallsInstance\xspace}
\newcommand{\mcon}{NewInstance\xspace}
\newcommand{\mpna}{PassNullArg\xspace}
\newcommand{\mnp}{NoParams\xspace}
\newcommand{\mnt}{NoThrow\xspace}
\newcommand{\mnc}{NoTryCatch\xspace}
\newcommand{\mnf}{NoFieldAccess\xspace}
\newcommand{\mdn}{NoNullDeref\xspace}
\newcommand{\mnpNeg}{$\overline{\text{NoParams}}$\xspace}
\newcommand{\mntNeg}{$\overline{\text{NoThrow}}$\xspace}
\newcommand{\mncNeg}{$\overline{\text{NoTryCatch}}$\xspace}
\newcommand{\mnfNeg}{$\overline{\text{NoFieldAccess}}$\xspace}
\newcommand{\mdnNeg}{$\overline{\text{NoNullDeref}}$\xspace}

\newcommand{\mnch}{NullCheck\xspace}
\newcommand{\mncp}{NullCheckOnParam\xspace}
\newcommand{\mmtb}{MatchingThen\xspace}
\newcommand{\mmtbNeg}{$\overline{\text{MatchingThen}}$\xspace}
\newcommand{\msco}{ShortCircuitOrder\xspace}
\newcommand{\mscoNeg}{$\overline{\text{ShortCircuitOrder}}$\xspace}


\newabbreviation{chatters}{CU}{chatbot-using participant}
\newcommand{\chatters}{\glspl{chatters}\xspace}

\newcommand{\syntax}{Syntax\xspace}
\newcommand{\logic}{Logic\xspace}
\newcommand{\intent}{Intent\xspace}
\newcommand{\reqs}{Reqs\xspace}
\newcommand{\conventions}{Style\xspace}
\newcommand{\curric}{Learn\xspace}
\newcommand{\fact}{Facts\xspace}
\newcommand{\answers}{Answers\xspace}


\newcommand{\application}{Application\xspace}
\newcommand{\comprehension}{Comprehension\xspace}
\newcommand{\transfer}{Transfer\xspace}
\newcommand{\analysis}{Analysis\xspace}
\newcommand{\knowledge}{Knowledge\xspace}

\newcommand{\result}[5][]{${#3}_{g#4}{#5}$}


\newabbreviation{uhh}{UHH}{Universit{\"a}t Hamburg}
\newcommand{\uhh}{\gls{uhh}\xspace}
\newabbreviation{se1}{SE1}{\textit{Softwareentwicklung 1}}
\newcommand{\se}{\gls{se1}\xspace}

\newabbreviation{llm}{LLM}{large language model}
\newcommand{\llm}{\gls{llm}\xspace}
\newcommand{\llms}{\glspl{llm}\xspace}

\newabbreviation{gai}{GenAI}{Generative AI}
\newcommand{\gai}{\gls{gai}\xspace}

\newabbreviation{rlhf}{RLHF}{Reinforcement Learning from Human Feedback}
\newcommand{\rlhf}{\gls{rlhf}\xspace}

\newabbreviation{ast}{AST}{abstract syntax tree}
\newcommand{\astree}{\gls{ast}\xspace}

\newabbreviation{npe}{NPE}{\texttt{NullPointerException}}
\newcommand{\npe}{\gls{npe}\xspace}
\newcommand{\ui}{UI\xspace}
\newabbreviation{nlp}{NLP}{natural language processing}
\newcommand{\nlp}{\gls{nlp}\xspace}
\newabbreviation{sota}{SOTA}{state-of-the-art}
\newcommand{\sota}{\gls{sota}\xspace}

\newcommand{\gpt}{GPT\xspace}
\newcommand{\gptv}[1]{\gpt-{#1}\xspace}
\newcommand{\gptt}{\gptv{3.5}}
\newcommand{\gptf}{\gptv{4}}

\newcommand{\cgpt}{ChatGPT\xspace}

\newcommand{\bluej}{BlueJ\xspace}
\newcommand{\moodle}{Moodle\xspace}
\newcommand{\coderunner}{CodeRunner\xspace}
\newcommand{\pdf}{PDF\xspace}


\newabbreviation{compcount}{Sub}{Submission Count}
\newcommand{\compcount}{\gls{compcount}\xspace}
\newabbreviation{chatcount}{Msg}{Message Count}
\newcommand{\chatcount}{\gls{chatcount}\xspace}
\newabbreviation{comperrs}{CER}{Compilation Error Rate}
\newcommand{\comperrs}{\gls{comperrs}\xspace}
\newabbreviation{testfail}{TFR}{Test Case Failure Rate}
\newcommand{\testfail}{\gls{testfail}\xspace}
\newabbreviation{ttfcomp}{TFS}{Time to First Submission}
\newcommand{\ttfcomp}{\gls{ttfcomp}\xspace}
\newabbreviation{tttpass}{TTPS}{Total Time to Passing Submission}
\newcommand{\tttpass}{\gls{tttpass}\xspace}
\newabbreviation{ttpass}{TPS}{Time to Passing Submission}
\newcommand{\ttpass}{\gls{ttpass}\xspace}
\newabbreviation{ttfchat}{TFM}{Time to First Message}
\newcommand{\ttfchat}{\gls{ttfchat}\xspace}
\newabbreviation{gssim}{GSSim}{Generation-Submission Similarity}
\newcommand{\gssim}{\gls{gssim}\xspace}
\newabbreviation{qpsim}{QPSim}{Question-Prompt Similarity}
\newcommand{\qpsim}{\gls{qpsim}\xspace}
\newabbreviation{gscopy}{GSCopy}{Generation-Submission Copy-Paste Events}
\newcommand{\gscopy}{\gls{gscopy}\xspace}
\newabbreviation{qpcopy}{QPCopy}{Question-Prompt Copy-Paste Events}
\newcommand{\qpcopy}{\gls{qpcopy}\xspace}


\newcommand{\plookup}{Inform\xspace}
\newcommand{\psolve}{Solve\xspace}
\newcommand{\pcompr}{Explain\xspace}
\newcommand{\pfix}{Fix\xspace}
\newcommand{\phint}{Hint\xspace}
\newcommand{\pothers}{Others\xspace}

\newcommand{\pgen}{Codegen\xspace}
\newcommand{\phelp}{Support\xspace}

\newcommand{\rfull}{Copy All\xspace}
\newcommand{\rsyntax}{Syntax\xspace}
\newcommand{\ridea}{Idea\xspace}
\newcommand{\rexplain}{Explanation\xspace}
\newcommand{\rnone}{None\xspace}


\newcommand{\dtooeasy}{too easy\xspace}
\newcommand{\deasy}{easy\xspace}
\newcommand{\dright}{fitting\xspace}
\newcommand{\dhard}{difficult\xspace}
\newcommand{\dtoohard}{too difficult\xspace}


\newcommand{\unever}{never\xspace}
\newcommand{\uspast}{past only\xspace}
\newcommand{\upast}{only in the past, not anymore\xspace}
\newcommand{\uslessmonthly}{$<$ m\xspace}
\newcommand{\ulessmonthly}{less than once a month\xspace}
\newcommand{\uslessweekly}{m – w\xspace}
\newcommand{\ulessweekly}{monthly through weekly\xspace}
\newcommand{\usweekly}{$>$ w\xspace}
\newcommand{\uweekly}{more than once a week\xspace}


\newcommand{\eg}{\textit{e.g.}\xspace}
\newcommand{\ie}{\textit{i.e.}\xspace}
\newcommand{\code}[1]{\texttt{#1}}

\newcommand{\id}[1]{{\small #1}}
\newcommand{\msg}[1]{[{\small #1}]}

\newcommand{\avgvals}[4]{{#1} {#4}\xspace(M = {#2}, $\sigma$ = {#3})}
\newcommand{\dataExpChatPresenceTotal}{37\xspace}
\newcommand{\dataExpChatPresenceTotalChatters}{23\xspace}
\newcommand{\dataExpChatPresenceRatio}{62.2\%\xspace}
\newcommand{\dataExpChatPresenceTotalOne}{23\xspace}
\newcommand{\dataExpChatPresenceTotalRatio}{62.2\%\xspace}
\newcommand{\dataExpChatPresenceTotalRatioChatters}{100\%\xspace}
\newcommand{\dataExpChatPresenceTotalReoccurrenceMean}{7.35\xspace}
\newcommand{\dataExpChatPresenceTotalReoccurrenceMedian}{6\xspace}
\newcommand{\dataExpChatPresenceTotalReoccurrenceStd}{6.06\xspace}
\newcommand{\dataExpChatPresenceLookup}{15\xspace}
\newcommand{\dataExpChatPresenceLookupRatio}{40.5\%\xspace}
\newcommand{\dataExpChatPresenceLookupRatioChatters}{65.2\%\xspace}
\newcommand{\dataExpChatPresenceLookupReoccurrenceMean}{2.13\xspace}
\newcommand{\dataExpChatPresenceLookupReoccurrenceMedian}{1\xspace}
\newcommand{\dataExpChatPresenceLookupReoccurrenceStd}{2.1\xspace}
\newcommand{\dataExpChatPresenceHint}{8\xspace}
\newcommand{\dataExpChatPresenceHintRatio}{21.6\%\xspace}
\newcommand{\dataExpChatPresenceHintRatioChatters}{34.8\%\xspace}
\newcommand{\dataExpChatPresenceHintReoccurrenceMean}{1.63\xspace}
\newcommand{\dataExpChatPresenceHintReoccurrenceMedian}{1\xspace}
\newcommand{\dataExpChatPresenceHintReoccurrenceStd}{0.74\xspace}
\newcommand{\dataExpChatPresenceOthers}{11\xspace}
\newcommand{\dataExpChatPresenceOthersRatio}{29.7\%\xspace}
\newcommand{\dataExpChatPresenceOthersRatioChatters}{47.8\%\xspace}
\newcommand{\dataExpChatPresenceOthersReoccurrenceMean}{2.36\xspace}
\newcommand{\dataExpChatPresenceOthersReoccurrenceMedian}{1\xspace}
\newcommand{\dataExpChatPresenceOthersReoccurrenceStd}{2.42\xspace}
\newcommand{\dataExpChatPresenceSolution}{17\xspace}
\newcommand{\dataExpChatPresenceSolutionRatio}{45.9\%\xspace}
\newcommand{\dataExpChatPresenceSolutionRatioChatters}{73.9\%\xspace}
\newcommand{\dataExpChatPresenceSolutionReoccurrenceMean}{2.24\xspace}
\newcommand{\dataExpChatPresenceSolutionReoccurrenceMedian}{2\xspace}
\newcommand{\dataExpChatPresenceSolutionReoccurrenceStd}{1.3\xspace}
\newcommand{\dataExpChatPresenceExplanation}{9\xspace}
\newcommand{\dataExpChatPresenceExplanationRatio}{24.3\%\xspace}
\newcommand{\dataExpChatPresenceExplanationRatioChatters}{39.1\%\xspace}
\newcommand{\dataExpChatPresenceExplanationReoccurrenceMean}{1.33\xspace}
\newcommand{\dataExpChatPresenceExplanationReoccurrenceMedian}{1\xspace}
\newcommand{\dataExpChatPresenceExplanationReoccurrenceStd}{0.71\xspace}
\newcommand{\dataExpChatPresenceFix}{13\xspace}
\newcommand{\dataExpChatPresenceFixRatio}{35.1\%\xspace}
\newcommand{\dataExpChatPresenceFixRatioChatters}{56.5\%\xspace}
\newcommand{\dataExpChatPresenceFixReoccurrenceMean}{4.31\xspace}
\newcommand{\dataExpChatPresenceFixReoccurrenceMedian}{4\xspace}
\newcommand{\dataExpChatPresenceFixReoccurrenceStd}{3.86\xspace}
\newcommand{\dataExpChatPresenceTotalReoccurrenceStats}[1]{\avgvals{7.35}{6}{6.06}{#1}\xspace}
\newcommand{\dataExpChatPresenceLookupReoccurrenceStats}[1]{\avgvals{2.13}{1}{2.1}{#1}\xspace}
\newcommand{\dataExpChatPresenceHintReoccurrenceStats}[1]{\avgvals{1.63}{1}{0.74}{#1}\xspace}
\newcommand{\dataExpChatPresenceOthersReoccurrenceStats}[1]{\avgvals{2.36}{1}{2.42}{#1}\xspace}
\newcommand{\dataExpChatPresenceSolutionReoccurrenceStats}[1]{\avgvals{2.24}{2}{1.3}{#1}\xspace}
\newcommand{\dataExpChatPresenceExplanationReoccurrenceStats}[1]{\avgvals{1.33}{1}{0.71}{#1}\xspace}
\newcommand{\dataExpChatPresenceFixReoccurrenceStats}[1]{\avgvals{4.31}{4}{3.86}{#1}\xspace}
\newcommand{\dataExpChatFirstLookup}{10\xspace}
\newcommand{\dataExpChatFirstHint}{4\xspace}
\newcommand{\dataExpChatFirstSolution}{9\xspace}
\newcommand{\dataExpChatFirstLookupRatio}{27\%\xspace}
\newcommand{\dataExpChatFirstLookupRatioChatters}{43.5\%\xspace}
\newcommand{\dataExpChatFirstHintRatio}{10.8\%\xspace}
\newcommand{\dataExpChatFirstHintRatioChatters}{17.4\%\xspace}
\newcommand{\dataExpChatFirstSolutionRatio}{24.3\%\xspace}
\newcommand{\dataExpChatFirstSolutionRatioChatters}{39.1\%\xspace}
\newcommand{\dataExpChatFirstNone}{14\xspace}
\newcommand{\dataExpChatFirstNoneRatio}{37.8\%\xspace}

\newcommand{\dataExpChatPrevalenceTotalTotalMean}{4.57\xspace}
\newcommand{\dataExpChatPrevalenceTotalTotalMedian}{2\xspace}
\newcommand{\dataExpChatPrevalenceTotalTotalStd}{5.96\xspace}
\newcommand{\dataExpChatPrevalenceTotalUsersMean}{7.35\xspace}
\newcommand{\dataExpChatPrevalenceTotalUsersMedian}{6\xspace}
\newcommand{\dataExpChatPrevalenceTotalUsersStd}{6.06\xspace}
\newcommand{\dataExpChatPrevalenceTotalTotalChattersMean}{7.35\xspace}
\newcommand{\dataExpChatPrevalenceTotalTotalChattersMedian}{6\xspace}
\newcommand{\dataExpChatPrevalenceTotalTotalChattersStd}{6.06\xspace}
\newcommand{\dataExpChatPrevalenceLookupTotalMean}{0.86\xspace}
\newcommand{\dataExpChatPrevalenceLookupTotalMedian}{0\xspace}
\newcommand{\dataExpChatPrevalenceLookupTotalStd}{1.69\xspace}
\newcommand{\dataExpChatPrevalenceLookupUsersMean}{2.13\xspace}
\newcommand{\dataExpChatPrevalenceLookupUsersMedian}{1\xspace}
\newcommand{\dataExpChatPrevalenceLookupUsersStd}{2.1\xspace}
\newcommand{\dataExpChatPrevalenceLookupTotalChattersMean}{1.39\xspace}
\newcommand{\dataExpChatPrevalenceLookupTotalChattersMedian}{1\xspace}
\newcommand{\dataExpChatPrevalenceLookupTotalChattersStd}{1.97\xspace}
\newcommand{\dataExpChatPrevalenceHintTotalMean}{0.35\xspace}
\newcommand{\dataExpChatPrevalenceHintTotalMedian}{0\xspace}
\newcommand{\dataExpChatPrevalenceHintTotalStd}{0.75\xspace}
\newcommand{\dataExpChatPrevalenceHintUsersMean}{1.63\xspace}
\newcommand{\dataExpChatPrevalenceHintUsersMedian}{1\xspace}
\newcommand{\dataExpChatPrevalenceHintUsersStd}{0.74\xspace}
\newcommand{\dataExpChatPrevalenceHintTotalChattersMean}{0.57\xspace}
\newcommand{\dataExpChatPrevalenceHintTotalChattersMedian}{0\xspace}
\newcommand{\dataExpChatPrevalenceHintTotalChattersStd}{0.9\xspace}
\newcommand{\dataExpChatPrevalenceOthersTotalMean}{0.7\xspace}
\newcommand{\dataExpChatPrevalenceOthersTotalMedian}{0\xspace}
\newcommand{\dataExpChatPrevalenceOthersTotalStd}{1.68\xspace}
\newcommand{\dataExpChatPrevalenceOthersUsersMean}{2.36\xspace}
\newcommand{\dataExpChatPrevalenceOthersUsersMedian}{1\xspace}
\newcommand{\dataExpChatPrevalenceOthersUsersStd}{2.42\xspace}
\newcommand{\dataExpChatPrevalenceOthersTotalChattersMean}{1.13\xspace}
\newcommand{\dataExpChatPrevalenceOthersTotalChattersMedian}{0\xspace}
\newcommand{\dataExpChatPrevalenceOthersTotalChattersStd}{2.03\xspace}
\newcommand{\dataExpChatPrevalenceSolutionTotalMean}{1.03\xspace}
\newcommand{\dataExpChatPrevalenceSolutionTotalMedian}{0\xspace}
\newcommand{\dataExpChatPrevalenceSolutionTotalStd}{1.42\xspace}
\newcommand{\dataExpChatPrevalenceSolutionUsersMean}{2.24\xspace}
\newcommand{\dataExpChatPrevalenceSolutionUsersMedian}{2\xspace}
\newcommand{\dataExpChatPrevalenceSolutionUsersStd}{1.3\xspace}
\newcommand{\dataExpChatPrevalenceSolutionTotalChattersMean}{1.65\xspace}
\newcommand{\dataExpChatPrevalenceSolutionTotalChattersMedian}{1\xspace}
\newcommand{\dataExpChatPrevalenceSolutionTotalChattersStd}{1.5\xspace}
\newcommand{\dataExpChatPrevalenceExplanationTotalMean}{0.32\xspace}
\newcommand{\dataExpChatPrevalenceExplanationTotalMedian}{0\xspace}
\newcommand{\dataExpChatPrevalenceExplanationTotalStd}{0.67\xspace}
\newcommand{\dataExpChatPrevalenceExplanationUsersMean}{1.33\xspace}
\newcommand{\dataExpChatPrevalenceExplanationUsersMedian}{1\xspace}
\newcommand{\dataExpChatPrevalenceExplanationUsersStd}{0.71\xspace}
\newcommand{\dataExpChatPrevalenceExplanationTotalChattersMean}{0.52\xspace}
\newcommand{\dataExpChatPrevalenceExplanationTotalChattersMedian}{0\xspace}
\newcommand{\dataExpChatPrevalenceExplanationTotalChattersStd}{0.79\xspace}
\newcommand{\dataExpChatPrevalenceFixTotalMean}{1.51\xspace}
\newcommand{\dataExpChatPrevalenceFixTotalMedian}{0\xspace}
\newcommand{\dataExpChatPrevalenceFixTotalStd}{3.05\xspace}
\newcommand{\dataExpChatPrevalenceFixUsersMean}{4.31\xspace}
\newcommand{\dataExpChatPrevalenceFixUsersMedian}{4\xspace}
\newcommand{\dataExpChatPrevalenceFixUsersStd}{3.86\xspace}
\newcommand{\dataExpChatPrevalenceFixTotalChattersMean}{2.43\xspace}
\newcommand{\dataExpChatPrevalenceFixTotalChattersMedian}{1\xspace}
\newcommand{\dataExpChatPrevalenceFixTotalChattersStd}{3.59\xspace}
\newcommand{\dataExpChatPrevalenceLookupRatio}{27.7\%\xspace}
\newcommand{\dataExpChatPrevalenceHintRatio}{6.4\%\xspace}
\newcommand{\dataExpChatPrevalenceOthersRatio}{11\%\xspace}
\newcommand{\dataExpChatPrevalenceSolutionRatio}{26.6\%\xspace}
\newcommand{\dataExpChatPrevalenceExplanationRatio}{6.8\%\xspace}
\newcommand{\dataExpChatPrevalenceFixRatio}{25.6\%\xspace}
\newcommand{\dataExpChatPrevalenceTotalTotalStats}[1]{\avgvals{4.57}{2}{5.96}{#1}\xspace}
\newcommand{\dataExpChatPrevalenceTotalUsersStats}[1]{\avgvals{7.35}{6}{6.06}{#1}\xspace}
\newcommand{\dataExpChatPrevalenceTotalTotalChattersStats}[1]{\avgvals{7.35}{6}{6.06}{#1}\xspace}
\newcommand{\dataExpChatPrevalenceLookupTotalStats}[1]{\avgvals{0.86}{0}{1.69}{#1}\xspace}
\newcommand{\dataExpChatPrevalenceLookupUsersStats}[1]{\avgvals{2.13}{1}{2.1}{#1}\xspace}
\newcommand{\dataExpChatPrevalenceLookupTotalChattersStats}[1]{\avgvals{1.39}{1}{1.97}{#1}\xspace}
\newcommand{\dataExpChatPrevalenceHintTotalStats}[1]{\avgvals{0.35}{0}{0.75}{#1}\xspace}
\newcommand{\dataExpChatPrevalenceHintUsersStats}[1]{\avgvals{1.63}{1}{0.74}{#1}\xspace}
\newcommand{\dataExpChatPrevalenceHintTotalChattersStats}[1]{\avgvals{0.57}{0}{0.9}{#1}\xspace}
\newcommand{\dataExpChatPrevalenceOthersTotalStats}[1]{\avgvals{0.7}{0}{1.68}{#1}\xspace}
\newcommand{\dataExpChatPrevalenceOthersUsersStats}[1]{\avgvals{2.36}{1}{2.42}{#1}\xspace}
\newcommand{\dataExpChatPrevalenceOthersTotalChattersStats}[1]{\avgvals{1.13}{0}{2.03}{#1}\xspace}
\newcommand{\dataExpChatPrevalenceSolutionTotalStats}[1]{\avgvals{1.03}{0}{1.42}{#1}\xspace}
\newcommand{\dataExpChatPrevalenceSolutionUsersStats}[1]{\avgvals{2.24}{2}{1.3}{#1}\xspace}
\newcommand{\dataExpChatPrevalenceSolutionTotalChattersStats}[1]{\avgvals{1.65}{1}{1.5}{#1}\xspace}
\newcommand{\dataExpChatPrevalenceExplanationTotalStats}[1]{\avgvals{0.32}{0}{0.67}{#1}\xspace}
\newcommand{\dataExpChatPrevalenceExplanationUsersStats}[1]{\avgvals{1.33}{1}{0.71}{#1}\xspace}
\newcommand{\dataExpChatPrevalenceExplanationTotalChattersStats}[1]{\avgvals{0.52}{0}{0.79}{#1}\xspace}
\newcommand{\dataExpChatPrevalenceFixTotalStats}[1]{\avgvals{1.51}{0}{3.05}{#1}\xspace}
\newcommand{\dataExpChatPrevalenceFixUsersStats}[1]{\avgvals{4.31}{4}{3.86}{#1}\xspace}
\newcommand{\dataExpChatPrevalenceFixTotalChattersStats}[1]{\avgvals{2.43}{1}{3.59}{#1}\xspace}
\newcommand{\dataExpChatResponseTotal}{172\xspace}
\newcommand{\dataExpChatResponseCorrectRatio}{38.8\%\xspace}
\newcommand{\dataExpChatResponseExplainsTotal}{143\xspace}
\newcommand{\dataExpChatResponseExplainsRatio}{83.1\%\xspace}
\newcommand{\dataExpChatResponseExplainsCorrectRatio}{35.5\%\xspace}
\newcommand{\dataExpChatResponseSolvesTotal}{105\xspace}
\newcommand{\dataExpChatResponseSolvesRatio}{61\%\xspace}
\newcommand{\dataExpChatResponseSolvesCorrectRatio}{18.4\%\xspace}
\newcommand{\dataExpChatResponseInformsTotal}{42\xspace}
\newcommand{\dataExpChatResponseInformsRatio}{24.4\%\xspace}
\newcommand{\dataExpChatResponseInformsCorrectRatio}{88.1\%\xspace}
\newcommand{\dataExpChatResponseSolutionExplanationTotal}{102\xspace}
\newcommand{\dataExpChatResponseSolutionExplanationRatio}{97.1\%\xspace}
\newcommand{\dataExpChatResponsePOneSolutionTotal}{80\xspace}
\newcommand{\dataExpChatResponsePOneSolutionRatio}{46.5\%\xspace}
\newcommand{\dataExpChatResponsePOneSolutionCorrectRatio}{8.8\%\xspace}
\newcommand{\dataExpChatResponsePTwoSolutionTotal}{13\xspace}
\newcommand{\dataExpChatResponsePTwoSolutionRatio}{7.6\%\xspace}
\newcommand{\dataExpChatResponsePTwoSolutionCorrectRatio}{76.9\%\xspace}

\newcommand{\dataExpChatUseFrequentlyTotalChattersMean}{6.44\xspace}
\newcommand{\dataExpChatUseFrequentlyTotalChattersMedian}{4\xspace}
\newcommand{\dataExpChatUseFrequentlyTotalChattersStd}{7.41\xspace}
\newcommand{\dataExpChatUseFrequentlyTotalChattersRatio}{72.2\%\xspace}
\newcommand{\dataExpChatUseFrequentlyTotalChattersTotal}{13\xspace}
\newcommand{\dataExpChatUseFrequentlyTotal}{18\xspace}
\newcommand{\dataExpChatUseFrequentlyTotalRatio}{48.6\%\xspace}
\newcommand{\dataExpChatUseFrequentlyTotalChattersStats}[1]{\avgvals{6.44}{4}{7.41}{#1}\xspace}
\newcommand{\dataExpChatUseNeverTotalChattersMean}{2.5\xspace}
\newcommand{\dataExpChatUseNeverTotalChattersMedian}{0\xspace}
\newcommand{\dataExpChatUseNeverTotalChattersStd}{4.18\xspace}
\newcommand{\dataExpChatUseNeverTotalChattersRatio}{33.3\%\xspace}
\newcommand{\dataExpChatUseNeverTotalChattersTotal}{2\xspace}
\newcommand{\dataExpChatUseNeverTotal}{6\xspace}
\newcommand{\dataExpChatUseNeverTotalRatio}{16.2\%\xspace}
\newcommand{\dataExpChatUseNeverTotalChattersStats}[1]{\avgvals{2.5}{0}{4.18}{#1}\xspace}
\newcommand{\dataExpChatUseRarelyTotalChattersMean}{2.92\xspace}
\newcommand{\dataExpChatUseRarelyTotalChattersMedian}{1\xspace}
\newcommand{\dataExpChatUseRarelyTotalChattersStd}{3.33\xspace}
\newcommand{\dataExpChatUseRarelyTotalChattersRatio}{61.5\%\xspace}
\newcommand{\dataExpChatUseRarelyTotalChattersTotal}{8\xspace}
\newcommand{\dataExpChatUseRarelyTotal}{13\xspace}
\newcommand{\dataExpChatUseRarelyTotalRatio}{35.1\%\xspace}
\newcommand{\dataExpChatUseRarelyTotalChattersStats}[1]{\avgvals{2.92}{1}{3.33}{#1}\xspace}
\newcommand{\dataExpChatUseMean}{3.22\xspace}
\newcommand{\dataExpChatUseMedian}{3\xspace}
\newcommand{\dataExpChatUseStd}{1.51\xspace}
\newcommand{\dataExpChatUseStats}[1]{\avgvals{3.22}{3}{1.51}{#1}\xspace}
\newcommand{\dataExpChatPrevalenceRawLookupTotal}{32\xspace}
\newcommand{\dataExpChatPrevalenceRawLookupRatio}{18.1\%\xspace}
\newcommand{\dataExpChatPrevalenceRawHintTotal}{13\xspace}
\newcommand{\dataExpChatPrevalenceRawHintRatio}{7.3\%\xspace}
\newcommand{\dataExpChatPrevalenceRawOthersTotal}{26\xspace}
\newcommand{\dataExpChatPrevalenceRawOthersRatio}{14.7\%\xspace}
\newcommand{\dataExpChatPrevalenceRawSolutionTotal}{38\xspace}
\newcommand{\dataExpChatPrevalenceRawSolutionRatio}{21.5\%\xspace}
\newcommand{\dataExpChatPrevalenceRawExplanationTotal}{12\xspace}
\newcommand{\dataExpChatPrevalenceRawExplanationRatio}{6.8\%\xspace}
\newcommand{\dataExpChatPrevalenceRawFixTotal}{56\xspace}
\newcommand{\dataExpChatPrevalenceRawFixRatio}{31.6\%\xspace}
\newcommand{\dataExpOutcomesPOneTotal}{495\xspace}
\newcommand{\dataExpOutcomesPOneCompilationErrorPresenceTotal}{33\xspace}
\newcommand{\dataExpOutcomesPOneCompilationErrorPresenceRatio}{89.2\%\xspace}
\newcommand{\dataExpOutcomesPOneCompilationErrorPrevalenceTotal}{334\xspace}
\newcommand{\dataExpOutcomesPOneCompilationErrorPrevalenceRatio}{67.5\%\xspace}
\newcommand{\dataExpOutcomesPOneCompilationErrorPrevalenceRatioByAttemptMean}{57.6\%\xspace}
\newcommand{\dataExpOutcomesPOneCompilationErrorPrevalenceRatioByAttemptMedian}{60\%\xspace}
\newcommand{\dataExpOutcomesPOneCompilationErrorPrevalenceRatioByAttemptStd}{30.4\%\xspace}
\newcommand{\dataExpOutcomesPOneNoNPEPresenceTotal}{27\xspace}
\newcommand{\dataExpOutcomesPOneNoNPEPresenceRatio}{73\%\xspace}
\newcommand{\dataExpOutcomesPOneNoNPEPrevalenceTotal}{111\xspace}
\newcommand{\dataExpOutcomesPOneNoNPEPrevalenceRatio}{22.4\%\xspace}
\newcommand{\dataExpOutcomesPOneNoNPEPrevalenceRatioByAttemptMean}{22\%\xspace}
\newcommand{\dataExpOutcomesPOneNoNPEPrevalenceRatioByAttemptMedian}{25\%\xspace}
\newcommand{\dataExpOutcomesPOneNoNPEPrevalenceRatioByAttemptStd}{18.2\%\xspace}
\newcommand{\dataExpOutcomesPOneManualThrowPresenceTotal}{1\xspace}
\newcommand{\dataExpOutcomesPOneManualThrowPresenceRatio}{2.7\%\xspace}
\newcommand{\dataExpOutcomesPOneManualThrowPrevalenceTotal}{2\xspace}
\newcommand{\dataExpOutcomesPOneManualThrowPrevalenceRatio}{0.4\%\xspace}
\newcommand{\dataExpOutcomesPOneManualThrowPrevalenceRatioByAttemptMean}{0.2\%\xspace}
\newcommand{\dataExpOutcomesPOneManualThrowPrevalenceRatioByAttemptMedian}{0\%\xspace}
\newcommand{\dataExpOutcomesPOneManualThrowPrevalenceRatioByAttemptStd}{1.2\%\xspace}
\newcommand{\dataExpOutcomesPOneCorrectPresenceTotal}{24\xspace}
\newcommand{\dataExpOutcomesPOneCorrectPresenceRatio}{64.9\%\xspace}
\newcommand{\dataExpOutcomesPOneCorrectPrevalenceTotal}{28\xspace}
\newcommand{\dataExpOutcomesPOneCorrectPrevalenceRatio}{5.7\%\xspace}
\newcommand{\dataExpOutcomesPOneCorrectPrevalenceRatioByAttemptMean}{17.8\%\xspace}
\newcommand{\dataExpOutcomesPOneCorrectPrevalenceRatioByAttemptMedian}{7.1\%\xspace}
\newcommand{\dataExpOutcomesPOneCorrectPrevalenceRatioByAttemptStd}{27.7\%\xspace}
\newcommand{\dataExpOutcomesPOneInvalidMethodDefPresenceTotal}{3\xspace}
\newcommand{\dataExpOutcomesPOneInvalidMethodDefPresenceRatio}{8.1\%\xspace}
\newcommand{\dataExpOutcomesPOneInvalidMethodDefPrevalenceTotal}{5\xspace}
\newcommand{\dataExpOutcomesPOneInvalidMethodDefPrevalenceRatio}{1\%\xspace}
\newcommand{\dataExpOutcomesPOneInvalidMethodDefPrevalenceRatioByAttemptMean}{0.6\%\xspace}
\newcommand{\dataExpOutcomesPOneInvalidMethodDefPrevalenceRatioByAttemptMedian}{0\%\xspace}
\newcommand{\dataExpOutcomesPOneInvalidMethodDefPrevalenceRatioByAttemptStd}{2.2\%\xspace}
\newcommand{\dataExpOutcomesPOneManualTriggerPresenceTotal}{4\xspace}
\newcommand{\dataExpOutcomesPOneManualTriggerPresenceRatio}{10.8\%\xspace}
\newcommand{\dataExpOutcomesPOneManualTriggerPrevalenceTotal}{7\xspace}
\newcommand{\dataExpOutcomesPOneManualTriggerPrevalenceRatio}{1.4\%\xspace}
\newcommand{\dataExpOutcomesPOneManualTriggerPrevalenceRatioByAttemptMean}{0.7\%\xspace}
\newcommand{\dataExpOutcomesPOneManualTriggerPrevalenceRatioByAttemptMedian}{0\%\xspace}
\newcommand{\dataExpOutcomesPOneManualTriggerPrevalenceRatioByAttemptStd}{2.3\%\xspace}
\newcommand{\dataExpOutcomesPOneStaticMethodDefPresenceTotal}{1\xspace}
\newcommand{\dataExpOutcomesPOneStaticMethodDefPresenceRatio}{2.7\%\xspace}
\newcommand{\dataExpOutcomesPOneStaticMethodDefPrevalenceTotal}{8\xspace}
\newcommand{\dataExpOutcomesPOneStaticMethodDefPrevalenceRatio}{1.6\%\xspace}
\newcommand{\dataExpOutcomesPOneStaticMethodDefPrevalenceRatioByAttemptMean}{1.1\%\xspace}
\newcommand{\dataExpOutcomesPOneStaticMethodDefPrevalenceRatioByAttemptMedian}{0\%\xspace}
\newcommand{\dataExpOutcomesPOneStaticMethodDefPrevalenceRatioByAttemptStd}{6.9\%\xspace}
\newcommand{\dataExpOutcomesPOneCompilationErrorPrevalenceRatioByAttemptStats}[1]{\avgvals{57.6\%}{60\%}{30.4\%}{#1}\xspace}
\newcommand{\dataExpOutcomesPOneNoNPEPrevalenceRatioByAttemptStats}[1]{\avgvals{22\%}{25\%}{18.2\%}{#1}\xspace}
\newcommand{\dataExpOutcomesPOneManualThrowPrevalenceRatioByAttemptStats}[1]{\avgvals{0.2\%}{0\%}{1.2\%}{#1}\xspace}
\newcommand{\dataExpOutcomesPOneCorrectPrevalenceRatioByAttemptStats}[1]{\avgvals{17.8\%}{7.1\%}{27.7\%}{#1}\xspace}
\newcommand{\dataExpOutcomesPOneInvalidMethodDefPrevalenceRatioByAttemptStats}[1]{\avgvals{0.6\%}{0\%}{2.2\%}{#1}\xspace}
\newcommand{\dataExpOutcomesPOneManualTriggerPrevalenceRatioByAttemptStats}[1]{\avgvals{0.7\%}{0\%}{2.3\%}{#1}\xspace}
\newcommand{\dataExpOutcomesPOneStaticMethodDefPrevalenceRatioByAttemptStats}[1]{\avgvals{1.1\%}{0\%}{6.9\%}{#1}\xspace}
\newcommand{\dataExpOutcomesPTwoTotal}{61\xspace}
\newcommand{\dataExpOutcomesPTwoCorrectPresenceTotal}{28\xspace}
\newcommand{\dataExpOutcomesPTwoCorrectPresenceRatio}{75.7\%\xspace}
\newcommand{\dataExpOutcomesPTwoCorrectPrevalenceTotal}{32\xspace}
\newcommand{\dataExpOutcomesPTwoCorrectPrevalenceRatio}{52.5\%\xspace}
\newcommand{\dataExpOutcomesPTwoCorrectPrevalenceRatioByAttemptMean}{77.3\%\xspace}
\newcommand{\dataExpOutcomesPTwoCorrectPrevalenceRatioByAttemptMedian}{100\%\xspace}
\newcommand{\dataExpOutcomesPTwoCorrectPrevalenceRatioByAttemptStd}{34.5\%\xspace}
\newcommand{\dataExpOutcomesPTwoCompilationErrorPresenceTotal}{5\xspace}
\newcommand{\dataExpOutcomesPTwoCompilationErrorPresenceRatio}{13.5\%\xspace}
\newcommand{\dataExpOutcomesPTwoCompilationErrorPrevalenceTotal}{17\xspace}
\newcommand{\dataExpOutcomesPTwoCompilationErrorPrevalenceRatio}{27.9\%\xspace}
\newcommand{\dataExpOutcomesPTwoCompilationErrorPrevalenceRatioByAttemptMean}{7.5\%\xspace}
\newcommand{\dataExpOutcomesPTwoCompilationErrorPrevalenceRatioByAttemptMedian}{0\%\xspace}
\newcommand{\dataExpOutcomesPTwoCompilationErrorPrevalenceRatioByAttemptStd}{21\%\xspace}
\newcommand{\dataExpOutcomesPTwoNullNotCheckedPresenceTotal}{7\xspace}
\newcommand{\dataExpOutcomesPTwoNullNotCheckedPresenceRatio}{18.9\%\xspace}
\newcommand{\dataExpOutcomesPTwoNullNotCheckedPrevalenceTotal}{8\xspace}
\newcommand{\dataExpOutcomesPTwoNullNotCheckedPrevalenceRatio}{13.1\%\xspace}
\newcommand{\dataExpOutcomesPTwoNullNotCheckedPrevalenceRatioByAttemptMean}{12.5\%\xspace}
\newcommand{\dataExpOutcomesPTwoNullNotCheckedPrevalenceRatioByAttemptMedian}{0\%\xspace}
\newcommand{\dataExpOutcomesPTwoNullNotCheckedPrevalenceRatioByAttemptStd}{27.5\%\xspace}
\newcommand{\dataExpOutcomesPTwoAffectedNormalBehaviorPresenceTotal}{3\xspace}
\newcommand{\dataExpOutcomesPTwoAffectedNormalBehaviorPresenceRatio}{8.1\%\xspace}
\newcommand{\dataExpOutcomesPTwoAffectedNormalBehaviorPrevalenceTotal}{4\xspace}
\newcommand{\dataExpOutcomesPTwoAffectedNormalBehaviorPrevalenceRatio}{6.6\%\xspace}
\newcommand{\dataExpOutcomesPTwoAffectedNormalBehaviorPrevalenceRatioByAttemptMean}{2.8\%\xspace}
\newcommand{\dataExpOutcomesPTwoAffectedNormalBehaviorPrevalenceRatioByAttemptMedian}{0\%\xspace}
\newcommand{\dataExpOutcomesPTwoAffectedNormalBehaviorPrevalenceRatioByAttemptStd}{8.8\%\xspace}
\newcommand{\dataExpOutcomesPTwoCorrectPrevalenceRatioByAttemptStats}[1]{\avgvals{77.3\%}{100\%}{34.5\%}{#1}\xspace}
\newcommand{\dataExpOutcomesPTwoCompilationErrorPrevalenceRatioByAttemptStats}[1]{\avgvals{7.5\%}{0\%}{21\%}{#1}\xspace}
\newcommand{\dataExpOutcomesPTwoNullNotCheckedPrevalenceRatioByAttemptStats}[1]{\avgvals{12.5\%}{0\%}{27.5\%}{#1}\xspace}
\newcommand{\dataExpOutcomesPTwoAffectedNormalBehaviorPrevalenceRatioByAttemptStats}[1]{\avgvals{2.8\%}{0\%}{8.8\%}{#1}\xspace}
\newcommand{\dataExpOutcomesCompErrorsOneReason}{';' expected\xspace}
\newcommand{\dataExpOutcomesCompErrorsOneTotal}{29\xspace}
\newcommand{\dataExpOutcomesCompErrorsOnePresence}{14\xspace}

\newcommand{\dataExpOutcomesCompErrorsTwoReason}{cannot find symbol – symbol: variable \_stock\xspace}
\newcommand{\dataExpOutcomesCompErrorsTwoTotal}{26\xspace}
\newcommand{\dataExpOutcomesCompErrorsTwoPresence}{14\xspace}

\newcommand{\dataExpOutcomesCompErrorsThreeReason}{cannot find symbol – symbol: variable stock\xspace}
\newcommand{\dataExpOutcomesCompErrorsThreeTotal}{33\xspace}
\newcommand{\dataExpOutcomesCompErrorsThreePresence}{12\xspace}

\newcommand{\dataExpOutcomesCompErrorsFourReason}{cannot find symbol – symbol: method legeStockAb()\xspace}
\newcommand{\dataExpOutcomesCompErrorsFourTotal}{32\xspace}
\newcommand{\dataExpOutcomesCompErrorsFourPresence}{11\xspace}

\newcommand{\dataExpOutcomesCompErrorsFiveReason}{method nimmStock in class Hund cannot be applied to given types; – required: Stock\xspace}
\newcommand{\dataExpOutcomesCompErrorsFiveTotal}{9\xspace}
\newcommand{\dataExpOutcomesCompErrorsFivePresence}{8\xspace}

\newcommand{\dataExpOutcomesCompErrorsSixReason}{')' expected\xspace}
\newcommand{\dataExpOutcomesCompErrorsSixTotal}{13\xspace}
\newcommand{\dataExpOutcomesCompErrorsSixPresence}{7\xspace}

\newcommand{\dataExpOutcomesCompErrorsSevenReason}{not a statement\xspace}
\newcommand{\dataExpOutcomesCompErrorsSevenTotal}{13\xspace}
\newcommand{\dataExpOutcomesCompErrorsSevenPresence}{7\xspace}

\newcommand{\dataExpOutcomesCompErrorsEightReason}{cannot find symbol – symbol: method nimmStock(<nulltype>)\xspace}
\newcommand{\dataExpOutcomesCompErrorsEightTotal}{7\xspace}
\newcommand{\dataExpOutcomesCompErrorsEightPresence}{7\xspace}

\newcommand{\dataExpOutcomesCompErrorsNineReason}{illegal start of expression\xspace}
\newcommand{\dataExpOutcomesCompErrorsNineTotal}{9\xspace}
\newcommand{\dataExpOutcomesCompErrorsNinePresence}{6\xspace}

\newcommand{\dataExpOutcomesCompErrorsOneZeroReason}{constructor Hund in class Hund cannot be applied to given types; – required: java.lang.String\xspace}
\newcommand{\dataExpOutcomesCompErrorsOneZeroTotal}{8\xspace}
\newcommand{\dataExpOutcomesCompErrorsOneZeroPresence}{6\xspace}

\newcommand{\dataExpOutcomesCompErrorsPTwoOneReason}{not a statement\xspace}
\newcommand{\dataExpOutcomesCompErrorsPTwoOneTotal}{3\xspace}
\newcommand{\dataExpOutcomesCompErrorsPTwoOnePresence}{2\xspace}

\newcommand{\dataExpOutcomesCompErrorsPTwoTwoReason}{'(' expected\xspace}
\newcommand{\dataExpOutcomesCompErrorsPTwoTwoTotal}{11\xspace}
\newcommand{\dataExpOutcomesCompErrorsPTwoTwoPresence}{1\xspace}

\newcommand{\dataExpOutcomesCompErrorsPTwoThreeReason}{incompatible types: Stock cannot be converted to boolean\xspace}
\newcommand{\dataExpOutcomesCompErrorsPTwoThreeTotal}{1\xspace}
\newcommand{\dataExpOutcomesCompErrorsPTwoThreePresence}{1\xspace}

\newcommand{\dataExpOutcomesCompErrorsPTwoFourReason}{';' expected\xspace}
\newcommand{\dataExpOutcomesCompErrorsPTwoFourTotal}{1\xspace}
\newcommand{\dataExpOutcomesCompErrorsPTwoFourPresence}{1\xspace}

\newcommand{\dataExpOutcomesCompErrorsPTwoFiveReason}{bad operand types for binary operator '==' – first type: Stock\xspace}
\newcommand{\dataExpOutcomesCompErrorsPTwoFiveTotal}{1\xspace}
\newcommand{\dataExpOutcomesCompErrorsPTwoFivePresence}{1\xspace}
\newcommand{\dataExpPerfMetricsCorrectMean}{1.41\xspace}
\newcommand{\dataExpPerfMetricsCorrectMedian}{2\xspace}
\newcommand{\dataExpPerfMetricsCorrectStd}{0.83\xspace}
\newcommand{\dataExpPerfMetricsDurationMean}{1310.8s\xspace}
\newcommand{\dataExpPerfMetricsDurationMedian}{1380s\xspace}
\newcommand{\dataExpPerfMetricsDurationStd}{713.3s\xspace}
\newcommand{\dataExpPerfMetricsRatioCorrectAll}{62.2 \%\xspace}
\newcommand{\dataExpPerfMetricsRatioCorrectAny}{78.4 \%\xspace}
\newcommand{\dataExpPerfMetricsDurationCompleteMean}{941.9s\xspace}
\newcommand{\dataExpPerfMetricsDurationCompleteMedian}{859s\xspace}
\newcommand{\dataExpPerfMetricsDurationCompleteStd}{559.2s\xspace}
\newcommand{\dataExpPerfMetricsDurationIncompleteMean}{1916.9s\xspace}
\newcommand{\dataExpPerfMetricsDurationIncompleteMedian}{1784s\xspace}
\newcommand{\dataExpPerfMetricsDurationIncompleteStd}{493.7s\xspace}
\newcommand{\dataExpPerfMetricsRatioCorrectLongAll}{30 \%\xspace}
\newcommand{\dataExpPerfMetricsRatioCorrectLongAny}{60 \%\xspace}
\newcommand{\dataExpPerfDifficultyOneOne}{1\xspace}
\newcommand{\dataExpPerfDifficultyOneTwo}{5\xspace}
\newcommand{\dataExpPerfDifficultyOneOneRatio}{16.7 \%\xspace}
\newcommand{\dataExpPerfDifficultyOneTwoRatio}{83.3 \%\xspace}
\newcommand{\dataExpPerfDifficultyOneTotalRatio}{16.2 \%\xspace}
\newcommand{\dataExpPerfDifficultyOneTotal}{6\xspace}
\newcommand{\dataExpPerfDifficultyOneZero}{0\xspace}
\newcommand{\dataExpPerfDifficultyOneZeroRatio}{0 \%\xspace}
\newcommand{\dataExpPerfDifficultyTwoOne}{1\xspace}
\newcommand{\dataExpPerfDifficultyTwoTwo}{14\xspace}
\newcommand{\dataExpPerfDifficultyTwoOneRatio}{5.6 \%\xspace}
\newcommand{\dataExpPerfDifficultyTwoTwoRatio}{77.8 \%\xspace}
\newcommand{\dataExpPerfDifficultyTwoTotalRatio}{48.6 \%\xspace}
\newcommand{\dataExpPerfDifficultyTwoTotal}{18\xspace}
\newcommand{\dataExpPerfDifficultyTwoZero}{3\xspace}
\newcommand{\dataExpPerfDifficultyTwoZeroRatio}{16.7 \%\xspace}
\newcommand{\dataExpPerfDifficultyThreeOne}{2\xspace}
\newcommand{\dataExpPerfDifficultyThreeTwo}{4\xspace}
\newcommand{\dataExpPerfDifficultyThreeOneRatio}{22.2 \%\xspace}
\newcommand{\dataExpPerfDifficultyThreeTwoRatio}{44.4 \%\xspace}
\newcommand{\dataExpPerfDifficultyThreeTotalRatio}{24.3 \%\xspace}
\newcommand{\dataExpPerfDifficultyThreeTotal}{9\xspace}
\newcommand{\dataExpPerfDifficultyThreeZero}{3\xspace}
\newcommand{\dataExpPerfDifficultyThreeZeroRatio}{33.3 \%\xspace}
\newcommand{\dataExpPerfDifficultyFourOne}{2\xspace}
\newcommand{\dataExpPerfDifficultyFourTwo}{0\xspace}
\newcommand{\dataExpPerfDifficultyFourOneRatio}{50 \%\xspace}
\newcommand{\dataExpPerfDifficultyFourTwoRatio}{0 \%\xspace}
\newcommand{\dataExpPerfDifficultyFourTotalRatio}{10.8 \%\xspace}
\newcommand{\dataExpPerfDifficultyFourTotal}{4\xspace}
\newcommand{\dataExpPerfDifficultyFourZero}{2\xspace}
\newcommand{\dataExpPerfDifficultyFourZeroRatio}{50 \%\xspace}
\newcommand{\dataExpPerfDurationFalseDistZero}{1380s\xspace}
\newcommand{\dataExpPerfDurationFalseDistTwoFive}{1482s\xspace}
\newcommand{\dataExpPerfDurationFalseDistMedian}{1784.5s\xspace}
\newcommand{\dataExpPerfDurationFalseDistSevenFive}{2317.5s\xspace}
\newcommand{\dataExpPerfDurationFalseDistOneZeroZero}{2806s\xspace}
\newcommand{\dataExpPerfDurationFalseDistMean}{1916.9s\xspace}
\newcommand{\dataExpPerfDurationFalseDistStd}{493.7s\xspace}
\newcommand{\dataExpPerfDurationFalseTotal}{14s\xspace}
\newcommand{\dataExpPerfDurationTrueDistZero}{293s\xspace}
\newcommand{\dataExpPerfDurationTrueDistTwoFive}{556s\xspace}
\newcommand{\dataExpPerfDurationTrueDistMedian}{859s\xspace}
\newcommand{\dataExpPerfDurationTrueDistSevenFive}{1289s\xspace}
\newcommand{\dataExpPerfDurationTrueDistOneZeroZero}{2526s\xspace}
\newcommand{\dataExpPerfDurationTrueDistMean}{941.9s\xspace}
\newcommand{\dataExpPerfDurationTrueDistStd}{559.2s\xspace}
\newcommand{\dataExpPerfDurationTrueTotal}{23s\xspace}
\newcommand{\dataExpPerfDifficultyDifficultyMean}{2.3\xspace}
\newcommand{\dataExpPerfDifficultyDifficultyMedian}{2\xspace}
\newcommand{\dataExpPerfDifficultyDifficultyStd}{0.88\xspace}
\newcommand{\dataExpPerfChatFiveCorrectMean}{1.36\xspace}
\newcommand{\dataExpPerfChatFiveCorrectMedian}{2\xspace}
\newcommand{\dataExpPerfChatFiveCorrectStd}{0.81\xspace}
\newcommand{\dataExpPerfChatFourCorrectMean}{0.71\xspace}
\newcommand{\dataExpPerfChatFourCorrectMedian}{0\xspace}
\newcommand{\dataExpPerfChatFourCorrectStd}{0.95\xspace}
\newcommand{\dataExpPerfChatOneCorrectMean}{1.5\xspace}
\newcommand{\dataExpPerfChatOneCorrectMedian}{2\xspace}
\newcommand{\dataExpPerfChatOneCorrectStd}{0.84\xspace}
\newcommand{\dataExpPerfChatTwoCorrectMean}{1.78\xspace}
\newcommand{\dataExpPerfChatTwoCorrectMedian}{2\xspace}
\newcommand{\dataExpPerfChatTwoCorrectStd}{0.67\xspace}
\newcommand{\dataExpPerfChatThreeCorrectMean}{1.75\xspace}
\newcommand{\dataExpPerfChatThreeCorrectMedian}{2\xspace}
\newcommand{\dataExpPerfChatThreeCorrectStd}{0.5\xspace}
\newcommand{\dataExpChatPlagSimilarityPOneNoneRatio}{32.5\%\xspace}
\newcommand{\dataExpChatPlagSimilarityPOneNone}{161\xspace}
\newcommand{\dataExpChatPlagSimilarityPOneTotalMean}{0.44\xspace}
\newcommand{\dataExpChatPlagSimilarityPOneTotalMedian}{0.45\xspace}
\newcommand{\dataExpChatPlagSimilarityPOneTotalStd}{0.37\xspace}
\newcommand{\dataExpChatPlagSimilarityPOneChatterNoneRatio}{18.1\%\xspace}
\newcommand{\dataExpChatPlagSimilarityPOneChatterNone}{74\xspace}
\newcommand{\dataExpChatPlagSimilarityPOneChatterMean}{0.53\xspace}
\newcommand{\dataExpChatPlagSimilarityPOneChatterMedian}{0.57\xspace}
\newcommand{\dataExpChatPlagSimilarityPOneChatterStd}{0.34\xspace}
\newcommand{\dataExpChatPlagSimilarityPOneAccessMean}{0.65\xspace}
\newcommand{\dataExpChatPlagSimilarityPOneAccessMedian}{0.68\xspace}
\newcommand{\dataExpChatPlagSimilarityPOneAccessStd}{0.26\xspace}
\newcommand{\dataExpChatPlagSimilarityPOneClose}{68\xspace}
\newcommand{\dataExpChatPlagSimilarityPOneCloseRatio}{20.4\%\xspace}
\newcommand{\dataExpChatPlagSimilarityPOneClosePresence}{15\xspace}
\newcommand{\dataExpChatPlagSimilarityPOneClosePresenceRatio}{40.5\%\xspace}
\newcommand{\dataExpChatPlagSimilarityPOneClosePresenceRatioChatters}{65.2\%\xspace}
\newcommand{\dataExpChatPlagSimilarityPOneClosePositionMean}{6.67\xspace}
\newcommand{\dataExpChatPlagSimilarityPOneClosePositionMedian}{5\xspace}
\newcommand{\dataExpChatPlagSimilarityPOneClosePositionStd}{5.7\xspace}
\newcommand{\dataExpChatPlagSimilarityPOneIdent}{46\xspace}
\newcommand{\dataExpChatPlagSimilarityPOneIdentRatio}{13.8\%\xspace}
\newcommand{\dataExpChatPlagSimilarityPOneIdentPresence}{13\xspace}
\newcommand{\dataExpChatPlagSimilarityPOneIdentPresenceRatio}{35.1\%\xspace}
\newcommand{\dataExpChatPlagSimilarityPOneIdentPresenceRatioChatters}{56.5\%\xspace}
\newcommand{\dataExpChatPlagSimilarityPOneIdentPositionMean}{7.31\xspace}
\newcommand{\dataExpChatPlagSimilarityPOneIdentPositionMedian}{5\xspace}
\newcommand{\dataExpChatPlagSimilarityPOneIdentPositionStd}{6.17\xspace}
\newcommand{\dataExpChatPlagSimilarityPOneFirstMean}{0.62\xspace}
\newcommand{\dataExpChatPlagSimilarityPOneFirstMedian}{0.62\xspace}
\newcommand{\dataExpChatPlagSimilarityPOneFirstStd}{0.28\xspace}
\newcommand{\dataExpChatPlagSimilarityPTwoNoneRatio}{65.6\%\xspace}
\newcommand{\dataExpChatPlagSimilarityPTwoNone}{40\xspace}
\newcommand{\dataExpChatPlagSimilarityPTwoTotalMean}{0.23\xspace}
\newcommand{\dataExpChatPlagSimilarityPTwoTotalMedian}{0\xspace}
\newcommand{\dataExpChatPlagSimilarityPTwoTotalStd}{0.36\xspace}
\newcommand{\dataExpChatPlagSimilarityPTwoChatterNoneRatio}{27.6\%\xspace}
\newcommand{\dataExpChatPlagSimilarityPTwoChatterNone}{8\xspace}
\newcommand{\dataExpChatPlagSimilarityPTwoChatterMean}{0.48\xspace}
\newcommand{\dataExpChatPlagSimilarityPTwoChatterMedian}{0.55\xspace}
\newcommand{\dataExpChatPlagSimilarityPTwoChatterStd}{0.39\xspace}
\newcommand{\dataExpChatPlagSimilarityPTwoAccessMean}{0.66\xspace}
\newcommand{\dataExpChatPlagSimilarityPTwoAccessMedian}{0.71\xspace}
\newcommand{\dataExpChatPlagSimilarityPTwoAccessStd}{0.29\xspace}
\newcommand{\dataExpChatPlagSimilarityPTwoClose}{6\xspace}
\newcommand{\dataExpChatPlagSimilarityPTwoCloseRatio}{28.6\%\xspace}
\newcommand{\dataExpChatPlagSimilarityPTwoClosePresence}{6\xspace}
\newcommand{\dataExpChatPlagSimilarityPTwoClosePresenceRatio}{16.2\%\xspace}
\newcommand{\dataExpChatPlagSimilarityPTwoClosePresenceRatioChatters}{26.1\%\xspace}
\newcommand{\dataExpChatPlagSimilarityPTwoClosePositionMean}{1\xspace}
\newcommand{\dataExpChatPlagSimilarityPTwoClosePositionMedian}{1\xspace}
\newcommand{\dataExpChatPlagSimilarityPTwoClosePositionStd}{0\xspace}
\newcommand{\dataExpChatPlagSimilarityPTwoIdent}{6\xspace}
\newcommand{\dataExpChatPlagSimilarityPTwoIdentRatio}{28.6\%\xspace}
\newcommand{\dataExpChatPlagSimilarityPTwoIdentPresence}{6\xspace}
\newcommand{\dataExpChatPlagSimilarityPTwoIdentPresenceRatio}{16.2\%\xspace}
\newcommand{\dataExpChatPlagSimilarityPTwoIdentPresenceRatioChatters}{26.1\%\xspace}
\newcommand{\dataExpChatPlagSimilarityPTwoIdentPositionMean}{1\xspace}
\newcommand{\dataExpChatPlagSimilarityPTwoIdentPositionMedian}{1\xspace}
\newcommand{\dataExpChatPlagSimilarityPTwoIdentPositionStd}{0\xspace}
\newcommand{\dataExpChatPlagSimilarityPTwoFirstMean}{0.67\xspace}
\newcommand{\dataExpChatPlagSimilarityPTwoFirstMedian}{0.71\xspace}
\newcommand{\dataExpChatPlagSimilarityPTwoFirstStd}{0.31\xspace}
\newcommand{\dataExpChatPlagSimilarityPOneTotalStats}[1]{\avgvals{0.44}{0.45}{0.37}{#1}\xspace}
\newcommand{\dataExpChatPlagSimilarityPOneChatterStats}[1]{\avgvals{0.53}{0.57}{0.34}{#1}\xspace}
\newcommand{\dataExpChatPlagSimilarityPOneAccessStats}[1]{\avgvals{0.65}{0.68}{0.26}{#1}\xspace}
\newcommand{\dataExpChatPlagSimilarityPOneClosePositionStats}[1]{\avgvals{6.67}{5}{5.7}{#1}\xspace}
\newcommand{\dataExpChatPlagSimilarityPOneIdentPositionStats}[1]{\avgvals{7.31}{5}{6.17}{#1}\xspace}
\newcommand{\dataExpChatPlagSimilarityPOneFirstStats}[1]{\avgvals{0.62}{0.62}{0.28}{#1}\xspace}
\newcommand{\dataExpChatPlagSimilarityPTwoTotalStats}[1]{\avgvals{0.23}{0}{0.36}{#1}\xspace}
\newcommand{\dataExpChatPlagSimilarityPTwoChatterStats}[1]{\avgvals{0.48}{0.55}{0.39}{#1}\xspace}
\newcommand{\dataExpChatPlagSimilarityPTwoAccessStats}[1]{\avgvals{0.66}{0.71}{0.29}{#1}\xspace}
\newcommand{\dataExpChatPlagSimilarityPTwoClosePositionStats}[1]{\avgvals{1}{1}{0}{#1}\xspace}
\newcommand{\dataExpChatPlagSimilarityPTwoIdentPositionStats}[1]{\avgvals{1}{1}{0}{#1}\xspace}
\newcommand{\dataExpChatPlagSimilarityPTwoFirstStats}[1]{\avgvals{0.67}{0.71}{0.31}{#1}\xspace}
\newcommand{\dataExpSubmissionsCountTotal}{556\xspace}
\newcommand{\dataExpSubmissionsCountRatio}{100\%\xspace}
\newcommand{\dataExpSubmissionsCountMean}{15.03\xspace}
\newcommand{\dataExpSubmissionsCountMedian}{14\xspace}
\newcommand{\dataExpSubmissionsCountStd}{9.43\xspace}
\newcommand{\dataExpSubmissionsPOneTotal}{495\xspace}
\newcommand{\dataExpSubmissionsPOneRatio}{89\%\xspace}
\newcommand{\dataExpSubmissionsPOneMean}{13.38\xspace}
\newcommand{\dataExpSubmissionsPOneMedian}{12\xspace}
\newcommand{\dataExpSubmissionsPOneStd}{9.5\xspace}
\newcommand{\dataExpSubmissionsPTwoTotal}{61\xspace}
\newcommand{\dataExpSubmissionsPTwoRatio}{11\%\xspace}
\newcommand{\dataExpSubmissionsPTwoMean}{1.65\xspace}
\newcommand{\dataExpSubmissionsPTwoMedian}{1\xspace}
\newcommand{\dataExpSubmissionsPTwoStd}{2.32\xspace}
\newcommand{\dataExpAttempts}{37\xspace}
\newcommand{\dataExpSubmissionsCountStats}[1]{\avgvals{15.03}{14}{9.43}{#1}\xspace}
\newcommand{\dataExpSubmissionsPOneStats}[1]{\avgvals{13.38}{12}{9.5}{#1}\xspace}
\newcommand{\dataExpSubmissionsPTwoStats}[1]{\avgvals{1.65}{1}{2.32}{#1}\xspace}
\newcommand{\dataExpMessagesCount}{349\xspace}
\newcommand{\dataExpMessagesCountPrompts}{177\xspace}
\newcommand{\dataExpMessagesCountResponses}{172\xspace}

\begin{abstract}

Programming students have a widespread access to powerful Generative AI tools like ChatGPT. 
While this can help understand the learning material and assist with exercises, educators are voicing more and more concerns about an overreliance on generated outputs and lack of critical thinking skills. 
It is thus important to understand how students actually use generative AI and what impact this could have on their learning behavior. 
To this end, we conducted a study including an exploratory experiment with 37 programming students, giving them monitored access to ChatGPT while solving a code authoring exercise. 
The task was not directly solvable by ChatGPT and required code comprehension and reasoning.
While only 23 of the students actually opted to use the chatbot, the majority of those eventually prompted it to simply generate a full solution.
We observed two prevalent usage strategies: to seek knowledge about  general concepts and to directly generate solutions.
Instead of using the bot to comprehend the code and their own mistakes, students often got trapped in a vicious cycle of submitting wrong generated code and then asking the bot for a fix.
Those who self-reported using generative AI regularly were more likely to prompt the bot to generate a solution.
Our findings indicate that concerns about potential decrease in programmers' agency and productivity with Generative AI are justified.
We discuss how researchers and educators can respond to the potential risk of students uncritically over-relying on Generative AI.
We also discuss potential modifications to our study design for  large-scale replications.
\end{abstract}





\maketitle


\section{Introduction}\label{introduction}

With the public release of \gai tools such as \cgpt \cite{O22_ChatGPT} and GitHub Copilot \cite{GH22_Copilot}, students in programming courses now have access to code authoring tools capable of solving coding exercises \cite{C21_Codex, BGK23_Overcome, S23_Progress} and exam questions entirely \cite{F23_Exam}.
However, such tools also tend to confidently present incorrect information \cite{C21_Guess, K24_StackOverflow, T24_DebugBench} or generate subtly incorrect code \cite{D23_Liability} which may be difficult to detect for beginners \cite{P23_Classroom, Stanik:ICSME:18}.
Furthermore, reliance on code generation can negatively impact code authoring skills \cite{K23_Support, J24_Impact}.

Due to the relative novelty of these \gai tools, \cgpt for example released in November 2022, their effects are not yet fully understood.
Initial studies have shown conflicting results regarding the impact of \gai on learners, with some reporting neutral or slightly positive changes \cite{K23_Support, V24_CS1, X24_Intro}, while others even found adverse effects \cite{J24_Impact, P24_Novices}. 
Nonetheless, the widespread use of these tools among students \cite{P23_GAI4Ed} increasingly requires educators to respond.

A recent interview study with individual students revealed interest in using the technology for purposes like generating supplementary learning materials, but also skipping coursework they don't find engaging enough \cite{Z23_GAI4Ed}.
Meanwhile, university instructors are largely unsure about how many of their students are using \cgpt and to what extent \cite{P23_GAI4Ed}.
Across multiple surveys and position papers, educators have expressed  concerns, particularly about over-reliance on generated content and about the ease of cheating \cite{CCS23_Cheating, P23_GAI4Ed, Z23_GAI4Ed, B23_Hard}.

In response, even educators with a positive sentiment towards \gai have begun implementing restrictions and bans in the classroom.
Two major approaches appear to emerge from this development: either a) embracing \gai as a tool not only for teaching but also as a core software development tool, or b) preventing and restricting its use \cite{GL23_Ban}. 
Both approaches involve considerable work, as the curriculum and course design need to be modified to either include or exclude these tools from the classroom and the students' overall learning journey.

Before taking such impactful decisions, it is particularly important to understand how programming students are actually using \gai tools and whether the usage strategies would primarily help them learn or rather avoid coursework. 
To this end, we report on a study with programming beginners who had  monitored access to \cgpt, focusing on the following research questions:
\begin{itemize}
    \item \textbf{\rqeval:} Could a student pass our introductory programming course using only \cgpt-generated answers?
    \item \textbf{\rqstrat:} What strategies do students employ when using \cgpt while solving a programming exercise?
    \item \textbf{\rqauto:} How much work do students delegate to \cgpt and how much autonomous thinking do they keep within a programming exercise?
\end{itemize}

Our study is unique concerning the task we gave to participants, since it was not directly solvable by \cgpt. Students had to comprehend and reason about the code \cite{Maalej:TOSEM:2014} to arrive at the solution or to formulate a relevant prompt for getting effective assistance from the bot. 
By analyzing the interactions of participants with the bot as well as their previous \gai experience, we observed a widespread ``lazy'' cycle of asking the bot to solve the task, executing its wrong or misleading suggestion, and then copying the error message and asking again.
By analyzing the diff of the submitted code, we found that students relying on the bot were rather following the bot suggestions even with completely different solution approaches, while students without the bot were rather trying to incrementally improve their solutions. 
In 62\% of cases with generations, the modified submission was more similar to the generated code than to the previous submission.
We report on a batch of qualitative and quantitative analyses and discuss potential implications on educational assessment as well as effective usage of \gai by programming students and novice developers.

The remainder of the paper is structured as follows:
Section \ref{approach} describes the design of our study to answer these RQs, including research methods, participants, and tasks.
Section \ref{results} presents the results along the RQs. 
Then, Section \ref{discussion} summarizes the findings, highlights a few observations, and discusses the study implications and limitations.
Finally, Section \ref{related} discusses related work while 
Section \ref{conclusion} concludes the paper.

\section{Study Design}\label{approach}
We first present the research methodology for the task-solving evaluation and student experiment. Then, we introduce the study setting, including the programming course and the participants. 

\subsection{Research Methodology}

To answer \rqeval, we evaluated the task-solving performance of \gpt models on all exercise assignments of our large first-semester introductory programming course.
To answer \rqstrat and \rqauto, we conducted an experimental study with students in the same course, who volunteered to solve an exercise while having access to a monitored \cgpt interface.
We recorded and labeled the students' interactions with \cgpt, and performed pattern analysis on the resulting chat logs.

\subsubsection{Task-Solving Evaluation}\label{sec:a_eval}

For all assignments that must to be solved to pass the programming course, we created individual prompts to be submitted to the \gpt API.
Some exercises are split into subtasks with different levels of granularity.
We separated these tasks into minimal congruent subtask groups, \ie subtasks that shared the same project context, such that only immediately related subtasks were joined into a single prompt.
Additionally, if the exercise required knowledge of an existing template project for students to build off of, that project was also appended to the prompt text.
The full input to the language model consisted of a short system prompt instructing it to solve a Java programming exercise, the exercise text as given to students, and – if applicable – a code snippet containing the project context.

At time of the study, \gptf was only available through paid access and with strict rate limitations within \cgpt.
We thus decided to evaluate both the latest model and the one most accessible to students.
For \gptt, this was the model \code{gpt-3.5-turbo-1106}. For \gptf, we used the \code{gpt-4-1106-preview} model.
The temperature parameter for both models was set to \code{0.0} to reduce randomness in the output.
We inserted the generated solution into the appropriate project files, performed the usual tests human tutors do for the exercises to pass, and recorded the results alongside the generated code.


As the course relies heavily on manual reviews by course tutors \cite{H19_SE1}, any submitted code not only needs to meet the functional requirements of the exercises, but also match the paradigms and coding conventions taught in the course.
In fact, the majority of in-person exercises included in the studied course did not have automated unit tests to verify the student solutions.
Instead, in addition to code reviews, course tutors performed a series of manual tests, by asking students to demonstrate specific behaviors using the interactive \bluej object interface.
These code reviews and manual tests, which are required to pass, are specified in the tutor instructions for each exercise.

The criteria applied by course tutors can be divided into the following categories:
\begin{itemize}
    \item \textbf{Task Requirements.} The student answer must fulfill the requirements of the task itself. For code authoring tasks, this includes being syntactically and logically correct (\textit{\syntax} and \textit{\logic}), as well as meeting all functional requirements (\textit{\reqs}). 
    For other knowledge tasks (i.e.~writing tasks), the answers should address the actual question (\textit{\answers}), be factually correct (\textit{\fact}), and address all aspects of the question (\textit{\reqs}).
    \item \textbf{Course Requirements.} The answer must fulfill the requirements shared across all exercises of the course, most importantly the adherence to the course's coding conventions (\textit{\conventions}).
    These are not explicitly stated in each question text, but still expected from the students.    
    \item \textbf{Authenticity.} A student must plausibly be able to derive (and explain) the provided answer from the learning material (\textit{\curric}). Students utilizing more advanced concepts must demonstrate a strong understanding of their solution, otherwise their answer is considered likely plagiarized and thus rejected.
    Additionally, any comments and explanations must be consistent with the code they describe (\textit{\intent}), or the authenticity of the presented solution is questioned.
\end{itemize}

We evaluated the generated results according to these criteria as either \textit{pass} or \textit{fail}.
If there was no code in the response, we marked the corresponding criteria as \textit{undetermined}.

\subsubsection{Student Experiment}

In the experimental study, we presented students with an optional  code understanding and authoring exercise that they were asked to solve to the best of their ability.
Additionally, participants were given access to a \cgpt-like chatbot to assist them with the exercise. The use of the bot was not required.
We recorded participants code changes, chatbot conversations, copy/paste events, and their solution submissions.
The exercise was realized as a quiz activity on the \moodle e-learning platform, with two introductory self-assessment semantic scale questions and a \textit{\coderunner} coding task. 
Students were used to this e-learning platform as about 20\% of their usual assignments were handled there.

\begin{figure}
    \includegraphics[width=0.6\textwidth]{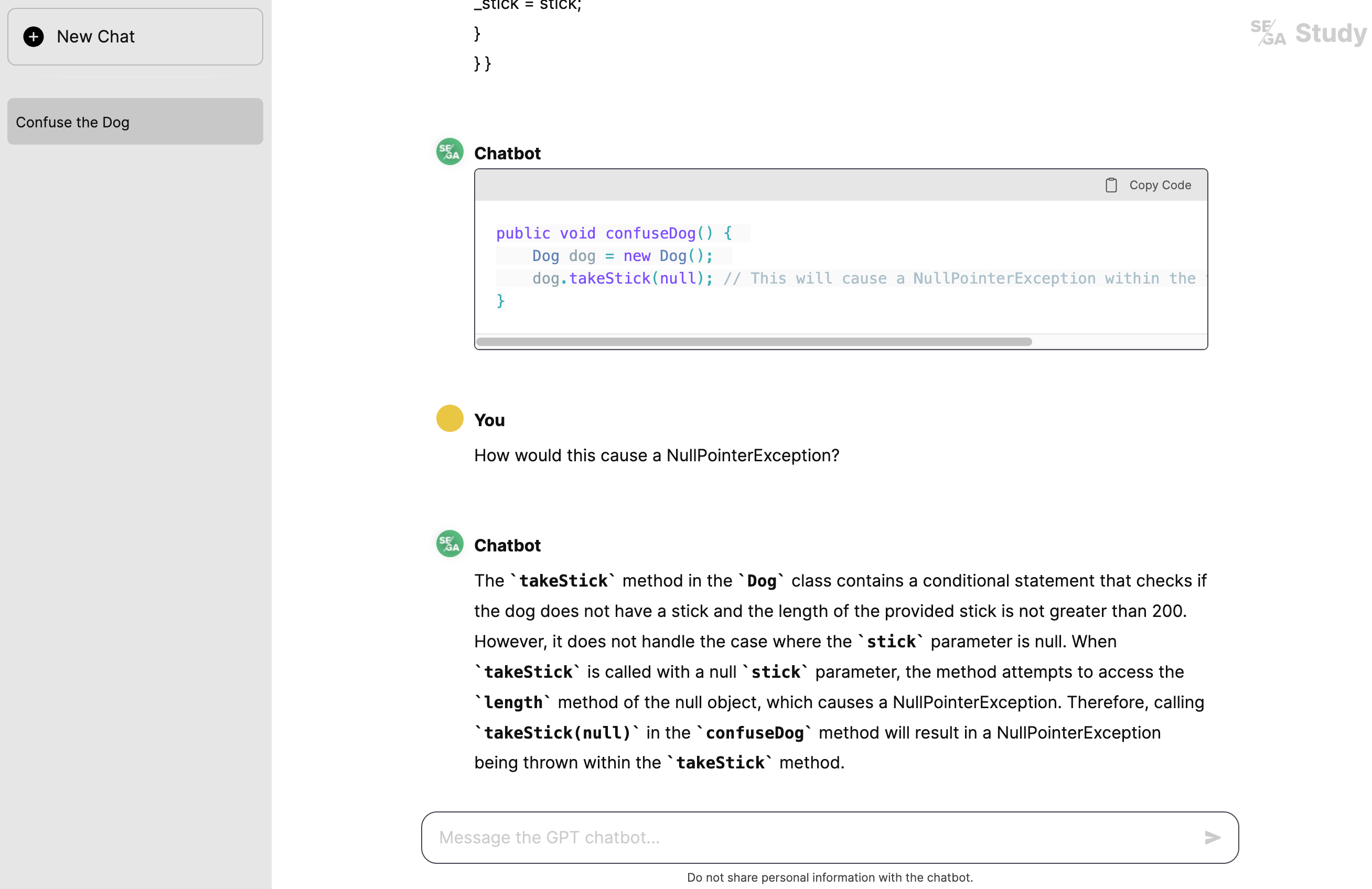}
    \caption{The chatbot \ui participants were shown during the study.}
    \label{fig:a_chatbot_sample}
    \Description{An illustrative image of the chatbot user interface. The left side shows a sidebar with a button titled ``New Chat'' and a single conversation tab titled ``NullpointerException Unleashed''. The right side shows a chat conversation with three messages. The first is a message by the chatbot with a syntax-highlighted code snippet. The second is a message by the user, labeled ``You''. The third is another message by the chatbot with regular text. Below there is a text field for text entry.}
\end{figure}

The chatbot \ui featured a subset of the functionality of \cgpt.
Participants could create conversation threads and hold a back-and-forth conversation with the chatbot.
Responses were gradually revealed as they were generated, and rendered with full Markdown support.
Overall the bot very closely mimics the look of \cgpt as a generic \gai tool.
Figure \ref{fig:a_chatbot_sample} shows a screenshot of the chatbot. It uses the \gpt API to generate responses to participants' prompts.
In the experiment, we used the \code{gpt-3.5-turbo-1106} model, which was the latest publicly available  at the time.

\paragraph{Data Analysis}\label{sec:a_exp_labels}

Our data analysis includes a quantitative and a qualitative component.
We collected aggregate activity metrics on the students' attempts, performance, interactions, and the timeline of events.
Additionally, two authors \textbf{manually labeled} independently each student prompt and each chatbot response to identify types of requested information and interaction patterns.

With \textbf{\rqstrat}, we set out to  understand  the kinds of prompts students  submit to an AI assistant.
For this, we coded the general objective of each prompt based on labels listed in Table~\ref{tab:a_prompt_labels}. 
The labels were initially created through a deductive approach \cite{M13_Docs, Neuendorf:2017}, and then consolidated after reviewing the actual  chat logs.
After independent labeling, the first annotator reviewed all disagreements and resolved those where one label was clearly inapplicable by definition.
The remaining disagreements were individually discussed and mutually resolved by the annotators.
For the prompt labels, the annotators had an initial agreement of 72\% ($\kappa = 0.65$).

Initially, we had outlined a label \textit{Retry} for participants requesting a corrected solution specifically without describing the preceding failure, \ie asking the bot to simply ``try again''.
We found this to be a very rare occurrence and very similar to the label \textit{\pfix}. Therefore, we decided to merge both labels. 
For coarser analysis, we also grouped multiple prompt labels into a category \textit{\pgen}, which represents all prompts asking for code output.
Similarly, the category \textit{\phelp} represents prompts asking for knowledge or assistance in understanding or completing the exercise.

\begin{table}
\caption{Labels for prompts submitted by participants, with descriptions and examples.}
\label{tab:a_prompt_labels}
\begin{tabular}{ l p{11.5cm} }
\textbf{Label} & \textbf{Description} \\
\hline
\hline
\hspace{2mm}& \textit{\textbf{\pgen}} (Code Generation)\\
\hline
\textit{\psolve} & Request a solution to the problem directly, including by pasting the question text.\\
& \textit{Example:} ``[question text] How do I produce the NullPointerException here?''\\
\hline
\textit{\pfix} & Request a corrected version of a non-working solution.\\
& \textit{Example:} ``I tried your solution but I got this error: [error message]''\\
\hline
\hline
& \textit{\textbf{\phelp}} (Knowledge \& Comprehension)\\
\hline
\textit{\phint} & Ask for a hint or partial answer to the problem, given which more work is still required to arrive at the solution.\\
& \textit{Example:} ``[question code] Which of these methods might produce a null error?''\\
\hline
\textit{\plookup} & Ask for information about a concept from the exercise, without sharing the specific problem the participant is trying to solve.\\
& \textit{Example:} ``When do NullPointerExceptions occur in Java?''\\
\hline
\textit{\pcompr} & Ask for an explanation for observed behavior or for a proposed solution.\\
& \textit{Example:} ``Why does this code cause a NullPointerException?''\\
\hline
\hline
\textit{\pothers} & Any prompts not assignable to the remaining labels, \eg off-topic remarks or incomplete messages.\\
\end{tabular}

\end{table}

We also coded the chatbot responses following the labels listed in Table \ref{tab:a_response_labels}.
These labels are non-exclusive, \ie a response can simultaneously be assigned two labels.
For example, a generated code solution that is additionally explained in text would be assigned both \textit{Solves} and \textit{Explains}.
After independent labeling, the annotators had 80\% agreement on the labels for this task.

\begin{table}
\caption{Labels for the responses generated by the \gptt chatbot.}
\begin{tabular}{ l p{11.5cm} }
\textbf{Label} & \textbf{Description} \\
\hline
\hline
\textit{Informs} & Provides general information about a concept, with no reference to the exercise problem.\\
\hline
\textit{Solves} & Provides a code snippet that attempts to solve one of the exercise problems.\\
\hline
\textit{Explains} & Provides an explanation for a code snippet.\\
\end{tabular}
\label{tab:a_response_labels}
\end{table}

Lastly, for both generated factual claims and code solutions, we labeled whether they were correct or incorrect. For solution attempts, this means whether the code fulfills the stated requirements.
In cases where it was ambiguous whether the response was correct given the intent of the request, such as when the preceding prompt did not contain any instructions, neither label was applied.

\label{sec:a_ex_auto}
To answer \textbf{\rqauto}, inspired by the concerns of ``blindly copying and pasting solutions'' \cite[p.15]{P23_GAI4Ed}, we tracked clipboard events – \ie cut, copy and paste actions – both on the chatbot and exercise pages.
To trace the flow of information, we then grouped the recorded actions into pairs of a cut or copy event directly followed by a paste event. We compared the text contents to ensure the events in the pair were actually related.

To understand the degree to which submission attempts were based on chatbot-generated code, we compared each submission to all code snippets previously generated in the chat history using Ratcliff/Obershelp \cite{R88_Matching}.
We selected this similarity measure based on its modeling of common substrings, which is less affected by modifications common in source code, such as the deletion or reordering of entire statements.
To reduce noise in the similarity results, we extracted the function body relevant to the exercise from both the generated solution and the student submission, normalized the whitespace and removed line comments before comparing.

Due to the small space of possible solutions, even semantically heavily modified submissions would have some similarity to previously generated code.
Through manual review, we found that submissions with a similarity score above 90\% were unambiguously close, structurally and semantically, to the generated code solution.
We thus consider this our threshold for a \textit{close match}.

Additionally, we analyzed the similarity between the current and previous submission, as well as to any code generations between two submissions.
However, this measure does not differentiate between functional and non-functional modifications.
For example, changing a local variable name has the same impact on similarity as changing out a method call, despite the former not having any impact on the solution.
To address this issue, we manually reviewed all code diffs generated by consecutive submissions with a code generation in between, and labeled them according to the labels listed in Table  \ref{tab:a_change_labels}.
The agreement on this labeling task was 82\% ($\kappa = 0.72$).

\begin{table}
\caption{Labels for the code changes between  consecutive submissions with chatbot interactions in between.}
\begin{tabular}{ l p{11cm} }
\textbf{Label} & \textbf{Description} \\
\hline
\hline
\textit{\rfull} & The generated code was submitted with no semantic modifications.\\
\hline
\textit{\ridea} & The participant incorporated an approach from the generated solution into the code.\\
\hline
\textit{\rsyntax} & The participant applied the syntax of a generic example snippet into the code.\\
\hline
\textit{\rexplain} & The participant modified their code according to a generated textual explanation or instruction.\\
\hline
\textit{\rnone} & There is no discernible connection between the generation and the code.\\
\end{tabular}
\label{tab:a_change_labels}
\end{table}

We used \textbf{sequence pattern mining} to identify common patterns of actions in the students' interaction logs. 
We treated each student interaction as a single event, and represented individual prompts on the category level.
We merged consecutive copy-paste actions, generated by copying separate passages of text from the same page, into one event.

For this analysis, we used the NOSEP algorithm by \citet{W17_NOSEP}, as we were looking for consecutive, non-overlapping subsequences that may occur multiple times in one sequence.
We specified a gap range of $[0,2]$ to allow for minor tolerance against actions such as \textit{\pothers} prompts within a pattern, and a minimum support of $10$.
To identify each pattern occurrence and link it back to the participant, we used the NETLAP algorithm \cite{W17_NETLAP} on each mined pattern.

\subsection{Setting and Participants}

\subsubsection{Course Structure}\label{course}
Both the task-solving evaluation and the experiment were conducted as part of the introductory course \textit{Softwareentwicklung 1} (SE1) at the University of Hamburg.
The course typically accommodates between 500 and 700 students each year.
The course is designed to accommodate programming novices without any prior experience. 
The course takes place over a 14-week semester, covering the basics of programming, control flow and logic, object-oriented programming, data structures, and code quality. 

Each week, a worksheet with 2-3 major exercises is released to students, which they solve through in-person pair programming during two-hour lab sessions.
The exercises include code authoring, answering questions, or creating diagrams.
The exercises are inspired by \citet{B09_Java}.

Upon completing a major exercise, students request a tutor to review and accept their work.
During this review, tutors ask additional questions to test the students' understanding of the material, allowing them to ``identify and correct misunderstandings early and provide immediate feedback with personalized explanations'' \cite{H19_SE1}.
Additionally, tutors can verify the authenticity of the presented work.
Tutors are instructed to reject work that either does not meet the course requirements or that the students cannot adequately explain.
While this does not prevent plagiarism outright, students are disincentivized from presenting plagiarized work unless they have a thorough understanding of the concepts and reasoning behind it.

An interactive quiz hosted on the \moodle platform is released alongside the worksheet each week.
The course requires students to achieve a grade of at least 90\% averaged across all quizzes.
Students can attempt the quiz as often as they like, but only within a week of its release. 
The assessments also include code authoring questions using the \moodle \textit{\coderunner} \cite{LH16_CodeRunner} plugin.
In these questions, students get the problem statement and a minimal code editor in their browser.
Their code is automatically compiled and tested whenever they use the ''Check'' button to submit an attempt.
Immediate feedback from the test runner is provided, showing compilation errors, runtime exceptions, or test failures.
There is no penalty for repeated attempts.

\subsubsection{Experimental Task}\label{sec:a_exp_task}
\begin{figure}
    \centering
    \includegraphics[width=0.95\textwidth]{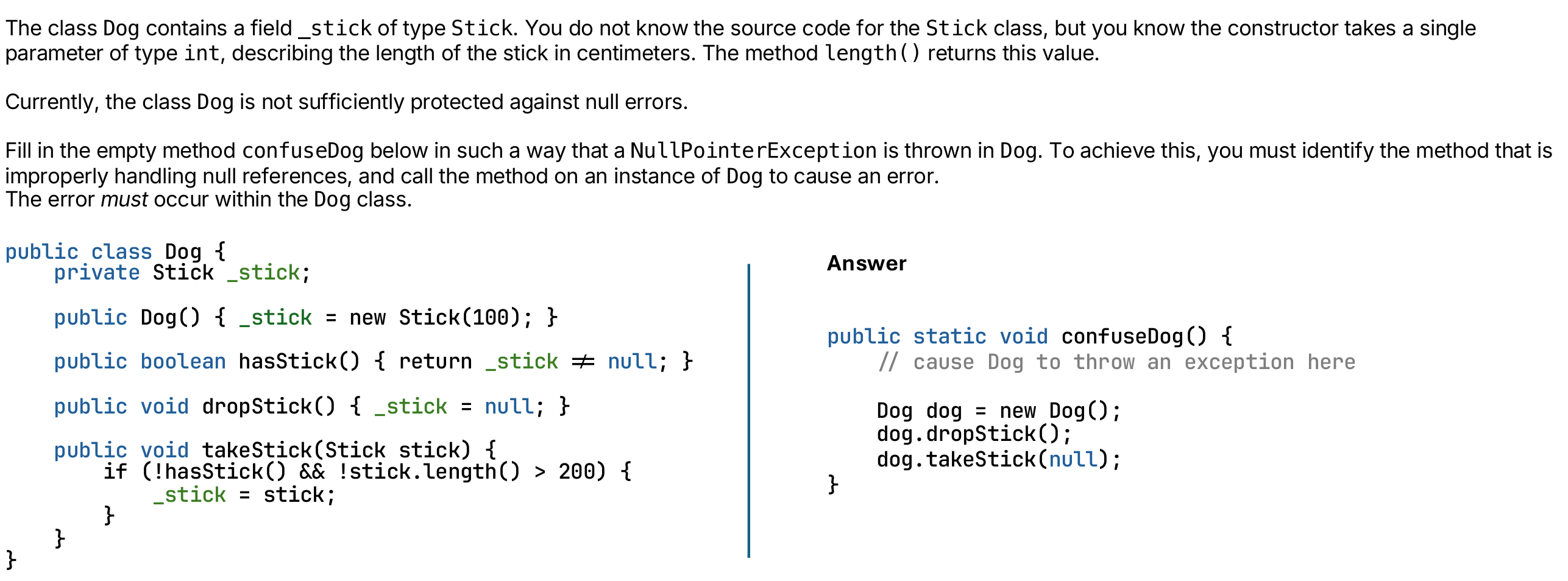}
    \caption{The exercise text, the provided source code (left) and the solution (right) to problem \exfind. The source code has been abridged for brevity.}
    \label{fig:/exp:exercise}
\end{figure}

For the purposes of generating meaningful student-chatbot interactions, it was important that the task could not be reliably solved by \cgpt.
We found that at the experience level we targeted, isolated code authoring exercises, including most of the existing SE1 course materials, were unsuitable for this reason.
Prior work found \cgpt to struggle with identifying issues in code, especially for logic errors \cite{T24_DebugBench, H23_Responses}.
We thus designed a code reading and comprehension exercise that would be particularly challenging for the \llm, which we confirmed in initial testing.
Still, we did not increase the complexity or context size beyond what the students were accustomed to.

The exercise is presented in Figure \ref{fig:/exp:exercise}.
It consisted of two problems.
In \textbf{\exfind}, participants should  identify and trigger a null-related bug in a provided class.
Crucially, in our problem setup, the ``obvious'' solution of passing null wherever possible was insufficient, such that students could not solve the exercise by simply applying a recently learned pattern without tracing the provided code.
Instead, they had to combine their knowledge of multiple concepts, such as object state, null dereferencing and operator short-circuiting.
Additionally, purposefully producing failure states was a novel task for the students who had so far only learned why they occur and how to prevent them.
In \textbf{\exfix}, participants were then asked to add the missing \code{null} check to the original code, to prevent such errors.
During the experiment, participants were not allowed to reference other course material, look up information online, or ask other students for help.

The experiment was conducted \textbf{in-person} in a controlled environment  between November and December 2023.
All participants were students of the same  course.
To accommodate their schedules, students could participate immediately following each of their respective lab sessions.
Students  were informed of the study during a course lecture as well as through a \moodle announcement.
Additionally, students were approached for recruitment in-person during the lab sessions.
As incentive,  participants who completed the experiment were able to participate in a gift card raffle.

Participation was completely voluntary and did not have any impact on the actual course  grading. 
We informed participants that they were required to either complete the exercise or spend at least 20 minutes on it.
Otherwise, they would not be eligible for the reward and their data would be discarded. There was no upper time limit.
While participants were not misled about the purpose of the study, it was kept intentionally vague to reduce potential subject expectancy effects.
It was described to them as a study of  tool support for learning and problem-solving.


\section{Results}\label{results}
We present the results along the three research questions.

\subsection{Task-Solving Evaluation (RQ1)}\label{eval}

\begin{figure}
\centering
\begin{minipage}[t]{.475\textwidth}
    \centering
    \includegraphics[width=.9\textwidth]{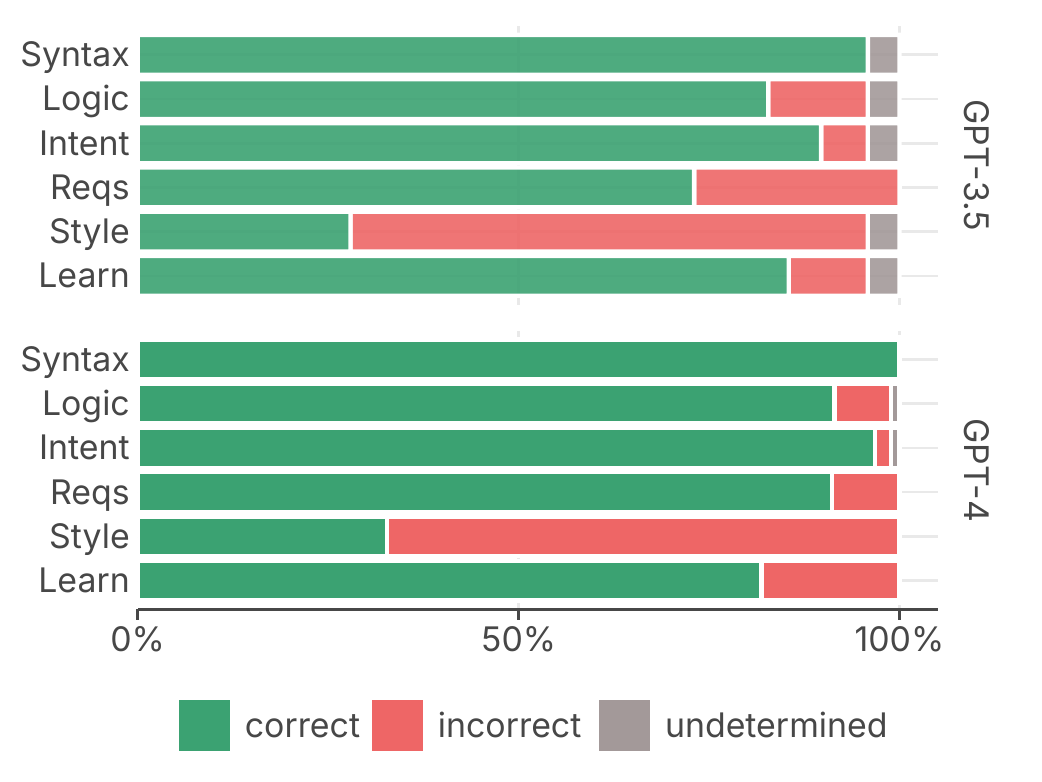}
    \captionof{figure}{Performance in coding exercises.}
    \label{fig:eval/results/code}
\end{minipage}\hspace*{\fill}
\begin{minipage}[t]{.5\textwidth}
    \centering
    \includegraphics[width=.9\textwidth]{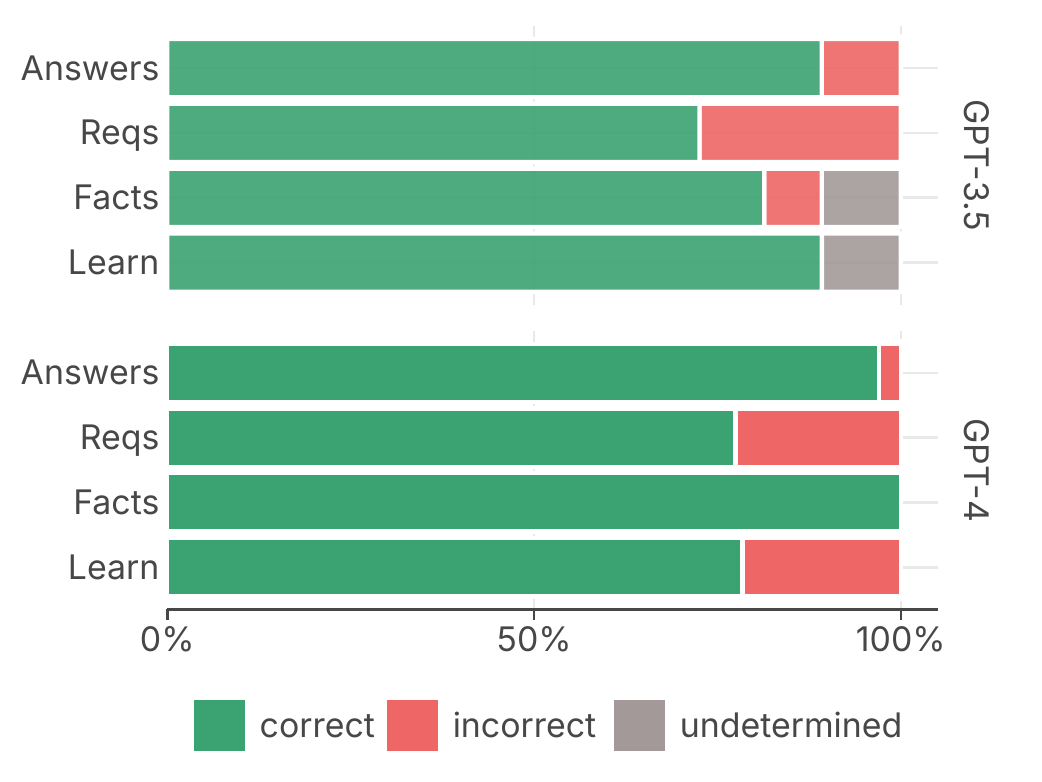}
    \captionof{figure}{Performance in question-answering exercises.}
    \label{fig:e_result_writing}
\end{minipage}
\end{figure}

The results of the task-solving evaluation are depicted on Figure \ref{fig:eval/results/code} and Figure \ref{fig:e_result_writing}. 

\subsubsection{Coding Exercises}\label{sec:eval/results/code}
Both \gpt models produced syntactically correct code in all responses.
Responses from \gptt had slightly more logical errors than \gptf, and failed to meet the exercise requirements more than twice as often.
Despite not being explicitly instructed to do so, a large majority of the responses were aligned with what a student of the course would already know at that time.
Inconsistencies between the described intent and the actual implementation were rare, though in one particularly obvious case, \gptt provided multiple line comments describing logic that was entirely absent from the code.

Adherence to the course coding conventions was low overall.
Neither language model's baseline code generation – \ie with no existing code provided as a reference – was in alignment with the  formatting guidelines. 
Even with context code, the majority of generated code did not follow the demonstrated bracketing style, nor were documentation comments added consistently.

\gptt also scored notably higher on \textit{\curric} in exercises where project code was provided, though this is similarly related to the specific course conventions.
Most of the failures to fulfill the \textit{\curric} criterion were caused by use of the keyword \code{this}, which is introduced separately and later in the course than classes and constructors.

\subsubsection{Writing Exercises}\label{sec:e_eval_writing}

We found that the responses generated by \gptf went beyond the scope of the course material several times.
These cases include mention of classes such as \code{BigDecimal} that students would not have heard of, referencing static/class methods for exercises prior to their introduction, and in one instance, discussing the thread safety of \code{HashMap} and \code{HashSet}.

The majority of factually incorrect responses were marked as such based on technicalities.
For example, \gptt claimed that insertion into a \code{LinkedList} at an index was an $\textrm{O}(1)$ operation.
This is only true if you already hold a reference to an adjacent node or ignore the traversal required to reach the a node first.
The course material states that the default insertion is an $\textrm{O}(n)$ operation.

\subsection{Chatbot Use (RQ2)}

\subsubsection{Exercise Completion}

In total, 42 students participated in the experiment.
Three students participated in an initial test run with an older version of the exercise. Their results were not included in the analysis.
Additionally, two students ended the experiment after a short period (4.5 min.~and 9 min.)  without completing the tasks. We thus concluded that they dropped the participation and discarded their data. 
This leaves \dataExpChatPresenceTotal participants for the analysis.

\dataExpPerfMetricsRatioCorrectAll of participants completed the entire exercise correctly, and \dataExpPerfMetricsRatioCorrectAny successfully solved at least one problem. 
The average duration of experiment sessions was \dataExpPerfMetricsDurationCompleteMean (M = \dataExpPerfMetricsDurationCompleteMedian, $\sigma$ = \dataExpPerfMetricsDurationCompleteStd) for students who completed both problems, and \dataExpPerfMetricsDurationIncompleteMean (M = \dataExpPerfMetricsDurationIncompleteMedian, $\sigma$ = \dataExpPerfMetricsDurationIncompleteStd) for those who did not. 
Across the experiment, \dataExpSubmissionsCountTotal submission attempts were recorded. The vast majority (\dataExpSubmissionsPOneRatio) were for problem \exfind.
The participants made an average of \dataExpSubmissionsPOneStats{submission attempts for \exfind}, of which \dataExpOutcomesPOneCompilationErrorPrevalenceRatio resulted in compilation errors and \dataExpOutcomesPOneNoNPEPrevalenceRatio in a test failure because their code ran without errors.
For \exfix, participants made \dataExpSubmissionsPTwoStats{attempts} on average, of which \dataExpOutcomesPTwoCompilationErrorPrevalenceRatio resulted in compilation errors and \dataExpOutcomesPTwoNullNotCheckedPrevalenceRatio failed because their code did not handle the error case.

Before starting the exercise, we asked participants to rate how challenging they found the course on a 5-point scale, with 1 being very easy and 5 being very difficult.
We observed a moderate negative correlation between perceived course difficulty and exercise completion ($r = -0.47, p = 0.003$), as well as a moderate positive correlation between perceived difficulty and experiment duration ($r = 0.4, p = 0.013$).
This indicates that the students' performance in the experiment aligned with their general experience in the course.

\subsubsection{Usage of \gai}

Before the exercise, we asked participants how often they use \gai tools such as \cgpt or GitHub Copilot. 
\dataExpChatUseNeverTotalRatio reported to have never used such tools before.
\dataExpChatUseFrequentlyTotalRatio reported using \gai tools at least monthly (\textit{\ulessweekly} or \textit{\uweekly}).
We observed a moderate positive correlation between the reported use of \gai tools and the frequency of chatbot interactions during the experiment ($r = 0.35, p = 0.03$).
We did not observe any significant relationship between reported \gai use and performance on the exercise.

All participants had access to the chatbot interface.
Still, \dataExpChatFirstNoneRatio did not use it at all during the experiment.
In total, \dataExpMessagesCountPrompts prompts were sent to the chatbot and \dataExpMessagesCountResponses responses were received.
The remaining 5 prompts went unanswered due to network interruptions.
On average, each participant submitted \dataExpChatPrevalenceTotalTotalStats{prompts}.
Among \chatters only, the average was \dataExpChatPrevalenceTotalTotalChattersStats{prompts}. 
We found that the exercise completion was lower, on average, among \chatters, though the difference was not statistically significant.

The average experiment duration varied significantly between students who did and did not use the chatbot ($t = 3.3, p = 0.003$).
For \chatters, the average duration was 26:15 (M = 26:03, $\sigma$ = 10:38). For those who did not use the chatbot, this was 14:36 (M = 10:30, $\sigma$ = 10:27).
To account for participants leaving at arbitrary times if they could not solve the exercise after the first 20 minutes, we additionally separated them by exercise completion.
When comparing the duration only among completed attempts, the difference is not significant ($t = 1.9, p > 0.05$).
We therefore did not find a relationship between chatbot use and the time to arrive at the correct solution.

\subsubsection{Student Prompts}\label{sec:e_chat_prompts}
\begin{figure}
\centering
    \begin{minipage}[t]{.49\textwidth}
    \centering
    \includegraphics[width=.98\textwidth]{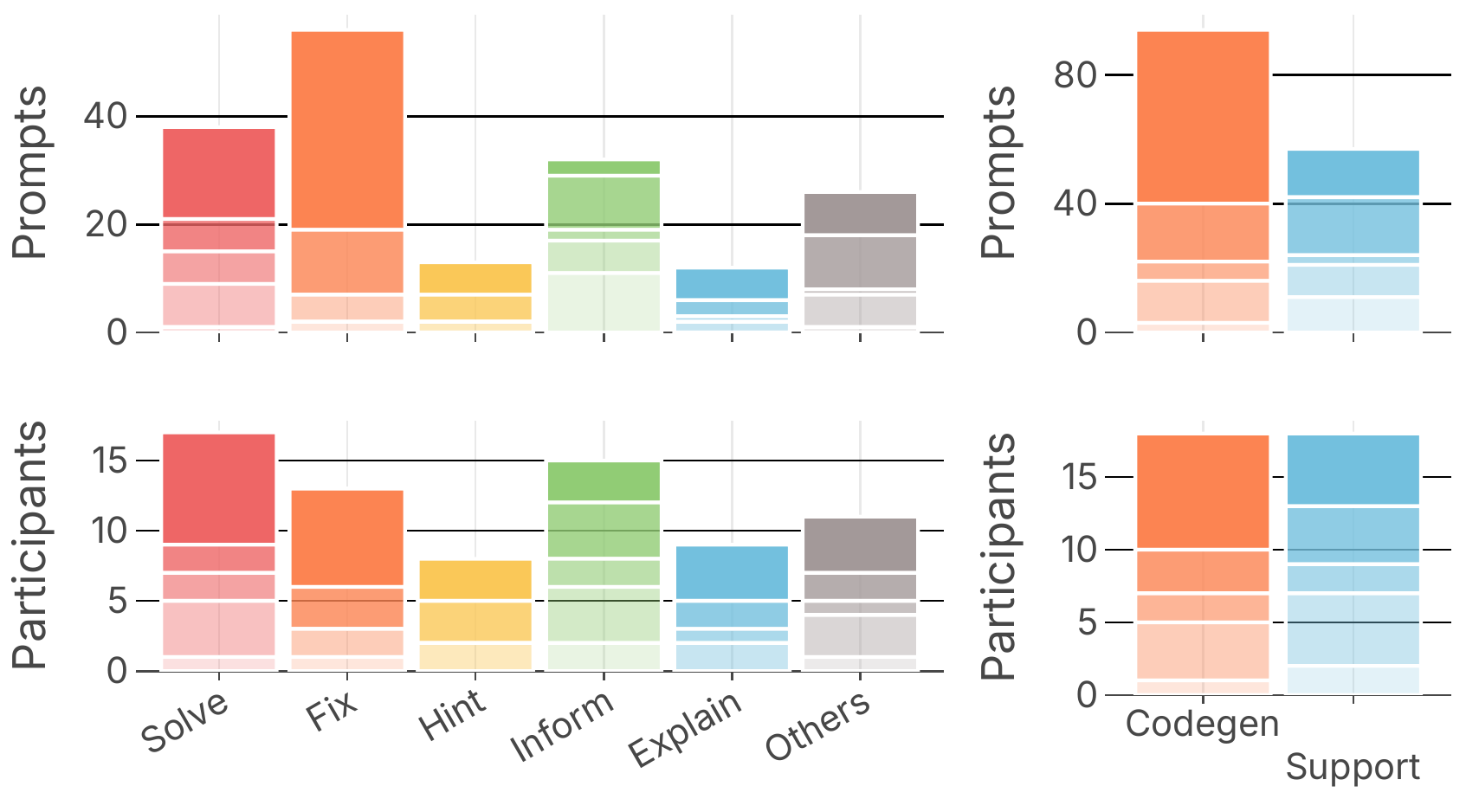}
    \captionof{figure}{Total number of prompts by type (top) and number of participants that submitted at least one prompt of that type (bottom). Opacity indicates self-reported \gai usage.}
    \label{fig:/exp:chat:presence/}
\end{minipage}\hspace*{\fill}
\begin{minipage}[t]{.49\textwidth}
    \centering
    \includegraphics[width=.98\textwidth]{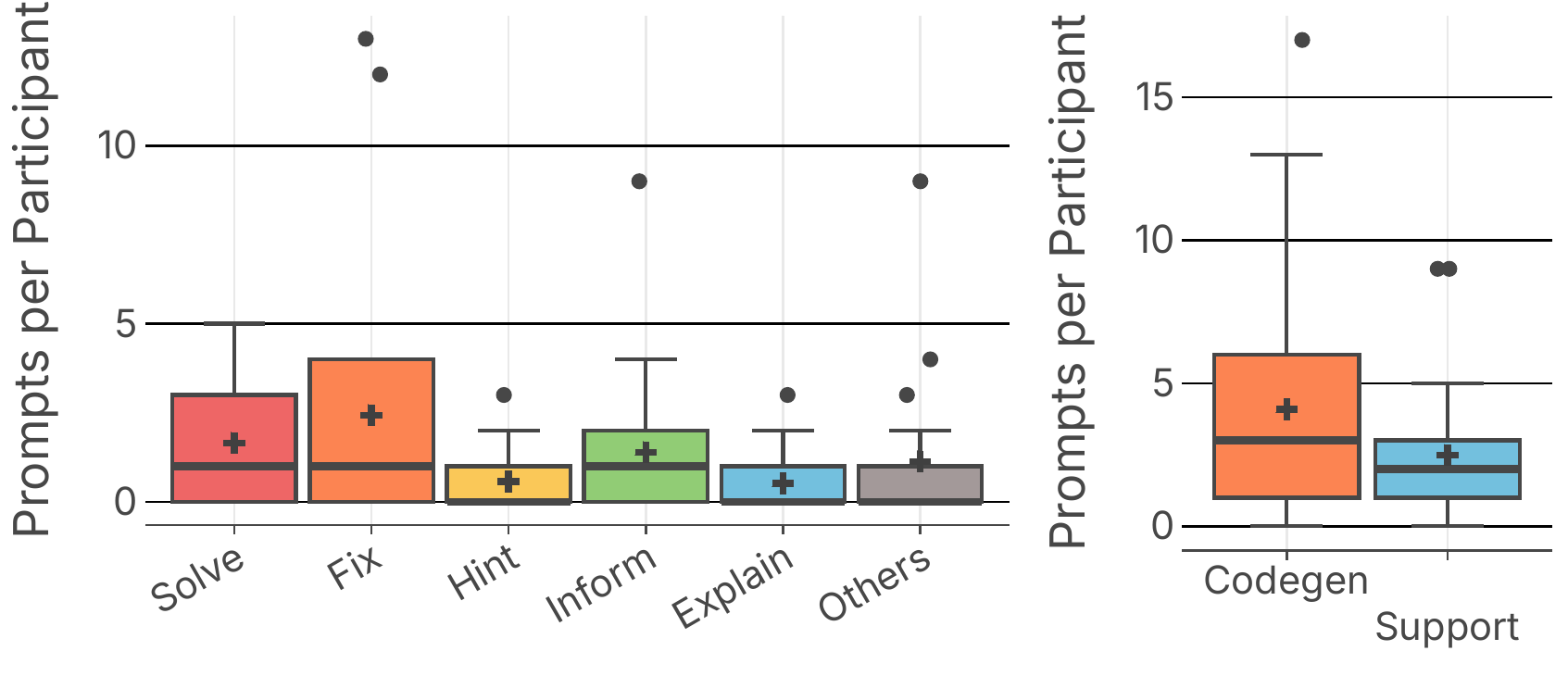}
    \captionof{figure}{Number of prompts submitted by each participant, aggregated across \chatters.}\label{fig:/exp:chat:occurrence/}
\end{minipage}
\end{figure}

In this section, all values are relative to the participants who used the chatbot during the experiment.
Figure \ref{fig:/exp:chat:presence/} shows the total number of prompts for each type, as well as the number of participants that have submitted such a prompt at least once.
Overall, we found that students submitted more \textit{\pgen} prompts than \textit{\phelp}, with a ratio of 62\% to 38\%\footnote{\textit{\pothers} prompts are excluded from the totals here, as they were exclusively either off-topic or erroneously submitted.}, confirming observations by \citet{M25_Patterns}.

Figure \ref{fig:/exp:chat:occurrence/} shows the number of occurrences of each prompt type per-participant.
We observed \textit{\psolve} prompts from  most of the participants, as \dataExpChatPresenceSolutionRatioChatters submitted a \textit{\psolve} prompt at some point during the experiment.
Interestingly, we also found that some participants submitted more than two \textit{\psolve} prompts.
Overall, the most common prompt type  was \textit{\pfix}, though this was driven in part by two outliers, who each submitted 13 \textit{\pfix} prompts during their session.
The least common prompt type was \textit{\pcompr}, observed in only \dataExpChatPresenceExplanationRatioChatters of conversations, constituting \dataExpChatPrevalenceRawExplanationRatio of prompts.
Of the \dataExpChatPrevalenceRawExplanationTotal \textit{\pcompr} prompts we found, five asked ``what is wrong with this code?'' after encountering an error, though without including the error message in the prompt.
Only in two prompts did a participant ask for an explanation of the generated answer after discovering it was incorrect.

Some participants started their chatbot conversations with questions related to the exercise, but without providing the necessary context.
In these cases, the chatbot would then ask for more information from the user.
Presumably, participants omitting context were under the impression that the chatbot had already been specifically instructed on their task.
Additionally, we observed participants prompting for a \textit{\pfix} without describing the problem, \ie prompting to the effect of ``that doesn't work, try again''.
This supports prior findings by \citet{M25_Patterns} and \citet{Kruse:ICSME:2024} that some students do not understand what information precisely needs to be shared with the chatbot.
More often, however, participants would repeat the problem statement instead, or slightly rephrase it.
One approach we observed was to repeat only the instruction itself, omitting the introduction text and context code, potentially in an attempt to ``remind'' the model of the task at hand.
Notably, this behavior seems to be uncommon among developers at large \cite{J24_Developers}.

\begin{figure}
\end{figure}

\begin{table}
\caption{Correlation analysis of prompt types to preceding survey responses using Kendall's $\tau$. Statistically significant values are bolded. For individual labels (top rows), significance was adjusted by Bonferroni correction ($\alpha = 0.05/5 = 0.01$).}
\label{fig:exp_chat_corr}
\begin{tabular}{p{2.75cm} r r}
\textbf{Label} & \textbf{Course Difficulty} & \textbf{\gai Usage}\\
\hline
\hline
\textit{\psolve} & -0.47 ($p$ = 0.02) & 0.12 ($p$ = 0.58)\\
\hline
\textit{\pfix} & 0.10 ($p$ = 0.63) & 0.42 ($p$ = 0.05)\\
\hline
\textit{\phint} & 0.14 ($p$ = 0.52) & 0.15 ($p$ = 0.48)\\
\hline
\textit{\plookup} & -0.23 ($p$ = 0.29) & \textbf{-0.61 ($p$ = 0.002)}\\
\hline
\textit{\pcompr} & 0.29 ($p$ = 0.17) & 0.13 ($p$ = 0.56)\\
\hline
\hline
\textit{\pgen} <> \textit{\phelp} & -0.07 ($p$ = 0.75) & \textbf{0.45 ($p$ = 0.03)}\\
    \end{tabular}
\end{table}

We also analyzed the distribution of prompt types among \chatters for relationships with the survey responses. 
Table  \ref{fig:exp_chat_corr} shows the results.
In particular, we found a significant negative correlation between the participants' self-reported usage of \gai tools and the occurrence of \textit{\plookup} prompts.

While not statistically significant, we observed a moderate negative correlation between perceived course difficulty and the share of \textit{\psolve} prompts.
In our experiment sessions, students who perceived the course as easier thus tended towards asking the chatbot for a code generated solution rather than hints or information.
Additional experimental verification is needed to determine whether this constitutes an actual relationship between student attitude and their \llm prompting behavior.

On the category level, we found that students who reported using \gai regularly submitted more \textit{\pgen} prompts as opposed to \textit{\phelp}.
This indicates notably different usage patterns and strategies for students that have more experience with \gai tools.

\subsubsection{Bot Responses}\label{sec:e_exp_responses}
\dataExpChatResponseSolvesRatio of the \dataExpChatResponseTotal chatbot responses contained an auto-generated solution to one of the exercise problems, of which \dataExpChatResponseSolutionExplanationRatio also contained a written explanation of the solution attempt.
In total, \dataExpChatResponseExplainsRatio of responses included an explanation, either for a generated solution, for an example code snippet, or for a general concept.

As intended by the study design, the \gptt model was unable to solve problem \exfind reliably.
Out of \dataExpChatResponsePOneSolutionTotal generated solutions, only \dataExpChatResponsePOneSolutionCorrectRatio were correct.
The model performed better on \exfix, where \dataExpChatResponsePTwoSolutionCorrectRatio of \dataExpChatResponsePTwoSolutionTotal generated solutions were correct.

\dataExpChatResponseInformsCorrectRatio of the \textit{\plookup} responses were correct, indicating that when requesting only general information about basic programming concepts, \gptt does not appear to ``hallucinate''.
These included questions about when \code{null}-related errors occur in Java and how to produce one.
Notably, one participant requested the documentation for the class \code{NullPointerException}, and the generated response correctly relayed it verbatim.

\subsection{Autonomous Thinking (RQ3)}\label{sec:e_exp_plag}

\subsubsection{Similarity of Consecutive Submissions}\label{sec:exp_plag_similarity}

Among submissions where participants had at least one \textit{\pgen} response available to them, the average similarity to the generated code was \dataExpChatPlagSimilarityPOneAccessStats{on \exfind} and \dataExpChatPlagSimilarityPTwoAccessStats{on \exfix}.
These numbers align with findings by \citet{K23_Support}, who observed an average similarity of 63\% ($\sigma$ = 42\%) between generated and submitted code, albeit using the Jaccard similarity measure.

We found that the average similarity of consecutive submissions was significantly lower if any chatbot interactions occurred between the submissions, at 64.1\% ($M$ = 73.5\%, $\sigma$ = 26.6\%) compared to 82.2\% ($M$ = 89.0\%, $\sigma$ = 20.7\%) without, across all participants ($t = 5.76, p < 0.0001$). 
This indicates that students relying on the bot were rather following the bot suggestions even with completely different solution approaches, while students without the bots were rather trying to incrementally improve their solutions. 
In 62\% of cases with generations, the modified submission was more similar to the generated code than to the previous submission.

In our manual analysis of code changes with chatbot interactions, we found 54.3\% of submissions to be semantically identical to generated code (\textit{\rfull}).
13.6\% of submissions reused part of the generated solution (\textit{\ridea}), while 11.1\% copied a language construct (\textit{\rsyntax}).
We observed 2.5\% of changes to be based on a generated textual description (\textit{\rexplain}) and the remaining 18.5\% to be entirely unrelated to the chatbot interactions.



\subsubsection{Preceding Activity}

\begin{figure}
    \centering
    \begin{minipage}[t]{.49\textwidth}
    \includegraphics[width=0.98\textwidth]{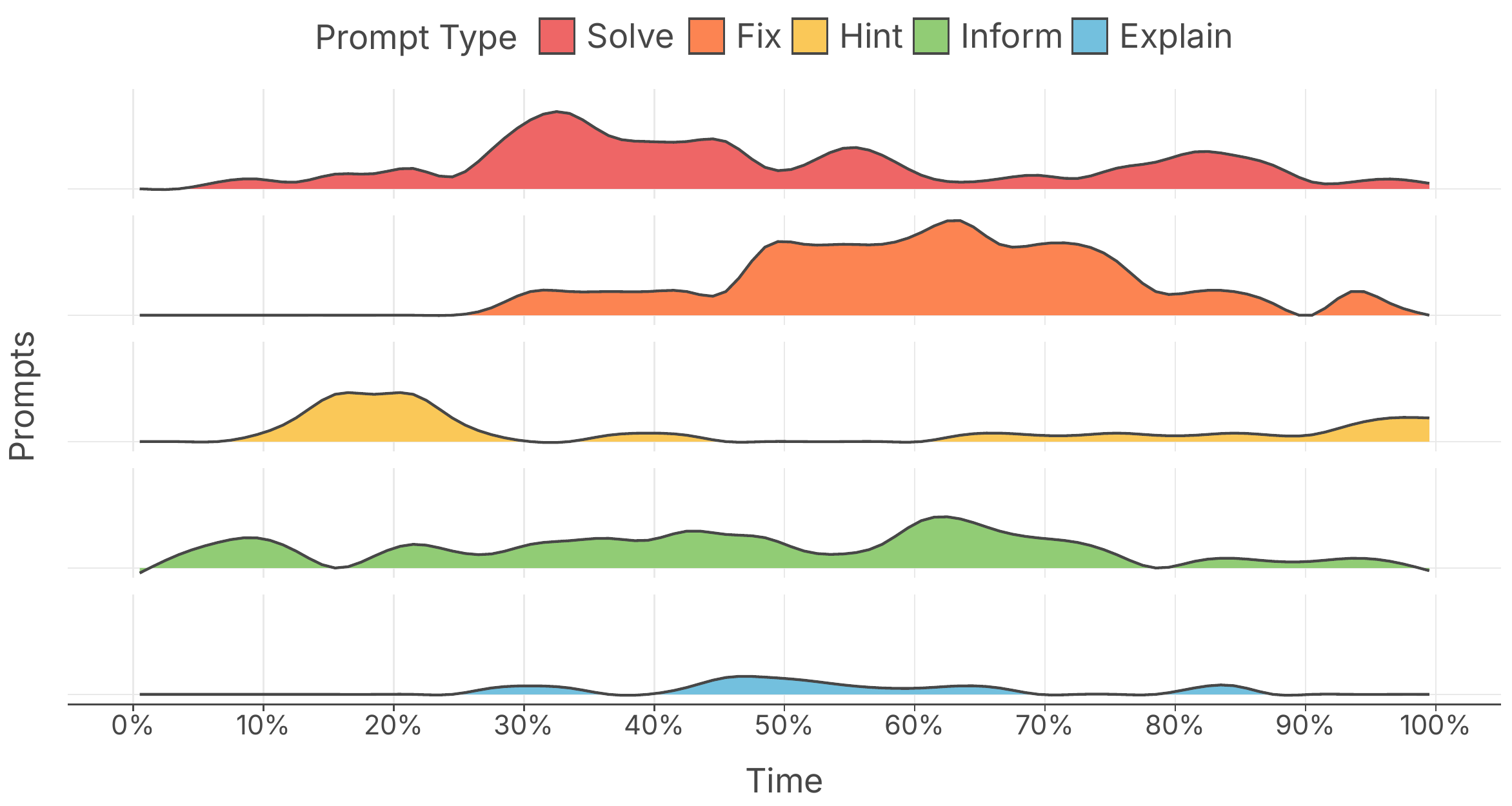}
    \captionof{figure}{Timeline showing the distribution of prompts throughout each experiment session.}
    \label{fig:/exp/timeline}
\end{minipage}\hspace*{\fill}
\begin{minipage}[t]{.49\textwidth}
    \centering
    \includegraphics[width=0.98\textwidth]{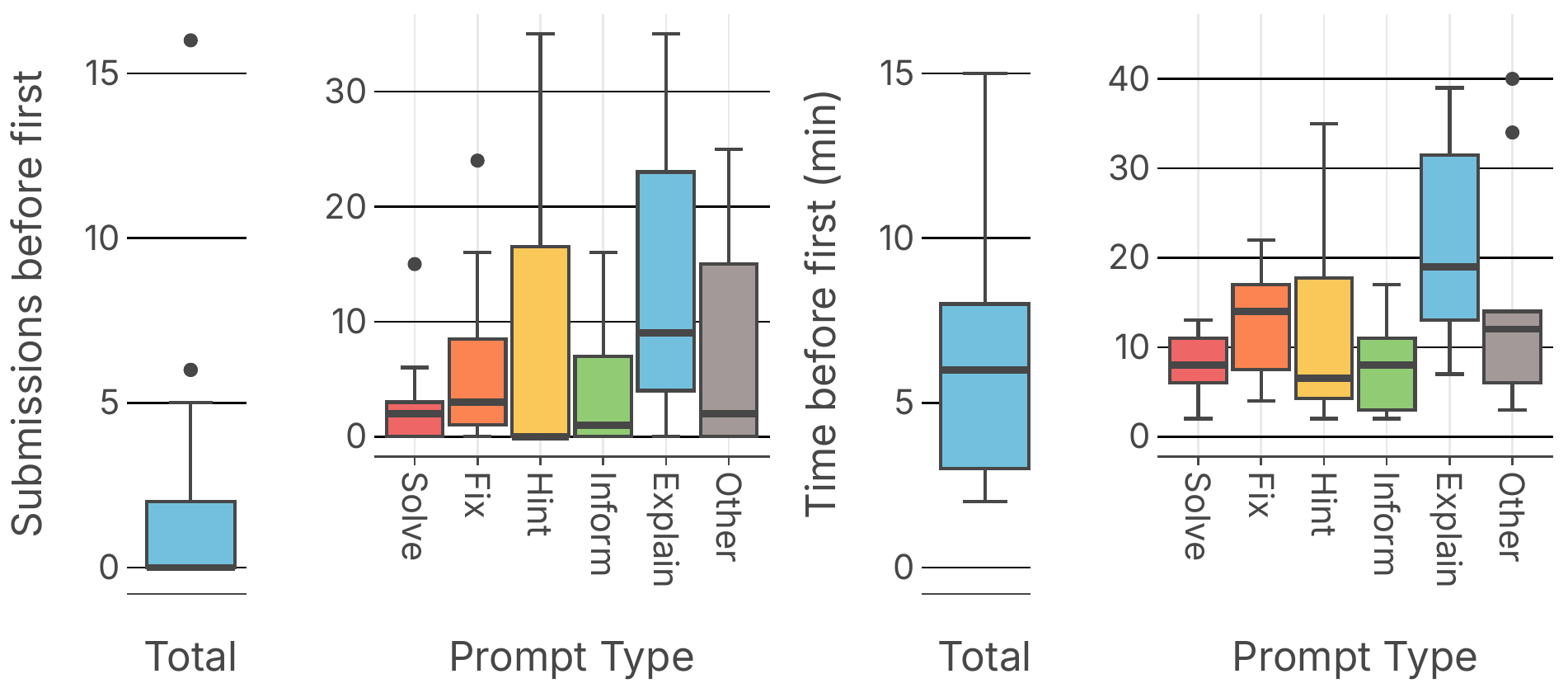}
    \caption{Submission attempts (left) and elapsed time (right) before the \textit{first} occurrence of a prompt type.}
    \label{fig:/exp:chat:prior}
\end{minipage}
\end{figure}

To understand when participants decided to use the chatbot, we analyzed the occurrences of prompt types over time in each experiment session.
Figure \ref{fig:/exp/timeline} overviews the prompt type distribution.
Additionally, we reviewed the number of submission attempts and the elapsed time before the first occurrence of each prompt type, which are shown in Figure \ref{fig:/exp:chat:prior}. 

We found that among participants who ended up using the chatbot, most started doing so relatively early in their session (but not immediately at the start).
On average, participants submitted their first chatbot prompt 6.4 minutes (M = 5, $\sigma$ = 3.99) after starting the session. 
\dataExpChatFirstLookupRatioChatters of \chatters began their chat with a \textit{\plookup} prompt, followed by \dataExpChatFirstSolutionRatioChatters with \textit{\psolve} and \dataExpChatFirstHintRatioChatters starting with \textit{\phint}.
Across \chatters, 43.5\% submitted at least one \textit{\phelp} prompt and 26.1\% submitted at least one \textit{\pgen} prompt before they made their first code submission attempt.
None of the students submitted a generated solution as their first code submission.

\subsubsection{Prompting Behavioral Patterns}
\begin{figure}
    \centering
    \includegraphics[width=\textwidth]{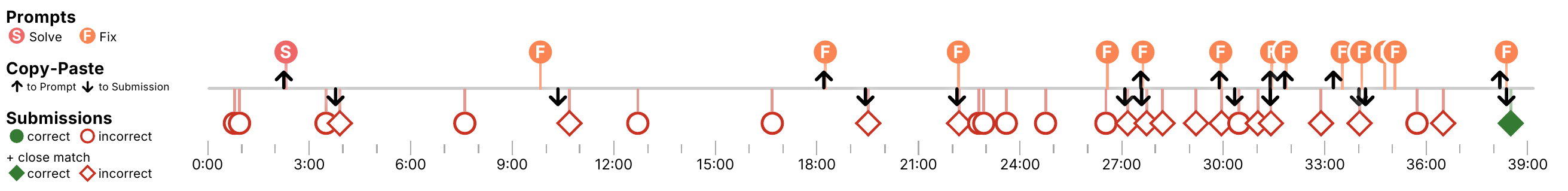}
    \caption{Interaction logs for the entire session of participant \msg{XKW}.}
    \label{fig:/exp:chat:timeline/sample}
\end{figure}

\begin{figure}[H]
    \centering
    \includegraphics[width=0.7\textwidth]{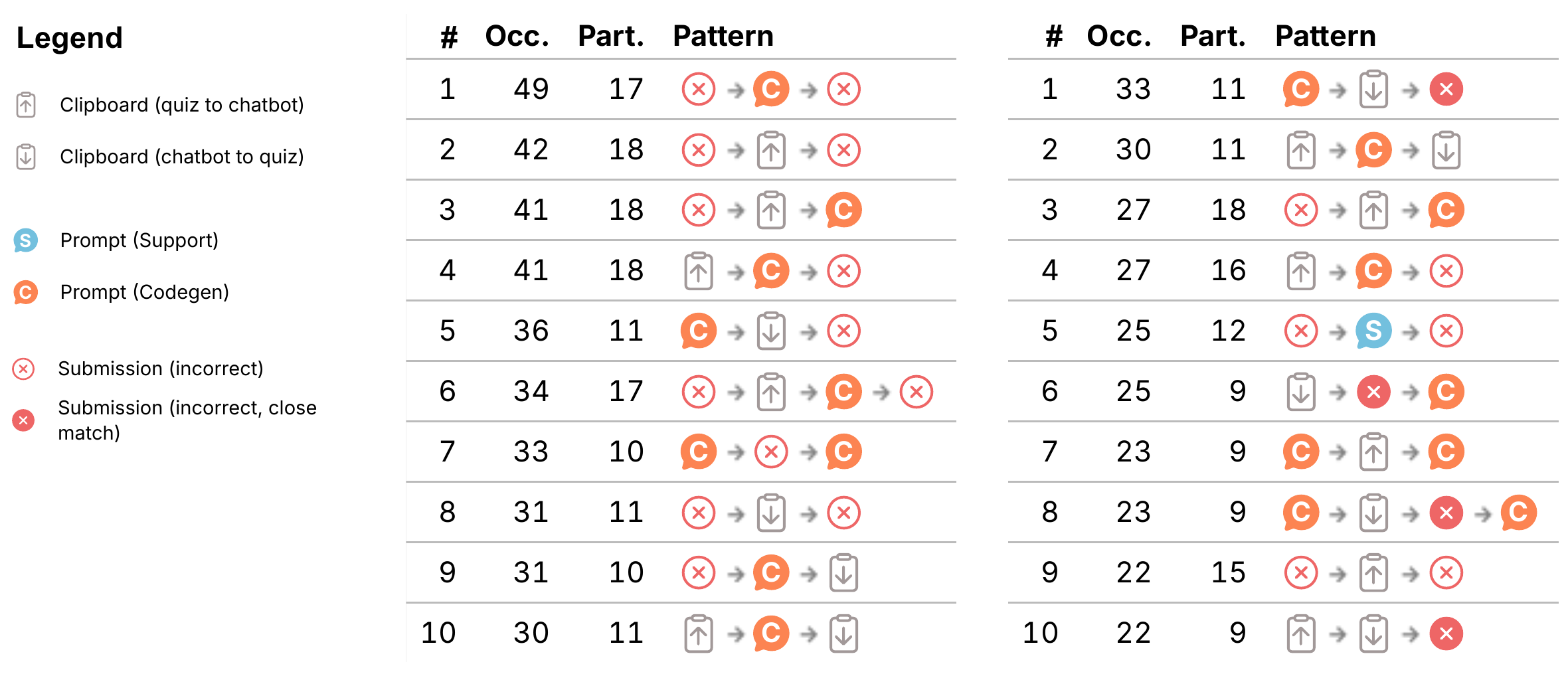}
    \caption{Most common closed interaction subsequences among participants, with all submissions as the same interaction type (left) and with submissions differentiated by whether they are a close match (right).}
    \label{fig:/exp:chat:patterns/top}
\end{figure}

Figure \ref{fig:/exp:chat:timeline/sample} presents an example of the interaction logs we extracted from participant sessions.
We used sequence pattern mining to identify common sequences of actions in these logs.
The most common patterns are reported in Figure \ref{fig:/exp:chat:patterns/top} (left).
We report both the total number of occurrences of each pattern, as well as how many participants have followed this pattern at least once during their session.
The results show frequent cycles of incorrect submissions, followed by code generation prompts or clipboard events, followed again by incorrect submissions.
We additionally performed the same analysis while differentiating submissions by whether they constitute a close match as per section \ref{sec:exp_plag_similarity}.
We report these results in Figure \ref{fig:/exp:chat:patterns/top} (right).

We found that while the most common sequences were present in the majority of \chatters interaction logs, some had comparable occurrence counts while being exhibited by far fewer participants.
None of the sequences contained interaction items for a correct submission.
We assume that any possible patterns with correct submissions were overshadowed, as students only needed one correct submission to pass, but made many incorrect submissions prior.

\section{Discussion}\label{discussion}

\subsection{Findings and Implications}\label{sec:d_exp_analysis}

\subsubsection{\gai in Programming Education and Assessment}

We found that a student could not pass our course by relying solely on \gai, but almost exclusively due to secondary assessments that go beyond code correctness and which require demonstration of conceptual understanding in a controlled environment. 
Both \gpt models were able to generate correct code and essay answers for most of the exercises.
In a non-personal setting, such as remote homework submissions, a student could therefore likely pass the course using \gptf, even with manually graded assignments.
Though the \llm occasionally missed individual requirements, such mistakes could also arise through honest misunderstandings by students.
They would be easily corrected based on tutor feedback.
In almost all cases, the \gpt output is thus nearly indistinguishable from that of a well-performing student.
This renders \textbf{submission-based} assignments significantly less effective for assessing students' understanding of submitted code or underlying concepts, despite the latter being a key requirement among educators to consider \gai use acceptable \cite{P23_GAI4Ed}.

However, with the format of our course, minor mistakes can become a significant challenge for a student presenting AI-generated work as their own.
Because the necessary corrections are also minor, the students---having presumably put significant thought into their answer already---are expected to apply them on the spot.
Inability to do so, or to at least discuss potential approaches with the tutor, would be a strong indicator of academic dishonesty.
Additionally, our students must be able to justify and explain the use of concepts, syntax, or APIs that have \textit{not} been covered in the course \textit{yet}.
Without a solid understanding of the existing course materials, it would be difficult to even identify the elements in the code that may lead to further questioning, and then convincingly explain them.
As this risk is unrelated to the reasoning capabilities of the \llms, it cannot be mitigated by using more advanced models.
This observation is supported by the lack of improvement in the \textit{Learn} category between \gptt and \gptf.

The students' self-reported \gai usage and the engagement with the chatbot in the experiment were mixed.
While this indicates that tools like \cgpt do not yet have universal acceptance among students, we found that self-reported regular \gai usage was associated with more frequent chatbot interactions and a higher propensity for code generation requests.
This suggests that as students \textbf{self-learn} and gain experience with \llms, they become more reliant on its code output.

We derive two major recommendations. 
First, the findings suggest GenAI should be \textbf{introduced as a tool} early in software development courses to avoid misperception through self-learning.  
Particularly, instructors should highlight various usage strategies beyond generating code, e.g.~to reason about and understand code---in combination with other tools like debuggers and visualizations. 
Limitations and tendencies of current models should be discussed in classes to create awareness, e.g.~about hallucinations or overconfidence. 
Second, we recommend, if possible, to shift towards \textbf{interview-like} assessments and code reviews instead of or in addition to homework.
This creates more opportunities to detect academic dishonesty and knowledge gaps without requiring a categorical ban on \gai tools.
Future work could explore whether the feedback and the evaluation in these assessments could itself be assisted by \gai, possibly using a tutor-in-the-loop approach to lessen the resource burden on educators.


\subsubsection{Learning to Think with \gai and to Develop, Compare, and Challenge Solution Paths}
In our experiment, the participants used the chatbot for general knowledge retrieval as well as to have the task solved (or partially solved) for them.
Requests for explanations were exceedingly rare, and we did not observe students asking \textit{any} comprehension questions about the exercise itself, or why their own solution wasn't working.
In contrast, we observed that 97.1\% of the generated solutions already included an explanation.
This may have been perceived by students as sufficient, reducing their need to explicitly ask for elaboration. 
Unless explicitly instructed otherwise, \gptt has a tendency to be verbose \cite{K24_StackOverflow} and confident \cite{H23_Responses}.

The most common behavioral pattern we identified in our sequence analysis appears to be a case of work avoidance: the chatbot is prompted to generate code, some or all of the code is copied and pasted into the answer box, and a submission---with very high similarity to the generated code---is made.
While this pattern occurred most often, it was exhibited by less than half of all \chatters.
This suggests that only some students engage in this direct form of work delegation, but those that do may try the approach multiple times.
More generally, we found that the participants changed the code significantly more between submissions if they requested code generation immediately prior.
In the majority of those cases, the result was more similar to the generated code than to the previous submission.
This indicates that students have a tendency to align their code with \gai output, even if they have to discard much of their existing attempt.
While it is generally good to explore various thoughts and solution approaches, it is also important to be able to focus and forward-reason about an emerging solution path. 
Methodological as well as tool-assisted approaches to develop and sharpen such skills are yet to be explored.  

The pattern we identified from most participants  consists of an attempt with no close match to generated code, then a paste into the chatbot prompt window, and the submission of a \textit{\pgen} prompt.
This indicates that many participants sought code generation from the \llm after making a submission they had at least partially thought up themselves.
This is also corroborated by the preceding activity, as the majority of students made at least two submission attempts before submitting their first \textit{\pgen} prompt.
Although \dataExpChatPlagSimilarityPOneClosePresenceRatioChatters of \chatters submitted code that constituted a close match to the \llm's output at some point in the experiment, we also find that none did so on their first attempt.
Our observations regarding the high prevalence of \textit{\pgen} requests therefore do not show upfront intent to ``cheat'' from the majority of \cgpt-using students. 
Rather, it appears that only after failing to solve the problem themselves---a situation developers encounter frequenter in their daily work---do the students turn to the chatbot to delegate the remaining work.

\subsubsection{On Usefulness and Usability of \gai Tools for Novices}

While reviewing the students' interactions, we noted that multiple participants ended up in cycles of seeing an error message, prompting the chatbot for assistance, implementing its suggestion, and receiving another error message.
For instance, the interaction timeline in Figure \ref{fig:/exp:chat:timeline/sample} shows an accelerating back-and-forth between prompts and submissions. 
Through our sequence analysis, we were able to quantify this interaction pattern, which turned out to be the most common across all \chatters.
As the \gptt model was unable to produce a correct solution to problem \exfind in most cases, even in response to \textit{\pfix} prompts, this was a notably ineffective strategy.
Nonetheless, some participants repeatedly tried ''coercing'' the model into generating a correct submission through a series of \textit{\pgen} prompts.
This suggests they were \textbf{unable to recognize} when a \gai tool is incapable of providing effective assistance, similar to observations by \citet{P24_Novices} on novices using inline code completion.

Using \llms for code generation shifts the developer's focus from ideation \cite{Wei:2024} to parsing and debugging unfamiliar code.
Prior work has shown that this is a challenge for professional developers \cite{S22_Programming, V22_Expectations}.
Without the technical experience, it is even more difficult for beginners to review \cgpt's output.
We expect this effect to be exacerbated if students also depend on \cgpt for knowledge and conceptual questions, which has been found to be negatively associated with performance in recent work \cite{M25_Patterns}.
Further, \gai tools can increase developers' \textbf{perceived} productivity \cite{Z24_Productivity} and level of understanding \cite{P24_Novices} without delivering actual improvements, potentially leading to a false sense of confidence \cite{L25_Thinking} among novices.
Research has yet to explore effective feedback strategies that make model uncertainty transparent to users and minimize overconfidence. 

We observed that \cgpt often apologized and occasionally even justified its mistakes before attempting to correct itself, which may have increased the students' trust in the system. 
This happened in 10 experiment sessions, in fact up to 14 times in one single session (participant 28X).
Explanations of potential errors in an automated system have been shown to increase trust and reliance on the system, largely mitigating the negative effect of observing an error \cite{D03_Trust}.
Thus, \gpt's acknowledgment of mistakes may actually make students less likely to break out of the error-prompting cycle.
This highlights the need to sensitize students to the risks and limitations of using \gai.
Our experiment task could be used as a practical example in class, allowing students to experience first-hand how consistently \cgpt can make mistakes, and how confidently it can appear to defend them.


\subsection{Examples of Student-Bot Interactions}\label{sec:d_exp_ex}
We highlight a few student-chatbot interactions which were substantially unique.

\subsubsection{\id{GCL}: The Impossible Exercise}\label{sec:e_exp_example_gcl}
As discussed in Section \ref{sec:a_exp_task}, the experiment task was designed to cause \gptt to produce incorrect outputs.
The chatbot responses to participant \id{GCL} were particularly misleading.
Over multiple attempts, the chatbot was unable to produce the correct solution to the coding problem \exfind.
Eventually, the chatbot claimed that the provided erroneous code sample from the exercise was actually already free of issues, and thus there was no solution to \exfind.
This led the participant to conclude that the goal of the study was to present an unsolvable task \msg{GCL-M33}, that they had been tricked \msg{GCL-M37}, and that there was a connection to the field of psychology \msg{GCL-M43}.
At this point, the participant closed the session and ended the experiment.

This interaction demonstrates a unique new challenge for learners: a conversational \gai agent may convince them that the task itself is flawed, rather than recognizing that the agent is incapable of generating a correct solution.
While only participant \id{GCL} was directly told by the chatbot that the exercise was impossible, other participants also received responses that contradicted the output of the test runner, such as reassuring a student that the generated solution would work even though it hadn't \msg{E6R-M15}.
We do not know whether other participants concluded the experiment under the impression that the task itself was faulty or unsolvable, or that they were unable to solve the problem.
Though this was an isolated incident, it raises the concerning possibility that an inexperienced student---already struggling to assert the quality of \cgpt's output---may be encouraged by the chatbot to trust it \textit{over their instructor} on conflicting statements.
While tools like \cgpt include warnings not to trust their generations and validate claims through external sources, whether students take them seriously or consider \cgpt to be speaking with authority warrants further investigation.

\subsubsection{\id{28X}: What Changed?}\label{sec:e_exp_example_28x}
In one of the very few conversations where the chatbot generated a fully correct solution to \exfind, the participant \id{28X} made a mistake in copying and pasting the code into the answer textbox, and subsequently dismissed the solution they were given.

After a brief interaction, the chatbot generated the correct solution to the problem and even correctly explained the mechanism behind it.
However, the generated method was declared \code{static}, a modification which was not accounted for in the test runner, as students hadn't yet been introduced to class methods.
After the participant had pasted the entire generated method definition – including the \code{static} modifier – into their answer box, the test runner rejected it with the message ''[...] is a class method, but it should be an instance method.'' \msg{28X-S107}.

The participant relayed the error message back to the chatbot \msg{28X-M21} and received a corrected solution without the \code{static} modifier \msg{28X-M22}.
From this point on, however, they only copied the \textit{body} of the updated method to insert back into the answer text box \msg{28X-C16}, missing the crucial change to the method head, and then told the chatbot repeatedly that the error had not been fixed yet \msg{28X-M24, 28X-M28}.
Only after several more iterations of asking for unrelated changes to the code did they finally submit the correct answer \msg{28X-S119}, nine minutes after they had initially been given the correct solution.

In copy-pasting the generated code, the student inadvertently used an unfamiliar concept, potentially without even realizing they were missing crucial context.
This is the same problem we observed in the \textit{Learn} category of our task-solving evaluation, demonstrating that it can negatively impact beginners beyond the context of interview assessments.
Here, the student had to reconcile an error message they could not understand with a chatbot claiming that the error had been resolved, without any guidance on what their mistake was.
It is unclear how much awareness of \gai limitations would have helped this particular student, as the generated output \textit{was} correct and iterating over it further was counterproductive.

\subsection{Limitations and Recommendations for Improvement}\label{threats}

\subsubsection{Threats to Validity}

In our task-solving evaluation, the referenced \llms may be updated and yield different outputs for the configuration and prompts used in this work.
The temperature parameter of the models was set to \code{0.0} to mitigate this as much as possible.  
However, some studies have reported significant changes in the behavior of \gptt and \gptf over time \cite{C23_Changing}, though it is not clear whether these results are based on architectural changes, continuous training, or simply the non-deterministic nature of the \llms.

Many of the exercises in the studied course are adaptations 
of those found in Barnes and K{\"o}lling \cite{B09_Java}, an often cited book for teaching object-oriented programming using Java and \bluej.
We expect the performance results on the writing and in-person coding exercises to be generalizable to other courses that have drawn from the same sources when designing their exercises, aside from the atypical code style conventions of the studied course.
The phrasing of the prompt and problem statement can significantly impact whether the \llm will solve it correctly \cite{D23_Prompting}.
The logical correctness of model responses may therefore be reflective of the way the problem statements in the course are written and not the \llm's actual ceiling for reasoning. 
Furthermore, the material used is entirely in German, which may affect the performance of the language models \cite{L23_Multilingual}.
To gauge whether this has a significant impact on our results, we reviewed some of the problems the \llms were unable to solve and manually prompted them with English translations.
This yielded responses of comparable quality with the same logical errors.

In the experiment, participants may have been influenced by the setting or the knowledge of their actions being recorded when deciding whether to use the chatbot.
The same potential risk exists for the self-reported use of generative AI, which students may have felt discouraged from sharing honestly. 
They may have thought it to be more socially acceptable to report slightly lower than true \cgpt usage numbers due to the perception of such activity as engaging in cheating \cite{Z23_GAI4Ed}.
However, as the results are fairly consistent with similar surveys around the time \cite{K23_Support, P23_GAI4Ed, Z23_GAI4Ed} and we were commonly able to observe behavior that could be considered unethical in an academic context \cite{P23_GAI4Ed}, we do not believe this to be a major limitation.

Participant self-selection poses a potential threat to the internal validity of the study.
Students had to be interested and comfortable with partaking in a programming study and have the time available to participate.
During recruitment, multiple students expressed general interest in the study, but feared they were not ``good enough'' to solve a programming task in a controlled environment.
Some students expressed interest but were behind on their coursework and thus needed the entire time of the lab session to work on mandatory exercises, after which point they would no longer be available.
This may have biased the participant pool towards students who performed better in the course, and would therefore need less assistance from a chatbot.

The participant pool consisted entirely of students of one university course in one year.
Sentiments and strategies on \gai use may vary across different populations and over time, limiting the generalizability of our findings. Thus, replicating our study in different universities is desirable. 

\subsubsection{Recommendations for Future Studies}

Our findings suggest that once learners decide turn to a \cgpt-like assistant for help, especially if they already have prior experience with generative AI, they predominantly seek code solutions rather than support in creatively working through the problem \cite{Wei:2024}.
In a future study, we recommend to test the students' understanding of their final code submission after exercise completion. 
This can help to determine what students were able to retain from the exercise, especially if their solutions were largely guided by the chatbot.

Additionally, we believe the perception of \gai and \cgpt in particular should be investigated further, to understand what motivates the students' behavior and how much authority and trust they place in these tools.
In our study, we created an environment that was as close to the students' regular working environment as possible, to elicit their typical behavior.
This means we did not capture a lot of information on the students' thoughts and impressions during the experiment.
Exit surveys or think-aloud observations can be used to gain more insight into student sentiment while using \gai tools for programming \cite{P24_Novices, Kruse:ICSME:2024}.

Our results suggest that students are either not aware of the risks and limitations of using \gai, or not sensitized enough to recognize them during regular use.
Whether guidance on prompting, \llm best practices and its limitations could meaningfully improve beginners' ability to handle confidently incorrect \llm outputs warrants investigation.
A follow-up study could replicate our experiment, using the same or a comparable task, with students who have received prior training or educational material related to \gai use.
Alternatively, different chatbot behaviors could encourage students to think through the problem themselves or ``nudge'' them towards finding the solution \cite{Pham:RE:21}, potentially even beyond the model's inherent reasoning capabilities.
Our \cgpt-like chatbot did not include any task-specific instructions in the system prompt to elicit certain behavior from the model.
A future study could compare the impact of different system prompts on the students' code reuse and prompting behaviors, as well as whether the students' own task-solving performance can be improved.

Finally, we only tested a single, fairly complex exercise, related to object states and references.
The propensity of learners to use \gai outputs has been shown to vary significantly by exercise topic \cite{K23_Support, M25_Patterns}.
Therefore, we recommend replicating this study with exercises involving different concepts, with a similar relative complexity level.
\section{Related Work}\label{related}

\subsection{Overall Sentiment}

\citet{K23_LLM4Ed} summarize various challenges posed by the use of \llms in education.
The authors particularly highlight the risk of learners relying too heavily on the model, noting that ``the model simplifies the acquisition of answers or information, which can amplify laziness and counteract the learners' interest to [...] come to their own conclusions and solutions'' \cite[p. 5]{K23_LLM4Ed}. Our work is the first to provide quantitative empirical evidence and detailed behavioral insights on the prevalence of chatbot overreliance specifically in programming education.

\citet{GL23_Ban} interviewed instructors of university introductory programming courses on their sentiment towards the use of \cgpt in education.
They found that most educators were very unsure how many of their students actually used \cgpt.
Notably, every single interviewee of their study independently brought up cheating concerns near the start of their interview, indicating that this is an important issue for educators in the space.
These concerns have been echoed by a smaller interview study by \citet{Z23_GAI4Ed}, where all instructors and a majority of the students reported anticipating an increase in academic dishonesty due to \gai tools.
Similar sentiment has also been shown in a survey by \citet{P23_GAI4Ed}.
They found 51\% of surveyed instructors to believe either \textit{many} or \textit{almost all} students ``are using GenAI Tools in ways that [they] would not approve of'' \cite[p.13]{P23_GAI4Ed}.
Almost every educator considered submission of generated code unethical if the student did not understand it, though only 60\% would have also considered it unethical if the student had taken the time to read and understand the auto-generated solution first. 

In response, some instructors have begun shifting their course grade composition more towards exam scores, explicitly showing students the limitations of AI code generation, or outright banning \cgpt and comparable tools in their class \cite[p. 113]{GL23_Ban}.
\citet{Z23_GAI4Ed} saw students universally disagreeing with banning \gai tools outright, though \citet{P23_GAI4Ed} found that the majority of students support at least some level of restrictions on its use.

\subsection{Task-Solving Performance}\label{sec:rel_llm_perf}

\citet{F22_Codex} analyzed the performance of the Codex model on educational programming problems of various difficulties.
For questions on a CS1 course assessment, they found that Codex solved 43.5\% of questions correctly on the first try, ignoring trivial formatting errors. 
When given multiple attempts with the same penalty scheme applied to students, the Codex model would have placed in the top quartile of students.
In a subsequent analysis of exam questions in a CS2 course, \citet{F23_Exam} also found Codex placing in the top 25\% of students.
\citet{P23_Classroom} found \cgpt able to solve 68\% of coding exercises of a functional programming course in the first try, expanding to 86\% after follow-ups.
They found similar performance for ``easy'' and ``medium'' difficulty problems, but fewer correct answers for ``hard'' problems.

\citet{BGK23_Overcome} analyzed \cgpt's performance on the programming tasks of their introductory university course.
They found that a student could pass the course using only \cgpt-generated answers, albeit with a grade of 55\%, only scoring slightly above the passing threshold.
They also analyzed the time required to copy-paste the question text into the \cgpt input window as well as inserting the response into the answer field.
They found a 91\% reduction in the time invested in solving the assignments naively, \ie simply submitting the first generated solution, and still a 74\% reduction in time when the code was then manually adjusted to receive full marks \cite{BGK23_Overcome}.

\subsection{Interactions of Programming Students and Programmers with \gai}\label{sec:rel_llm_ed_interact}

\citet{M22_Explanations} demonstrated a potential application for \llms in generating beginner-friendly explanations for source code.
In a subsequent work, \citet{M23_Explanations} found that the explanations were mostly rated useful by students, both for their own understanding and as a general learning tool \cite{M23_Explanations}. 
We observed that without prior guidance and instructions, students barely used the \gai bot for explanation. 
\citet{L23_Explanations} found   that automatically generated code explanations by \gptv{3} were rated as both a more accurate and easier to understand than explanations created by students. 
However, unlike \citet{M23_Explanations}, they found students to prefer line-by-line explanations \cite{L23_Explanations}. 

\citet{K23_Support} studied the behavior of learners when given access to the Codex model, and the impact of the tool on their learning progression.
For questions related to loops, 60\% of submitted answers in the Codex group were entirely AI-generated \cite[p. 10]{K23_Support}.
On follow-up tasks without access to a code generator, the authors found no significant difference in the correctness of solutions presented by learners who had and had not used Codex earlier. 
In an assessment a week later, learners in the Codex group exhibited significantly higher rates of errors on code authoring tasks with no starter code given, but had similar correctness scores and completion times \cite{K23_Support}. 

\citet{H23_Responses} prompted \cgpt and Codex with real help requests from students of an online programming course and assessed the models' ability to identify the problem and provide meaningful assistance.
While \cgpt-3.5 was able to identify all issues in the majority of code snippets, it notably also found non-existent issues in 40\% of cases.

\citet{P23_Interactions} studied the interactions of novice programming students with GitHub Copilot, an autocompletion-style code generation tool.
They observed a common pattern of ``drifting'', where participants would be led astray by incorrect generated solutions which they would have to correct, as well as ``sheperding'', where they would focus more of their attention on guiding the \llm towards suggesting a correct solution than writing it on their own.

\citet{P24_Novices} analyzed the metacognitive challenges students encountered when using GitHub Copilot.
While some students were able to solve their task faster with the tool, they found that metacognitive difficulties encountered by struggling students were not alleviated by \gai.
Instead, struggling students may even be faced with new difficulties, such as being unable to recognize when a code suggestion was helpful, or overestimating their understanding of the problem. 
Most recently, \citet{M25_Patterns} analyzed the prompting patterns of students in a CS2 data structures course.
They found solution generation to be both the most common prompt type and overall conversation intention.
Similar to our observations, they found that students lacking conceptual understanding struggled with interpreting and using the chatbot output.
They also found chatbot use and, more notably, code generation requests to be positively associated with solution correctness.
However, this includes submissions with copy-pasted generated code, the prevalence of which is not stated.


While much research has been conducted on use cases, performance evaluation, and limitations of \gai for code generation and software development in general, studies of how developers interact with \gai (including behavioral and prompting patterns) are still sparse. 
Most notable, \citet{J24_Developers} recently analyzed prompting patterns and code reuse in the DevGPT dataset \cite{X24_DevGPT} of \cgpt-conversations referenced on GitHub.
They found generated code to be present in subsequent commits from 17\% of conversations, and a further 26\% with some modifications.
In a user study with developers, \citet{V22_Expectations} found GitHub Copilot to have little impact on task completion time, but strong user preference for Copilot over the IDE's built-in suggestions.
Understanding, evaluating and debugging generated code was found to be a major challenge for participants. 
Analyzing \cgpt-generated answers to StackOverflow questions, \citet{K24_StackOverflow} found just over half of the generated answers to contain incorrect information, which was then overlooked by human evaluators in some cases.
Most recently, \citet{Kruse:ICSME:2024} conducted an experiment on prompting skills comparing experienced programmers with students. 
The authors observed that prompts, particularly by students, led to lower quality generated documentation than a prepared prompt executed with a single click. 
Similar to our finding, the authors concluded that prompting skills cannot be expected for effective use of \gai tools and need to be taught similarly as teaching how to use a debugger or a test suite.

%
%

\section{Conclusion}\label{conclusion}

The evaluation of the \gpt language models on the exercises of our introductory software development course validates previous findings that current-generation widely accessible \gai tools are capable of solving the vast majority of, though not all, first-semester programming assignments.
For courses where in-person code reviews of the assignments are an integral part---as in our setting---
we found that the answer to \textbf{\rqeval} is ``no'', due to practical challenges around justifying and correcting implementation details in AI-generated code.

In the experimental study, we demonstrated that the use of chatbot assistants is not universal across students. Over a third of our participants chose not to engage with the provided assistant at all.
Notably, whether the students used the chatbot had no measurable impact on task-solving performance or completion time.
To answer \textbf{\rqstrat}, we showed that among students who used a chatbot during the task-solving process, two primary strategies emerged: using it as a knowledge base for information retrieval, and using it to generate code solutions.
We also observed ineffective but common patterns of students repeatedly attempting to coax the chatbot into providing a correct solution, by relaying the output of its previous incorrect attempt back to the chatbot.
The students' continued interactions with the chatbot, despite it consistently generating incorrect solutions, suggests an inability to recognize the limitations of \llms.

Most students attempted to solve the problem by themselves first, and did not ask for a code solution in their first message to the chatbot.
However, a majority would end up eventually requesting a code solution and trying to submit it.
None of the participants asked the model to discuss why their own solution wasn't behaving as expected.
Therefore, to answer \textbf{\rqauto}, we found that students do not immediately delegate task-solving to a chatbot, but generally turn to it for a full solution upon encountering difficulties.
The willingness to reuse generated code solutions, even if only after failing to solve the problem independently, indicates that concerns around over-reliance on \cgpt and subsequent academic dishonesty are warranted.
The tendency of students who report using \gai tools regularly to more strongly favor code generation prompts particularly raises questions about the habits students build through repeated interactions with AI assistants.

In summary, \cgpt poses a challenge to learn programming and develop critically thinking skills.
If students have unguided and uncritical access to the tool, a significant fraction may use it to avoid autonomous work on tasks they find  challenging, without the expertise required to critically evaluate its output and avoid being misled.
Interview-like assessments could help uncover cases of AI-related plagiarism, or at least ensure students actually understand the solutions they are submitting. 
More research is needed on the effect of \gai on learning, long-term retention,  and critically thinking and reasoning skills. 


\section{Data Availability}
The full survey and exercise as well as the interactions and messages recorded from the experiment sessions are available at \url{https://figshare.com/s/d8b532a6ca49b80e6df7}.


\bibliographystyle{ACM-Reference-Format}
\bibliography{paper}


\begin{thebibliography}{52}


\ifx \showCODEN    \undefined \def \showCODEN     #1{\unskip}     \fi
\ifx \showISBNx    \undefined \def \showISBNx     #1{\unskip}     \fi
\ifx \showISBNxiii \undefined \def \showISBNxiii  #1{\unskip}     \fi
\ifx \showISSN     \undefined \def \showISSN      #1{\unskip}     \fi
\ifx \showLCCN     \undefined \def \showLCCN      #1{\unskip}     \fi
\ifx \shownote     \undefined \def \shownote      #1{#1}          \fi
\ifx \showarticletitle \undefined \def \showarticletitle #1{#1}   \fi
\ifx \showURL      \undefined \def \showURL       {\relax}        \fi
\providecommand\bibfield[2]{#2}
\providecommand\bibinfo[2]{#2}
\providecommand\natexlab[1]{#1}
\providecommand\showeprint[2][]{arXiv:#2}

\bibitem[Barnes and K{\"o}lling(2009)]%
        {B09_Java}
\bibfield{author}{\bibinfo{person}{David~J Barnes} {and} \bibinfo{person}{Michael K{\"o}lling}.} \bibinfo{year}{2009}\natexlab{}.
\newblock \bibinfo{booktitle}{\emph{Java lernen mit BlueJ: Eine Einf{\"u}hrung in die objektorientierte Programmierung}}.
\newblock \bibinfo{publisher}{Pearson Deutschland GmbH}.
\newblock


\bibitem[Becker et~al\mbox{.}(2023)]%
        {B23_Hard}
\bibfield{author}{\bibinfo{person}{Brett~A Becker}, \bibinfo{person}{Paul Denny}, \bibinfo{person}{James Finnie-Ansley}, \bibinfo{person}{Andrew Luxton-Reilly}, \bibinfo{person}{James Prather}, {and} \bibinfo{person}{Eddie~Antonio Santos}.} \bibinfo{year}{2023}\natexlab{}.
\newblock \showarticletitle{Programming is hard-or at least it used to be: Educational opportunities and challenges of ai code generation}. In \bibinfo{booktitle}{\emph{Proceedings of the 54th ACM Technical Symposium on Computer Science Education V. 1}}. \bibinfo{pages}{500--506}.
\newblock
\href{https://doi.org/10.1145/3545945.3569759}{doi:\nolinkurl{10.1145/3545945.3569759}}


\bibitem[Berrezueta-Guzman and Krusche(2023)]%
        {BGK23_Overcome}
\bibfield{author}{\bibinfo{person}{Jonnathan Berrezueta-Guzman} {and} \bibinfo{person}{Stephan Krusche}.} \bibinfo{year}{2023}\natexlab{}.
\newblock \showarticletitle{Recommendations to create programming exercises to overcome ChatGPT}. In \bibinfo{booktitle}{\emph{2023 IEEE 35th International Conference on Software Engineering Education and Training (CSEE\&T)}}. IEEE, \bibinfo{pages}{147--151}.
\newblock
\href{https://doi.org/10.1109/CSEET58097.2023.00031}{doi:\nolinkurl{10.1109/CSEET58097.2023.00031}}


\bibitem[Cao et~al\mbox{.}(2021)]%
        {C21_Guess}
\bibfield{author}{\bibinfo{person}{Boxi Cao}, \bibinfo{person}{Hongyu Lin}, \bibinfo{person}{Xianpei Han}, \bibinfo{person}{Le Sun}, \bibinfo{person}{Lingyong Yan}, \bibinfo{person}{Meng Liao}, \bibinfo{person}{Tong Xue}, {and} \bibinfo{person}{Jin Xu}.} \bibinfo{year}{2021}\natexlab{}.
\newblock \showarticletitle{Knowledgeable or educated guess? revisiting language models as knowledge bases}.
\newblock \bibinfo{journal}{\emph{arXiv preprint}} (\bibinfo{year}{2021}).
\newblock
\href{https://doi.org/10.48550/arXiv.2106.09231}{doi:\nolinkurl{10.48550/arXiv.2106.09231}}


\bibitem[Chen et~al\mbox{.}(2023)]%
        {C23_Changing}
\bibfield{author}{\bibinfo{person}{Lingjiao Chen}, \bibinfo{person}{Matei Zaharia}, {and} \bibinfo{person}{James Zou}.} \bibinfo{year}{2023}\natexlab{}.
\newblock \bibinfo{title}{How is ChatGPT's behavior changing over time?}
\newblock
\href{https://doi.org/10.48550/arXiv.2307.09009}{doi:\nolinkurl{10.48550/arXiv.2307.09009}}
\newblock
\shownote{arXiv preprint}.


\bibitem[Chen et~al\mbox{.}(2021)]%
        {C21_Codex}
\bibfield{author}{\bibinfo{person}{Mark Chen}, \bibinfo{person}{Jerry Tworek}, \bibinfo{person}{Heewoo Jun}, \bibinfo{person}{Qiming Yuan}, \bibinfo{person}{Henrique~Ponde de Oliveira~Pinto}, \bibinfo{person}{Jared Kaplan}, \bibinfo{person}{Harri Edwards}, \bibinfo{person}{Yuri Burda}, \bibinfo{person}{Nicholas Joseph}, \bibinfo{person}{Greg Brockman}, \bibinfo{person}{Alex Ray}, \bibinfo{person}{Raul Puri}, \bibinfo{person}{Gretchen Krueger}, \bibinfo{person}{Michael Petrov}, \bibinfo{person}{Heidy Khlaaf}, \bibinfo{person}{Girish Sastry}, \bibinfo{person}{Pamela Mishkin}, \bibinfo{person}{Brooke Chan}, \bibinfo{person}{Scott Gray}, \bibinfo{person}{Nick Ryder}, \bibinfo{person}{Mikhail Pavlov}, \bibinfo{person}{Alethea Power}, \bibinfo{person}{Lukasz Kaiser}, \bibinfo{person}{Mohammad Bavarian}, \bibinfo{person}{Clemens Winter}, \bibinfo{person}{Philippe Tillet}, \bibinfo{person}{Felipe~Petroski Such}, \bibinfo{person}{Dave Cummings}, \bibinfo{person}{Matthias Plappert}, \bibinfo{person}{Fotios Chantzis},
  \bibinfo{person}{Elizabeth Barnes}, \bibinfo{person}{Ariel Herbert-Voss}, \bibinfo{person}{William~Hebgen Guss}, \bibinfo{person}{Alex Nichol}, \bibinfo{person}{Alex Paino}, \bibinfo{person}{Nikolas Tezak}, \bibinfo{person}{Jie Tang}, \bibinfo{person}{Igor Babuschkin}, \bibinfo{person}{Suchir Balaji}, \bibinfo{person}{Shantanu Jain}, \bibinfo{person}{William Saunders}, \bibinfo{person}{Christopher Hesse}, \bibinfo{person}{Andrew~N. Carr}, \bibinfo{person}{Jan Leike}, \bibinfo{person}{Josh Achiam}, \bibinfo{person}{Vedant Misra}, \bibinfo{person}{Evan Morikawa}, \bibinfo{person}{Alec Radford}, \bibinfo{person}{Matthew Knight}, \bibinfo{person}{Miles Brundage}, \bibinfo{person}{Mira Murati}, \bibinfo{person}{Katie Mayer}, \bibinfo{person}{Peter Welinder}, \bibinfo{person}{Bob McGrew}, \bibinfo{person}{Dario Amodei}, \bibinfo{person}{Sam McCandlish}, \bibinfo{person}{Ilya Sutskever}, {and} \bibinfo{person}{Wojciech Zaremba}.} \bibinfo{year}{2021}\natexlab{}.
\newblock \bibinfo{title}{Evaluating Large Language Models Trained on Code}.
\newblock
\href{https://doi.org/10.48550/arXiv.2107.03374}{doi:\nolinkurl{10.48550/arXiv.2107.03374}}
\newblock
\shownote{arXiv preprint}.


\bibitem[Cotton et~al\mbox{.}(2024)]%
        {CCS23_Cheating}
\bibfield{author}{\bibinfo{person}{Debby R.~E. Cotton}, \bibinfo{person}{Peter~A. Cotton}, {and} \bibinfo{person}{J.~Reuben Shipway}.} \bibinfo{year}{2024}\natexlab{}.
\newblock \showarticletitle{Chatting and cheating: Ensuring academic integrity in the era of ChatGPT}.
\newblock \bibinfo{journal}{\emph{Innovations in Education and Teaching International}} \bibinfo{volume}{61}, \bibinfo{number}{2} (\bibinfo{year}{2024}), \bibinfo{pages}{228--239}.
\newblock
\href{https://doi.org/10.1080/14703297.2023.2190148}{doi:\nolinkurl{10.1080/14703297.2023.2190148}}


\bibitem[Dakhel et~al\mbox{.}(2023)]%
        {D23_Liability}
\bibfield{author}{\bibinfo{person}{Arghavan~Moradi Dakhel}, \bibinfo{person}{Vahid Majdinasab}, \bibinfo{person}{Amin Nikanjam}, \bibinfo{person}{Foutse Khomh}, \bibinfo{person}{Michel~C Desmarais}, {and} \bibinfo{person}{Zhen Ming~Jack Jiang}.} \bibinfo{year}{2023}\natexlab{}.
\newblock \showarticletitle{Github copilot ai pair programmer: Asset or liability?}
\newblock \bibinfo{journal}{\emph{Journal of Systems and Software}}  \bibinfo{volume}{203} (\bibinfo{year}{2023}), \bibinfo{pages}{111734}.
\newblock
\href{https://doi.org/10.1016/j.jss.2023.111734}{doi:\nolinkurl{10.1016/j.jss.2023.111734}}


\bibitem[Denny et~al\mbox{.}(2023)]%
        {D23_Prompting}
\bibfield{author}{\bibinfo{person}{Paul Denny}, \bibinfo{person}{Viraj Kumar}, {and} \bibinfo{person}{Nasser Giacaman}.} \bibinfo{year}{2023}\natexlab{}.
\newblock \showarticletitle{Conversing with Copilot: Exploring Prompt Engineering for Solving CS1 Problems Using Natural Language}. In \bibinfo{booktitle}{\emph{Proceedings of the 54th ACM Technical Symposium on Computer Science Education V. 1}} (Toronto, Canada) \emph{(\bibinfo{series}{SIGCSE 2023})}. \bibinfo{publisher}{Association for Computing Machinery}, \bibinfo{address}{New York, NY, USA}, \bibinfo{pages}{1136–1142}.
\newblock
\showISBNx{9781450394314}
\href{https://doi.org/10.1145/3545945.3569823}{doi:\nolinkurl{10.1145/3545945.3569823}}


\bibitem[Dzindolet et~al\mbox{.}(2003)]%
        {D03_Trust}
\bibfield{author}{\bibinfo{person}{Mary~T Dzindolet}, \bibinfo{person}{Scott~A Peterson}, \bibinfo{person}{Regina~A Pomranky}, \bibinfo{person}{Linda~G Pierce}, {and} \bibinfo{person}{Hall~P Beck}.} \bibinfo{year}{2003}\natexlab{}.
\newblock \showarticletitle{The role of trust in automation reliance}.
\newblock \bibinfo{journal}{\emph{International journal of human-computer studies}} \bibinfo{volume}{58}, \bibinfo{number}{6} (\bibinfo{year}{2003}), \bibinfo{pages}{697--718}.
\newblock
\href{https://doi.org/10.1016/S1071-5819(03)00038-7}{doi:\nolinkurl{10.1016/S1071-5819(03)00038-7}}


\bibitem[Finnie-Ansley et~al\mbox{.}(2022)]%
        {F22_Codex}
\bibfield{author}{\bibinfo{person}{James Finnie-Ansley}, \bibinfo{person}{Paul Denny}, \bibinfo{person}{Brett~A. Becker}, \bibinfo{person}{Andrew Luxton-Reilly}, {and} \bibinfo{person}{James Prather}.} \bibinfo{year}{2022}\natexlab{}.
\newblock \showarticletitle{The Robots Are Coming: Exploring the Implications of OpenAI Codex on Introductory Programming}. In \bibinfo{booktitle}{\emph{Proceedings of the 24th Australasian Computing Education Conference}} (Virtual Event, Australia) \emph{(\bibinfo{series}{ACE '22})}. \bibinfo{publisher}{Association for Computing Machinery}, \bibinfo{address}{New York, NY, USA}, \bibinfo{pages}{10–19}.
\newblock
\showISBNx{9781450396431}
\href{https://doi.org/10.1145/3511861.3511863}{doi:\nolinkurl{10.1145/3511861.3511863}}


\bibitem[Finnie-Ansley et~al\mbox{.}(2023)]%
        {F23_Exam}
\bibfield{author}{\bibinfo{person}{James Finnie-Ansley}, \bibinfo{person}{Paul Denny}, \bibinfo{person}{Andrew Luxton-Reilly}, \bibinfo{person}{Eddie~Antonio Santos}, \bibinfo{person}{James Prather}, {and} \bibinfo{person}{Brett~A. Becker}.} \bibinfo{year}{2023}\natexlab{}.
\newblock \showarticletitle{My AI Wants to Know If This Will Be on the Exam: Testing OpenAI’s Codex on CS2 Programming Exercises}. In \bibinfo{booktitle}{\emph{Proceedings of the 25th Australasian Computing Education Conference}} (Melbourne, Australia) \emph{(\bibinfo{series}{ACE '23})}. \bibinfo{publisher}{Association for Computing Machinery}, \bibinfo{address}{New York, NY, USA}, \bibinfo{pages}{97–104}.
\newblock
\showISBNx{9781450399418}
\href{https://doi.org/10.1145/3576123.3576134}{doi:\nolinkurl{10.1145/3576123.3576134}}


\bibitem[GitHub(2022)]%
        {GH22_Copilot}
\bibfield{author}{\bibinfo{person}{GitHub}.} \bibinfo{year}{2022}\natexlab{}.
\newblock \bibinfo{howpublished}{\url{https://github.com/features/copilot/} (archived 2023-12-03. \url{https://web.archive.org/web/20231203005848/https://github.com/features/copilot/})}.
\newblock
\newblock
\shownote{Accessed 2023-12-03}.


\bibitem[H{\"a}ring and Maalej(2019)]%
        {H19_SE1}
\bibfield{author}{\bibinfo{person}{Marlo H{\"a}ring} {and} \bibinfo{person}{Walid Maalej}.} \bibinfo{year}{2019}\natexlab{}.
\newblock \showarticletitle{A Socio-Technical Framework for Face-to-Face Teaching in Large Software Development Courses.}. In \bibinfo{booktitle}{\emph{Software Engineering (Workshops)}}. \bibinfo{pages}{3--6}.
\newblock


\bibitem[Hellas et~al\mbox{.}(2023)]%
        {H23_Responses}
\bibfield{author}{\bibinfo{person}{Arto Hellas}, \bibinfo{person}{Juho Leinonen}, \bibinfo{person}{Sami Sarsa}, \bibinfo{person}{Charles Koutcheme}, \bibinfo{person}{Lilja Kujanp\"{a}\"{a}}, {and} \bibinfo{person}{Juha Sorva}.} \bibinfo{year}{2023}\natexlab{}.
\newblock \showarticletitle{Exploring the Responses of Large Language Models to Beginner Programmers’ Help Requests}. In \bibinfo{booktitle}{\emph{Proceedings of the 2023 ACM Conference on International Computing Education Research - Volume 1}} (Chicago, IL, USA) \emph{(\bibinfo{series}{ICER '23})}. \bibinfo{publisher}{Association for Computing Machinery}, \bibinfo{address}{New York, NY, USA}, \bibinfo{pages}{93–105}.
\newblock
\showISBNx{9781450399760}
\href{https://doi.org/10.1145/3568813.3600139}{doi:\nolinkurl{10.1145/3568813.3600139}}


\bibitem[Jin et~al\mbox{.}(2024)]%
        {J24_Developers}
\bibfield{author}{\bibinfo{person}{Kailun Jin}, \bibinfo{person}{Chung-Yu Wang}, \bibinfo{person}{Hung~Viet Pham}, {and} \bibinfo{person}{Hadi Hemmati}.} \bibinfo{year}{2024}\natexlab{}.
\newblock \showarticletitle{Can ChatGPT Support Developers? An Empirical Evaluation of Large Language Models for Code Generation}. In \bibinfo{booktitle}{\emph{2024 IEEE/ACM 21st International Conference on Mining Software Repositories (MSR)}}. IEEE, \bibinfo{pages}{167--171}.
\newblock


\bibitem[Jo{\v{s}}t et~al\mbox{.}(2024)]%
        {J24_Impact}
\bibfield{author}{\bibinfo{person}{Gregor Jo{\v{s}}t}, \bibinfo{person}{Viktor Taneski}, {and} \bibinfo{person}{Sa{\v{s}}o Karakati{\v{c}}}.} \bibinfo{year}{2024}\natexlab{}.
\newblock \showarticletitle{The Impact of Large Language Models on Programming Education and Student Learning Outcomes}.
\newblock \bibinfo{journal}{\emph{Applied Sciences}} \bibinfo{volume}{14}, \bibinfo{number}{10} (\bibinfo{year}{2024}), \bibinfo{pages}{4115}.
\newblock
\href{https://doi.org/10.3390/app14104115}{doi:\nolinkurl{10.3390/app14104115}}


\bibitem[Kabir et~al\mbox{.}(2024)]%
        {K24_StackOverflow}
\bibfield{author}{\bibinfo{person}{Samia Kabir}, \bibinfo{person}{David~N. Udo-Imeh}, \bibinfo{person}{Bonan Kou}, {and} \bibinfo{person}{Tianyi Zhang}.} \bibinfo{year}{2024}\natexlab{}.
\newblock \showarticletitle{Is Stack Overflow Obsolete? An Empirical Study of the Characteristics of ChatGPT Answers to Stack Overflow Questions}. In \bibinfo{booktitle}{\emph{Proceedings of the CHI Conference on Human Factors in Computing Systems}} \emph{(\bibinfo{series}{CHI ’24})}. \bibinfo{publisher}{ACM}, \bibinfo{pages}{1–17}.
\newblock
\href{https://doi.org/10.1145/3613904.3642596}{doi:\nolinkurl{10.1145/3613904.3642596}}


\bibitem[Kasneci et~al\mbox{.}(2023)]%
        {K23_LLM4Ed}
\bibfield{author}{\bibinfo{person}{Enkelejda Kasneci}, \bibinfo{person}{Kathrin Se{\ss}ler}, \bibinfo{person}{Stefan K{\"u}chemann}, \bibinfo{person}{Maria Bannert}, \bibinfo{person}{Daryna Dementieva}, \bibinfo{person}{Frank Fischer}, \bibinfo{person}{Urs Gasser}, \bibinfo{person}{Georg Groh}, \bibinfo{person}{Stephan G{\"u}nnemann}, \bibinfo{person}{Eyke H{\"u}llermeier}, {et~al\mbox{.}}} \bibinfo{year}{2023}\natexlab{}.
\newblock \showarticletitle{ChatGPT for good? On opportunities and challenges of large language models for education}.
\newblock \bibinfo{journal}{\emph{Learning and individual differences}}  \bibinfo{volume}{103} (\bibinfo{year}{2023}), \bibinfo{pages}{102274}.
\newblock
\href{https://doi.org/10.1016/j.lindif.2023.102274}{doi:\nolinkurl{10.1016/j.lindif.2023.102274}}


\bibitem[Kazemitabaar et~al\mbox{.}(2023)]%
        {K23_Support}
\bibfield{author}{\bibinfo{person}{Majeed Kazemitabaar}, \bibinfo{person}{Justin Chow}, \bibinfo{person}{Carl Ka~To Ma}, \bibinfo{person}{Barbara~J. Ericson}, \bibinfo{person}{David Weintrop}, {and} \bibinfo{person}{Tovi Grossman}.} \bibinfo{year}{2023}\natexlab{}.
\newblock \showarticletitle{Studying the effect of AI Code Generators on Supporting Novice Learners in Introductory Programming}. In \bibinfo{booktitle}{\emph{Proceedings of the 2023 CHI Conference on Human Factors in Computing Systems}} (Hamburg, Germany) \emph{(\bibinfo{series}{CHI '23})}. \bibinfo{publisher}{Association for Computing Machinery}, \bibinfo{address}{New York, NY, USA}, Article \bibinfo{articleno}{455}, \bibinfo{numpages}{23}~pages.
\newblock
\showISBNx{9781450394215}
\href{https://doi.org/10.1145/3544548.3580919}{doi:\nolinkurl{10.1145/3544548.3580919}}


\bibitem[Kruse et~al\mbox{.}(2024)]%
        {Kruse:ICSME:2024}
\bibfield{author}{\bibinfo{person}{Hans-Alexander Kruse}, \bibinfo{person}{Tim Puhlf{\"u}r{\ss}}, {and} \bibinfo{person}{Walid Maalej}.} \bibinfo{year}{2024}\natexlab{}.
\newblock \showarticletitle{Can Developers Prompt? A Controlled Experiment for Code Documentation Generation}. In \bibinfo{booktitle}{\emph{40th International Conference on Software Maintenance and Evolution (ICSME)}}. \bibinfo{pages}{574--586}.
\newblock
\href{https://doi.org/10.1109/ICSME58944.2024.00058}{doi:\nolinkurl{10.1109/ICSME58944.2024.00058}}


\bibitem[Lai et~al\mbox{.}(2023)]%
        {L23_Multilingual}
\bibfield{author}{\bibinfo{person}{Viet~Dac Lai}, \bibinfo{person}{Nghia~Trung Ngo}, \bibinfo{person}{Amir Pouran~Ben Veyseh}, \bibinfo{person}{Hieu Man}, \bibinfo{person}{Franck Dernoncourt}, \bibinfo{person}{Trung Bui}, {and} \bibinfo{person}{Thien~Huu Nguyen}.} \bibinfo{year}{2023}\natexlab{}.
\newblock \bibinfo{title}{ChatGPT Beyond English: Towards a Comprehensive Evaluation of Large Language Models in Multilingual Learning}.
\newblock
\href{https://doi.org/10.48550/arXiv.2304.05613}{doi:\nolinkurl{10.48550/arXiv.2304.05613}}
\newblock
\shownote{arXiv preprint}.


\bibitem[Lau and Guo(2023)]%
        {GL23_Ban}
\bibfield{author}{\bibinfo{person}{Sam Lau} {and} \bibinfo{person}{Philip Guo}.} \bibinfo{year}{2023}\natexlab{}.
\newblock \showarticletitle{From "Ban It Till We Understand It" to "Resistance is Futile": How University Programming Instructors Plan to Adapt as More Students Use AI Code Generation and Explanation Tools such as ChatGPT and GitHub Copilot}. In \bibinfo{booktitle}{\emph{Proceedings of the 2023 ACM Conference on International Computing Education Research - Volume 1}} (Chicago, IL, USA) \emph{(\bibinfo{series}{ICER '23})}. \bibinfo{publisher}{Association for Computing Machinery}, \bibinfo{address}{New York, NY, USA}, \bibinfo{pages}{106–121}.
\newblock
\showISBNx{9781450399760}
\href{https://doi.org/10.1145/3568813.3600138}{doi:\nolinkurl{10.1145/3568813.3600138}}


\bibitem[Lee et~al\mbox{.}(2025)]%
        {L25_Thinking}
\bibfield{author}{\bibinfo{person}{Hao-Ping~Hank Lee}, \bibinfo{person}{Advait Sarkar}, \bibinfo{person}{Lev Tankelevitch}, \bibinfo{person}{Ian Drosos}, \bibinfo{person}{Sean Rintel}, \bibinfo{person}{Richard Banks}, {and} \bibinfo{person}{Nicholas Wilson}.} \bibinfo{year}{2025}\natexlab{}.
\newblock \showarticletitle{The Impact of Generative AI on Critical Thinking: Self-Reported Reductions in Cognitive Effort and Confidence Effects From a Survey of Knowledge Workers}. In \bibinfo{booktitle}{\emph{CHI Conference on Human Factors in Computing Systems}} (Yokohama, Japan) \emph{(\bibinfo{series}{CHI '25})}. \bibinfo{publisher}{Association for Computing Machinery}.
\newblock
\href{https://doi.org/10.1145/3706598.3713778}{doi:\nolinkurl{10.1145/3706598.3713778}}


\bibitem[Leinonen et~al\mbox{.}(2023)]%
        {L23_Explanations}
\bibfield{author}{\bibinfo{person}{Juho Leinonen}, \bibinfo{person}{Paul Denny}, \bibinfo{person}{Stephen MacNeil}, \bibinfo{person}{Sami Sarsa}, \bibinfo{person}{Seth Bernstein}, \bibinfo{person}{Joanne Kim}, \bibinfo{person}{Andrew Tran}, {and} \bibinfo{person}{Arto Hellas}.} \bibinfo{year}{2023}\natexlab{}.
\newblock \showarticletitle{Comparing Code Explanations Created by Students and Large Language Models}. In \bibinfo{booktitle}{\emph{Proceedings of the 2023 Conference on Innovation and Technology in Computer Science Education V. 1}} (Turku, Finland) \emph{(\bibinfo{series}{ITiCSE 2023})}. \bibinfo{publisher}{Association for Computing Machinery}, \bibinfo{address}{New York, NY, USA}, \bibinfo{pages}{124–130}.
\newblock
\showISBNx{9798400701382}
\href{https://doi.org/10.1145/3587102.3588785}{doi:\nolinkurl{10.1145/3587102.3588785}}


\bibitem[Lobb and Harlow(2016)]%
        {LH16_CodeRunner}
\bibfield{author}{\bibinfo{person}{Richard Lobb} {and} \bibinfo{person}{Jenny Harlow}.} \bibinfo{year}{2016}\natexlab{}.
\newblock \showarticletitle{Coderunner: A Tool for Assessing Computer Programming Skills}.
\newblock \bibinfo{journal}{\emph{ACM Inroads}} \bibinfo{volume}{7}, \bibinfo{number}{1} (\bibinfo{date}{feb} \bibinfo{year}{2016}), \bibinfo{pages}{47–51}.
\newblock
\showISSN{2153-2184}
\href{https://doi.org/10.1145/2810041}{doi:\nolinkurl{10.1145/2810041}}


\bibitem[Maalej and Robillard(2013)]%
        {M13_Docs}
\bibfield{author}{\bibinfo{person}{Walid Maalej} {and} \bibinfo{person}{Martin~P Robillard}.} \bibinfo{year}{2013}\natexlab{}.
\newblock \showarticletitle{Patterns of knowledge in API reference documentation}.
\newblock \bibinfo{journal}{\emph{IEEE Transactions on software Engineering}} \bibinfo{volume}{39}, \bibinfo{number}{9} (\bibinfo{year}{2013}), \bibinfo{pages}{1264--1282}.
\newblock
\href{https://doi.org/10.1109/TSE.2013.12}{doi:\nolinkurl{10.1109/TSE.2013.12}}


\bibitem[Maalej et~al\mbox{.}(2014)]%
        {Maalej:TOSEM:2014}
\bibfield{author}{\bibinfo{person}{Walid Maalej}, \bibinfo{person}{Rebecca Tiarks}, \bibinfo{person}{Tobias Roehm}, {and} \bibinfo{person}{Rainer Koschke}.} \bibinfo{year}{2014}\natexlab{}.
\newblock \showarticletitle{On the Comprehension of Program Comprehension}.
\newblock \bibinfo{journal}{\emph{ACM Trans. Softw. Eng. Methodol.}} \bibinfo{volume}{23}, \bibinfo{number}{4}, Article \bibinfo{articleno}{31} (\bibinfo{date}{Sept.} \bibinfo{year}{2014}), \bibinfo{numpages}{37}~pages.
\newblock
\showISSN{1049-331X}
\href{https://doi.org/10.1145/2622669}{doi:\nolinkurl{10.1145/2622669}}


\bibitem[MacNeil et~al\mbox{.}(2023)]%
        {M23_Explanations}
\bibfield{author}{\bibinfo{person}{Stephen MacNeil}, \bibinfo{person}{Andrew Tran}, \bibinfo{person}{Arto Hellas}, \bibinfo{person}{Joanne Kim}, \bibinfo{person}{Sami Sarsa}, \bibinfo{person}{Paul Denny}, \bibinfo{person}{Seth Bernstein}, {and} \bibinfo{person}{Juho Leinonen}.} \bibinfo{year}{2023}\natexlab{}.
\newblock \showarticletitle{Experiences from using code explanations generated by large language models in a web software development e-book}. In \bibinfo{booktitle}{\emph{Proceedings of the 54th ACM Technical Symposium on Computer Science Education V. 1}}. \bibinfo{pages}{931--937}.
\newblock
\href{https://doi.org/10.1145/3545945.3569785}{doi:\nolinkurl{10.1145/3545945.3569785}}


\bibitem[MacNeil et~al\mbox{.}(2022)]%
        {M22_Explanations}
\bibfield{author}{\bibinfo{person}{Stephen MacNeil}, \bibinfo{person}{Andrew Tran}, \bibinfo{person}{Dan Mogil}, \bibinfo{person}{Seth Bernstein}, \bibinfo{person}{Erin Ross}, {and} \bibinfo{person}{Ziheng Huang}.} \bibinfo{year}{2022}\natexlab{}.
\newblock \showarticletitle{Generating diverse code explanations using the GPT-3 large language model}. In \bibinfo{booktitle}{\emph{Proceedings of the 2022 ACM Conference on International Computing Education Research-Volume 2}}. \bibinfo{pages}{37--39}.
\newblock
\href{https://doi.org/10.1145/3501709.3544280}{doi:\nolinkurl{10.1145/3501709.3544280}}


\bibitem[Mailach et~al\mbox{.}(2025)]%
        {M25_Patterns}
\bibfield{author}{\bibinfo{person}{Alina Mailach}, \bibinfo{person}{Dominik Gorgosch}, \bibinfo{person}{Norbert Siegmund}, {and} \bibinfo{person}{Janet Siegmund}.} \bibinfo{year}{2025}\natexlab{}.
\newblock \showarticletitle{“Ok Pal, we have to code that now”: interaction patterns of programming beginners with a conversational chatbot}.
\newblock \bibinfo{journal}{\emph{Empirical Software Engineering}} \bibinfo{volume}{30}, \bibinfo{number}{1} (\bibinfo{year}{2025}), \bibinfo{pages}{34}.
\newblock
\href{https://doi.org/10.1007/s10664-024-10561-6}{doi:\nolinkurl{10.1007/s10664-024-10561-6}}


\bibitem[Neuendorf(2017)]%
        {Neuendorf:2017}
\bibfield{author}{\bibinfo{person}{Kimberly~A. Neuendorf}.} \bibinfo{year}{2017}\natexlab{}.
\newblock \bibinfo{booktitle}{\emph{The Content Analysis Guidebook}}.
\newblock \bibinfo{address}{Thousand Oaks, California}.
\newblock
\href{https://doi.org/10.4135/9781071802878}{doi:\nolinkurl{10.4135/9781071802878}}


\bibitem[OpenAI(2022)]%
        {O22_ChatGPT}
\bibfield{author}{\bibinfo{person}{OpenAI}.} \bibinfo{year}{2022}\natexlab{}.
\newblock \bibinfo{title}{Introducing ChatGPT}.
\newblock \bibinfo{howpublished}{\url{https://openai.com/blog/chatgpt} (archived 2023-11-21: \url{https://web.archive.org/web/20231121185200/https://openai.com/blog/chatgpt/})}.
\newblock
\newblock
\shownote{Accessed 2023-11-22}.


\bibitem[Pham et~al\mbox{.}(2021)]%
        {Pham:RE:21}
\bibfield{author}{\bibinfo{person}{Yen~Dieu Pham}, \bibinfo{person}{Abir Bouraffa}, \bibinfo{person}{Marleen Hillen}, {and} \bibinfo{person}{Walid Maalej}.} \bibinfo{year}{2021}\natexlab{}.
\newblock \showarticletitle{The Role of Linguistic Relativity on the Identification of Sustainability Requirements: An Empirical Study}. In \bibinfo{booktitle}{\emph{2021 IEEE 29th International Requirements Engineering Conference (RE)}}. \bibinfo{pages}{117--127}.
\newblock
\href{https://doi.org/10.1109/RE51729.2021.00018}{doi:\nolinkurl{10.1109/RE51729.2021.00018}}


\bibitem[Popovici(2023)]%
        {P23_Classroom}
\bibfield{author}{\bibinfo{person}{Matei-Dan Popovici}.} \bibinfo{year}{2023}\natexlab{}.
\newblock \showarticletitle{ChatGPT in the Classroom. Exploring Its Potential and Limitations in a Functional Programming Course}.
\newblock \bibinfo{journal}{\emph{International Journal of Human--Computer Interaction}} (\bibinfo{year}{2023}), \bibinfo{pages}{1--12}.
\newblock
\href{https://doi.org/10.1080/10447318.2023.2269006}{doi:\nolinkurl{10.1080/10447318.2023.2269006}}


\bibitem[Prather et~al\mbox{.}(2023a)]%
        {P23_GAI4Ed}
\bibfield{author}{\bibinfo{person}{James Prather}, \bibinfo{person}{Paul Denny}, \bibinfo{person}{Juho Leinonen}, \bibinfo{person}{Brett~A. Becker}, \bibinfo{person}{Ibrahim Albluwi}, \bibinfo{person}{Michelle Craig}, \bibinfo{person}{Hieke Keuning}, \bibinfo{person}{Natalie Kiesler}, \bibinfo{person}{Tobias Kohn}, \bibinfo{person}{Andrew Luxton-Reilly}, \bibinfo{person}{Stephen MacNeil}, \bibinfo{person}{Andrew Petersen}, \bibinfo{person}{Raymond Pettit}, \bibinfo{person}{Brent~N. Reeves}, {and} \bibinfo{person}{Jaromir Savelka}.} \bibinfo{year}{2023}\natexlab{a}.
\newblock \showarticletitle{The Robots Are Here: Navigating the Generative AI Revolution in Computing Education}. In \bibinfo{booktitle}{\emph{Proceedings of the 2023 Working Group Reports on Innovation and Technology in Computer Science Education}} (Turku, Finland) \emph{(\bibinfo{series}{ITiCSE-WGR '23})}. \bibinfo{publisher}{Association for Computing Machinery}, \bibinfo{address}{New York, NY, USA}, \bibinfo{pages}{108–159}.
\newblock
\showISBNx{9798400704055}
\href{https://doi.org/10.1145/3623762.3633499}{doi:\nolinkurl{10.1145/3623762.3633499}}


\bibitem[Prather et~al\mbox{.}(2023b)]%
        {P23_Interactions}
\bibfield{author}{\bibinfo{person}{James Prather}, \bibinfo{person}{Brent~N. Reeves}, \bibinfo{person}{Paul Denny}, \bibinfo{person}{Brett~A. Becker}, \bibinfo{person}{Juho Leinonen}, \bibinfo{person}{Andrew Luxton-Reilly}, \bibinfo{person}{Garrett Powell}, \bibinfo{person}{James Finnie-Ansley}, {and} \bibinfo{person}{Eddie~Antonio Santos}.} \bibinfo{year}{2023}\natexlab{b}.
\newblock \showarticletitle{“It’s Weird That it Knows What I Want”: Usability and Interactions with Copilot for Novice Programmers}.
\newblock \bibinfo{journal}{\emph{ACM Transactions on Computer-Human Interaction}} (\bibinfo{date}{Aug.} \bibinfo{year}{2023}).
\newblock
\showISSN{1557-7325}
\href{https://doi.org/10.1145/3617367}{doi:\nolinkurl{10.1145/3617367}}


\bibitem[Prather et~al\mbox{.}(2024)]%
        {P24_Novices}
\bibfield{author}{\bibinfo{person}{James Prather}, \bibinfo{person}{Brent~N Reeves}, \bibinfo{person}{Juho Leinonen}, \bibinfo{person}{Stephen MacNeil}, \bibinfo{person}{Arisoa~S Randrianasolo}, \bibinfo{person}{Brett~A. Becker}, \bibinfo{person}{Bailey Kimmel}, \bibinfo{person}{Jared Wright}, {and} \bibinfo{person}{Ben Briggs}.} \bibinfo{year}{2024}\natexlab{}.
\newblock \showarticletitle{The Widening Gap: The Benefits and Harms of Generative AI for Novice Programmers}. In \bibinfo{booktitle}{\emph{Proceedings of the 2024 ACM Conference on International Computing Education Research - Volume 1}} (Melbourne, VIC, Australia) \emph{(\bibinfo{series}{ICER '24})}. \bibinfo{publisher}{Association for Computing Machinery}, \bibinfo{address}{New York, NY, USA}, \bibinfo{pages}{469–486}.
\newblock
\showISBNx{9798400704758}
\href{https://doi.org/10.1145/3632620.3671116}{doi:\nolinkurl{10.1145/3632620.3671116}}


\bibitem[Ratcliff and Metzener(1988)]%
        {R88_Matching}
\bibfield{author}{\bibinfo{person}{John~W. Ratcliff} {and} \bibinfo{person}{DM Metzener}.} \bibinfo{year}{1988}\natexlab{}.
\newblock \showarticletitle{Gestalt: an introduction to the Ratcliff/Obershelp pattern matching algorithm}.
\newblock \bibinfo{journal}{\emph{Dr. Dobbs Journal}}  \bibinfo{volume}{7} (\bibinfo{year}{1988}), \bibinfo{pages}{46}.
\newblock


\bibitem[Sarkar et~al\mbox{.}(2022)]%
        {S22_Programming}
\bibfield{author}{\bibinfo{person}{Advait Sarkar}, \bibinfo{person}{Andrew~D Gordon}, \bibinfo{person}{Carina Negreanu}, \bibinfo{person}{Christian Poelitz}, \bibinfo{person}{Sruti~Srinivasa Ragavan}, {and} \bibinfo{person}{Ben Zorn}.} \bibinfo{year}{2022}\natexlab{}.
\newblock \showarticletitle{What is it like to program with artificial intelligence?}
\newblock \bibinfo{journal}{\emph{arXiv preprint arXiv:2208.06213}} (\bibinfo{year}{2022}).
\newblock


\bibitem[Savelka et~al\mbox{.}(2023)]%
        {S23_Progress}
\bibfield{author}{\bibinfo{person}{Jaromir Savelka}, \bibinfo{person}{Arav Agarwal}, \bibinfo{person}{Marshall An}, \bibinfo{person}{Chris Bogart}, {and} \bibinfo{person}{Majd Sakr}.} \bibinfo{year}{2023}\natexlab{}.
\newblock \showarticletitle{Thrilled by Your Progress! Large Language Models (GPT-4) No Longer Struggle to Pass Assessments in Higher Education Programming Courses}. In \bibinfo{booktitle}{\emph{Proceedings of the 2023 ACM Conference on International Computing Education Research - Volume 1}} (Chicago, IL, USA) \emph{(\bibinfo{series}{ICER '23})}. \bibinfo{publisher}{Association for Computing Machinery}, \bibinfo{address}{New York, NY, USA}, \bibinfo{pages}{78–92}.
\newblock
\showISBNx{9781450399760}
\href{https://doi.org/10.1145/3568813.3600142}{doi:\nolinkurl{10.1145/3568813.3600142}}


\bibitem[Stanik et~al\mbox{.}(2018)]%
        {Stanik:ICSME:18}
\bibfield{author}{\bibinfo{person}{Christoph Stanik}, \bibinfo{person}{Lloyd Montgomery}, \bibinfo{person}{Daniel Martens}, \bibinfo{person}{Davide Fucci}, {and} \bibinfo{person}{Walid Maalej}.} \bibinfo{year}{2018}\natexlab{}.
\newblock \showarticletitle{A Simple NLP-Based Approach to Support Onboarding and Retention in Open Source Communities}. In \bibinfo{booktitle}{\emph{2018 IEEE International Conference on Software Maintenance and Evolution (ICSME)}}. \bibinfo{pages}{172--182}.
\newblock
\href{https://doi.org/10.1109/ICSME.2018.00027}{doi:\nolinkurl{10.1109/ICSME.2018.00027}}


\bibitem[Tian et~al\mbox{.}(2024)]%
        {T24_DebugBench}
\bibfield{author}{\bibinfo{person}{Runchu Tian}, \bibinfo{person}{Yining Ye}, \bibinfo{person}{Yujia Qin}, \bibinfo{person}{Xin Cong}, \bibinfo{person}{Yankai Lin}, \bibinfo{person}{Zhiyuan Liu}, {and} \bibinfo{person}{Maosong Sun}.} \bibinfo{year}{2024}\natexlab{}.
\newblock \showarticletitle{Debugbench: Evaluating debugging capability of large language models}.
\newblock \bibinfo{journal}{\emph{arXiv preprint}} (\bibinfo{year}{2024}).
\newblock
\href{https://doi.org/10.48550/arXiv.2401.04621}{doi:\nolinkurl{10.48550/arXiv.2401.04621}}


\bibitem[Vadaparty et~al\mbox{.}(2024)]%
        {V24_CS1}
\bibfield{author}{\bibinfo{person}{Annapurna Vadaparty}, \bibinfo{person}{Daniel Zingaro}, \bibinfo{person}{David~H. Smith~IV}, \bibinfo{person}{Mounika Padala}, \bibinfo{person}{Christine Alvarado}, \bibinfo{person}{Jamie Gorson~Benario}, {and} \bibinfo{person}{Leo Porter}.} \bibinfo{year}{2024}\natexlab{}.
\newblock \showarticletitle{CS1-LLM: Integrating LLMs into CS1 Instruction}. In \bibinfo{booktitle}{\emph{Proceedings of the 2024 on Innovation and Technology in Computer Science Education V. 1}} (Milan, Italy) \emph{(\bibinfo{series}{ITiCSE 2024})}. \bibinfo{publisher}{Association for Computing Machinery}, \bibinfo{address}{New York, NY, USA}, \bibinfo{pages}{297–303}.
\newblock
\showISBNx{9798400706004}
\href{https://doi.org/10.1145/3649217.3653584}{doi:\nolinkurl{10.1145/3649217.3653584}}


\bibitem[Vaithilingam et~al\mbox{.}(2022)]%
        {V22_Expectations}
\bibfield{author}{\bibinfo{person}{Priyan Vaithilingam}, \bibinfo{person}{Tianyi Zhang}, {and} \bibinfo{person}{Elena~L. Glassman}.} \bibinfo{year}{2022}\natexlab{}.
\newblock \showarticletitle{Expectation vs. Experience: Evaluating the Usability of Code Generation Tools Powered by Large Language Models}. In \bibinfo{booktitle}{\emph{Extended Abstracts of the 2022 CHI Conference on Human Factors in Computing Systems}} (New Orleans, LA, USA) \emph{(\bibinfo{series}{CHI EA '22})}. \bibinfo{publisher}{Association for Computing Machinery}, \bibinfo{address}{New York, NY, USA}, Article \bibinfo{articleno}{332}, \bibinfo{numpages}{7}~pages.
\newblock
\showISBNx{9781450391566}
\href{https://doi.org/10.1145/3491101.3519665}{doi:\nolinkurl{10.1145/3491101.3519665}}


\bibitem[Wei et~al\mbox{.}(2024)]%
        {Wei:2024}
\bibfield{author}{\bibinfo{person}{Jialiang Wei}, \bibinfo{person}{Anne-Lise Courbis}, \bibinfo{person}{Thomas Lambolais}, \bibinfo{person}{Gérard Dray}, {and} \bibinfo{person}{Walid Maalej}.} \bibinfo{year}{2024}\natexlab{}.
\newblock \bibinfo{title}{On AI-Inspired UI-Design}.
\newblock
\href{https://doi.org/10.48550/arXiv.2406.13631}{doi:\nolinkurl{10.48550/arXiv.2406.13631}}
\showeprint[arxiv]{2406.13631}~[cs.HC]


\bibitem[Wu et~al\mbox{.}(2017a)]%
        {W17_NETLAP}
\bibfield{author}{\bibinfo{person}{Youxi Wu}, \bibinfo{person}{Cong Shen}, \bibinfo{person}{He Jiang}, {and} \bibinfo{person}{Xindong Wu}.} \bibinfo{year}{2017}\natexlab{a}.
\newblock \showarticletitle{Strict pattern matching under non-overlapping condition}.
\newblock \bibinfo{journal}{\emph{Science China. Information Sciences}} \bibinfo{volume}{60}, \bibinfo{number}{1} (\bibinfo{year}{2017}), \bibinfo{pages}{012101}.
\newblock


\bibitem[Wu et~al\mbox{.}(2017b)]%
        {W17_NOSEP}
\bibfield{author}{\bibinfo{person}{Youxi Wu}, \bibinfo{person}{Yao Tong}, \bibinfo{person}{Xingquan Zhu}, {and} \bibinfo{person}{Xindong Wu}.} \bibinfo{year}{2017}\natexlab{b}.
\newblock \showarticletitle{NOSEP: Nonoverlapping sequence pattern mining with gap constraints}.
\newblock \bibinfo{journal}{\emph{IEEE transactions on cybernetics}} \bibinfo{volume}{48}, \bibinfo{number}{10} (\bibinfo{year}{2017}), \bibinfo{pages}{2809--2822}.
\newblock
\href{https://doi.org/10.1109/TCYB.2017.2750691}{doi:\nolinkurl{10.1109/TCYB.2017.2750691}}


\bibitem[Xiao et~al\mbox{.}(2024)]%
        {X24_DevGPT}
\bibfield{author}{\bibinfo{person}{Tao Xiao}, \bibinfo{person}{Christoph Treude}, \bibinfo{person}{Hideaki Hata}, {and} \bibinfo{person}{Kenichi Matsumoto}.} \bibinfo{year}{2024}\natexlab{}.
\newblock \showarticletitle{DevGPT: Studying Developer-ChatGPT Conversations}. In \bibinfo{booktitle}{\emph{2024 IEEE/ACM 21st International Conference on Mining Software Repositories (MSR)}}. IEEE, \bibinfo{pages}{227--230}.
\newblock


\bibitem[Xue et~al\mbox{.}(2024)]%
        {X24_Intro}
\bibfield{author}{\bibinfo{person}{Yuankai Xue}, \bibinfo{person}{Hanlin Chen}, \bibinfo{person}{Gina~R. Bai}, \bibinfo{person}{Robert Tairas}, {and} \bibinfo{person}{Yu Huang}.} \bibinfo{year}{2024}\natexlab{}.
\newblock \showarticletitle{Does ChatGPT Help With Introductory Programming?An Experiment of Students Using ChatGPT in CS1}. In \bibinfo{booktitle}{\emph{Proceedings of the 46th International Conference on Software Engineering: Software Engineering Education and Training}} (Lisbon, Portugal) \emph{(\bibinfo{series}{ICSE-SEET '24})}. \bibinfo{publisher}{Association for Computing Machinery}, \bibinfo{address}{New York, NY, USA}, \bibinfo{pages}{331–341}.
\newblock
\showISBNx{9798400704987}
\href{https://doi.org/10.1145/3639474.3640076}{doi:\nolinkurl{10.1145/3639474.3640076}}


\bibitem[Zastudil et~al\mbox{.}(2023)]%
        {Z23_GAI4Ed}
\bibfield{author}{\bibinfo{person}{Cynthia Zastudil}, \bibinfo{person}{Magdalena Rogalska}, \bibinfo{person}{Christine Kapp}, \bibinfo{person}{Jennifer Vaughn}, {and} \bibinfo{person}{Stephen MacNeil}.} \bibinfo{year}{2023}\natexlab{}.
\newblock \showarticletitle{Generative AI in Computing Education: Perspectives of Students and Instructors}. In \bibinfo{booktitle}{\emph{2023 IEEE Frontiers in Education Conference (FIE)}}. \bibinfo{pages}{1--9}.
\newblock
\href{https://doi.org/10.1109/FIE58773.2023.10343467}{doi:\nolinkurl{10.1109/FIE58773.2023.10343467}}


\bibitem[Ziegler et~al\mbox{.}(2024)]%
        {Z24_Productivity}
\bibfield{author}{\bibinfo{person}{Albert Ziegler}, \bibinfo{person}{Eirini Kalliamvakou}, \bibinfo{person}{X.~Alice Li}, \bibinfo{person}{Andrew Rice}, \bibinfo{person}{Devon Rifkin}, \bibinfo{person}{Shawn Simister}, \bibinfo{person}{Ganesh Sittampalam}, {and} \bibinfo{person}{Edward Aftandilian}.} \bibinfo{year}{2024}\natexlab{}.
\newblock \showarticletitle{Measuring GitHub Copilot's Impact on Productivity}.
\newblock \bibinfo{journal}{\emph{Commun. ACM}} \bibinfo{volume}{67}, \bibinfo{number}{3} (\bibinfo{date}{Feb.} \bibinfo{year}{2024}), \bibinfo{pages}{54–63}.
\newblock
\showISSN{0001-0782}
\href{https://doi.org/10.1145/3633453}{doi:\nolinkurl{10.1145/3633453}}


\end{thebibliography}

\appendix

\end{document}